\documentclass[aps,pre,epsf,superscriptaddress,amsmath,amssymb,amsfonts,twocolumn,showpacs]{revtex4-1}
\usepackage[english,american]{babel}
\usepackage[T1]{fontenc}
\usepackage{lmodern}    
\usepackage{amsmath}    
\usepackage{mathrsfs}
\usepackage{graphicx}   
\usepackage{verbatim}   
\usepackage{subfigure}  
\usepackage{hyperref}   

\usepackage{epsfig}
\usepackage{dcolumn}
\usepackage{bm}
\usepackage{braket}
\usepackage{mathtools}
\usepackage{color}

\newcommand{\abs}[1]{\left| #1 \right|}

\begin{document}

\title{Correlated dynamics of fermionic impurities induced by the\\ counterflow of an ensemble of fermions}

\author{J. Kwasniok}
\author{S.I. Mistakidis}
\affiliation{Center for Optical Quantum Technologies, Department of Physics, University of Hamburg, 
Luruper Chaussee 149, 22761 Hamburg Germany}
\author{P. Schmelcher}
\affiliation{Center for Optical Quantum Technologies, Department of Physics, University of Hamburg, 
Luruper Chaussee 149, 22761 Hamburg Germany}
\affiliation{The Hamburg Centre for Ultrafast Imaging, University of Hamburg, Luruper Chaussee 149, 22761 Hamburg, Germany} 

\date{\today}

\begin{abstract} 

We investigate the nonequilibrium quantum dynamics of a single and two heavy fermionic impurities being harmonically trapped and repulsively interacting with 
a finite ensemble of majority fermions.  
A quench of the potential of the majority species from a double-well to a harmonic trap is applied, enforcing its counterflow which in turn perturbs the impurities. 
For weak repulsions it is shown that the mixture undergoes a periodic mixing and demixing dynamics, while stronger interactions lead to a more pronounced 
dynamical spatial separation. 
In the presence of correlations the impurity exhibits an expansion dynamics which is absent in the Hartree-Fock case resulting in an enhanced degree of miscibility. 
We generalize our results to different impurity masses and demonstrate that the expansion amplitude of the impurity reduces for a larger mass. 
Furthermore, we showcase that the majority species is strongly correlated and a phase separation occurs on the two-body level. 
Most importantly, signatures of attractive impurity-impurity induced interactions mediated by the majority species are identified in the time-evolution of the two-body 
correlations of the impurities, a result that is supported by inspecting their spatial size. 

\end{abstract}
	
\maketitle

\section{Introduction} \label{sec:introduction}

Ultracold atoms represent a unique platform for realizing and subsequently probing the properties of multicomponent systems such as Bose-Bose \cite{catani2008,thalhammer2008,petrov2015}, 
Bose-Fermi \cite{stan2004,ospelkaus2006,ahufinger2005} and Fermi-Fermi (FF) mixtures \cite{wille2008,kohl2005,hadzibabic2002}. 
Several system parameters are experimentally tunable such as the interparticle interaction strength with the aid of Feshbach 
resonances \cite{chin2010,inouye1998} and the external potential landscape \cite{greiner2002,bloch2008}. 
It is possible to trap with a high precision a desired number of particles \cite{wenz2013,serwane2011,zurn2012} especially in one spatial dimension. 
Lately, a focal point of studies has been the investigation of highly particle imbalanced systems \cite{Wu2012,Heo2012,Roati2007,Schirotzek2009,kohstall2012metastability,Koschorreck2012,zhang2012polaron,scazza2017} 
consisting of impurities immersed in a many-body environment. 
Consequently, the impurities are dressed by the excitations of their host forming quasiparticles \cite{Landau1933,Frohlich1954} 
such as polarons \cite{schmidt2018universal,Massignan2014} whose properties, e.g. their effective mass \cite{Khandekar1988}, 
mobility \cite{Feynman1962,schmidt2018universal} and induced interactions \cite{Massignan2014} are altered compared to the bare particle case. 
Most importantly, both Bose \cite{Nils2016,Hu2016,catani2009entropy,fukuhara2013quantum} and Fermi polarons \cite{scazza2017repulsive,Koschorreck2012,kohstall2012metastability} have been 
experimentally realized and a variety of their characteristics such as their excitation spectrum have been probed \cite{Koschorreck2012,kohstall2012metastability,cetina2015decoherence,cetina2016ultrafast}. 
These experimental findings paved the way for a multitude of theoretical investigations in order to understand first the stationary properties of these quasiparticles 
\cite{volosniev2017analytical,dehkharghani2018coalescence,Mistakidis2019,Nils2016,Ardila2018,ardila2015impurity,ardila2019analyzing,
Grusdt2018,Grusdt2017,grusdt2015new,Tempere2009,panochko2019two} and very recently their nonequilibrium dynamics \cite{mistakidis2019effective,mistakidis2019quench}. 

A crucial ingredient for the adequate description of these systems, especially during their nonequilibrium dynamics, is the involvement of interparticle correlations 
\cite{volosniev2015real,mistakidis2019quench,mistakidis2019correlated,mistakidis2019effective,Grusdt2018,shchadilova2016quantum,kamar2019dynamics,boyanovsky2019dynamics,li2019controlling,pasek2019induced}. 
The impurities are indeed few-body systems and correlation effects become well-pronounced. 
Central phenomena emerging in the dynamics of impurities include non-linear structure formation \cite{grusdt2017bose,mistakidis2019correlated}, 
their orthogonality catastrophe \cite{mistakidis2019quench,knap2012time,Anderson1967}, dissipation and relaxation dynamics 
\cite{mistakidis2019dissipative, boyanovsky2019dynamics,lausch2018prethermalization}, collisions with their host 
\cite{burovski2014momentum,lychkovskiy2018necessary,meinert2017bloch,knap2014quantum,gamayun2018impact} and their transport 
properties in lattice systems \cite{cai2010interaction,Johnson2011,Siegl2018,theel2020entanglement}. 
Most importantly, besides the aforementioned studies and first applications, the dynamics of impurities and their accompanying many-body effects are still far from being completely 
understood especially in the case of more than a single impurity. 

A main focus of the previous investigations regarding the nonequilibrium dynamics of impurities has been the scenario where an external perturbation is applied 
either on the entire system or solely on the impurity subsystem. 
Then, the resultant time-evolution of both the impurity and its environment is monitored. 
However, a largely unexplored direction is to perturb only the bath and inspect the dynamical response of the impurities which are indirectly perturbed. 
As a model system, we consider one or two heavy fermionic impurities harmonically trapped being immersed in a one-dimensional Fermi sea 
that is in turn confined in a double-well. 
The mixture is initialized to its ground state configuration and the dynamics is induced by ramping down the potential barrier of the double-well. 
This scheme enforces a counterflow of the Fermi sea which then perturbs the impurities. 
Note that very recently spin charge separation processes have been reported for a similar setting but restricted to the single impurity case while quenching the 
double-well of both components \cite{barfknecht2019dynamics}. 

An imperative prospect here is to unravel the dynamical response of the impurities for different impurity-bath interactions in order to inspect how the environment 
mediates its perturbation depending on the number of the impurities. 
Especially the role of induced impurity-impurity interactions \cite{dehkharghani2018coalescence} caused by the Fermi sea is of immediate interest. 
Recall that two fermionic impurities do not exhibit direct $s$-wave impurity-impurity interactions since they are spin-polarized. 
Furthermore, dynamical phase separation processes \cite{Mistakidis2018,erdmann2019phase,pecak2016two} and possible structure formation emerging in the bath 
are relevant problems. 
Such a counterflow quench-induced process has been widely used in bosonic setups in order to generate dark and dark-bright solitary waves \cite{hamner2011generation,weller2008experimental,kevrekidis2007emergent}. 
To address the quench dynamics of the impurities we rely on the Multi-Layer Multi-Configuration Time-Dependent Hartree Method for atomic mixtures 
(ML-MCTDHX)~\cite{cao2017unified,cao2013multi,kronke2013non} which is a variational method allowing us to capture all the relevant interparticle 
correlations of the FF mixture. 
The nonequilibrium dynamics is studied for a wide range of repulsive interspecies interactions both within the Hartree-Fock (HF) and in the many-body 
ML-MCTDHX theoretical framework. 

We find that the dynamical behavior of the systems including a single or two fermionic impurities is qualitatively similar on the one-body density level with 
each species performing a breathing motion.
For weak repulsions it is shown that in the HF approach the mixture undergoes a periodic mixing and demixing dynamics \cite{mistakidis2019quench,erdmann2019phase,Mistakidis2018}. 
Increasing the interspecies repulsion and both in the HF and the many-body scenaria we unveil a pronounced dynamical spatial interspecies separation. 
Importantly in the presence of correlations the impurity experiences a predominant expansion, which is absent in the HF case, resulting in a smaller degree of phase separation between 
the species. 
The density of the Fermi sea develops fragmented patterns which are found to be shallower in the correlated many-body case.  

Additionally, we show that the expansion amplitude of the impurity reduces for larger masses both for a single and two impurities. 
However, the response of the Fermi sea is impacted noticeably only for the two impurity case, showing a reduced expansion amplitude for lighter impurities \cite{Siegl2018}. 
Furthermore, we showcase that the Fermi sea is strongly correlated and a phase separation occurs on the two-body level. 
Most importantly, signatures of attractive impurity-impurity induced interactions mediated by the Fermi sea are identified in the time-evolution of their two-body correlation function \cite{mistakidis2019many} and by comparing the spatial size of two and one impurity atoms \cite{mistakidis2019induced}. 

This work is structured as follows. 
Section \ref{sec:theoretical-framework} explicates our setup and quench protocol, the many-body ansatz and the basic observables 
of interest used to interpret the dynamics. 
Subsequently, we discuss in detail the quench-induced dynamics for a single [Sec. \ref{sec:one_impurity}] and two fermionic 
impurities [Sec. \ref{sec:two_impurities}] repulsively interacting with the Fermi sea. 
We summarize our results giving an outlook and discussing future extensions in Section \ref{sec:conclusions}. 
In Appendix \ref{barrier_width} we exemplify the effect of the width of the double-well barrier on the dynamics. 
Appendix \ref{methodology_convergence} elaborates on the details of the many-body simulations 
and demonstrates their convergence.

\section{Theoretical Framework} \label{sec:theoretical-framework}

\subsection{Setup and quench protocol} \label{sec:setup} 

\begin{figure}[ht]
\includegraphics[width=0.45\textwidth]{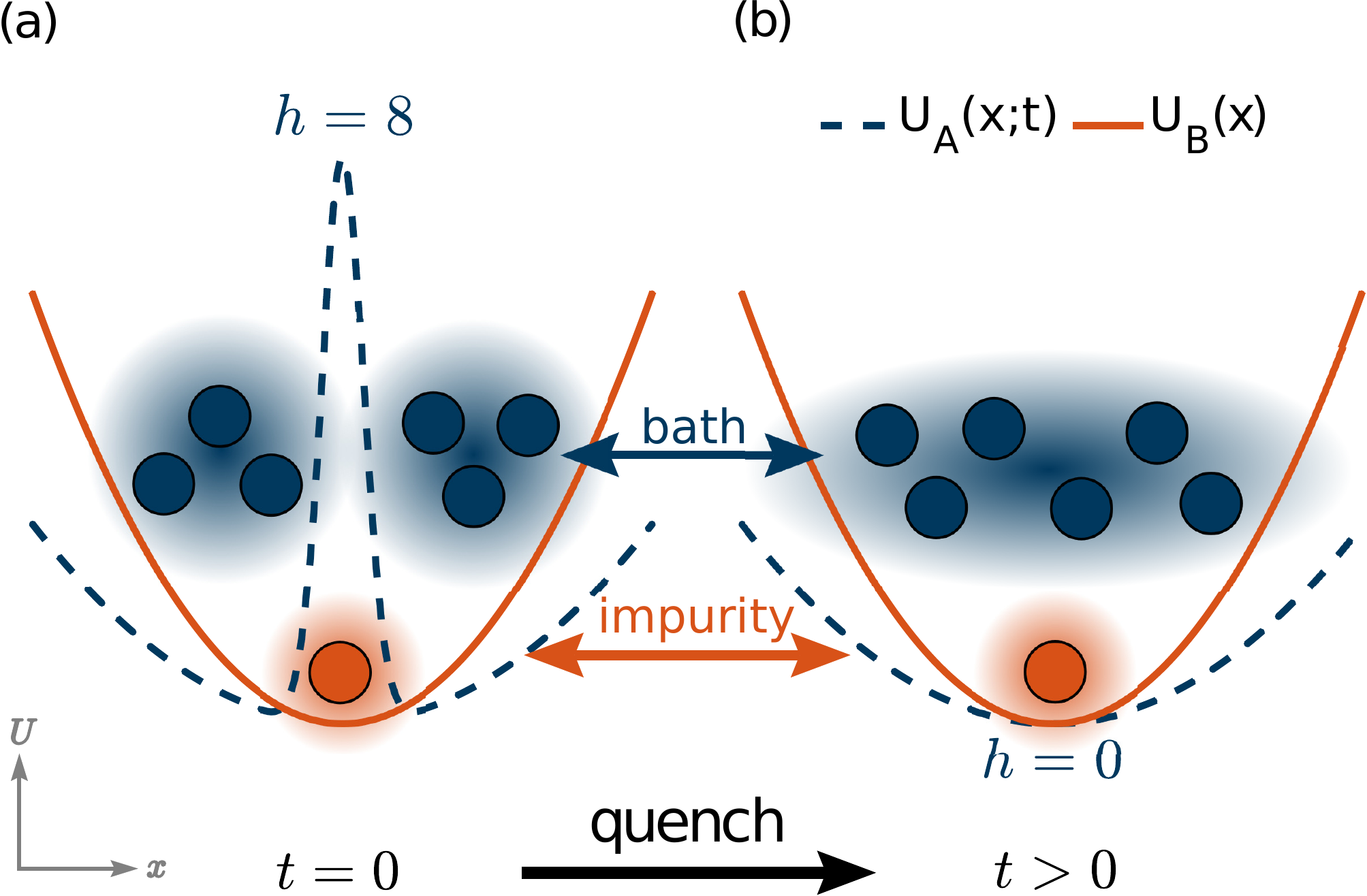}
\caption{Schematic representation of the setup and the considered quench protocol.
Here, a single impurity of mass $M_B$ (light orange circle) resides in a harmonic oscillator potential $U_B(x)=(1/2)M_B\omega_B^2(x^B)^2$ of frequency $\omega_B$ and it is immersed in a Fermi sea of 
six spin-polarized fermions with mass $M_A=(1/6)M_B$ (dark blue circles). 
The Fermi sea is initially trapped in a double-well $U_A(x;t=0)=(1/2)M_A\omega_A^2(x^A)^2 
+h/(w \sqrt{2\pi}) e^{-(x^A)^2/(2w^2)}$ composed of a harmonic trap with frequency $\omega_A$ and a Gaussian barrier of height 
$h$ and width $w$. 
To trigger the dynamics the Gaussian barrier is suddenly switched-off at $t=0$ and the system is left to evolve under the influence of $U_A(x;t)=(1/2)M_A\omega_A^2(x^A)^2$.}
\label{fig:setup}
\end{figure}

We examine a particle- and mass-imbalanced FF mixture consisting of either $N_B=1$ [Sec. \ref{sec:one_impurity}] or 
$N_B=2$ [Sec. \ref{sec:two_impurities}] impurity atoms immersed in a fermionic sea of $N_A=6$ particles. 
Each species is composed of spin-polarized fermions.
Furthermore the mass ratio between the species is $M_B=6M_A$, unless it is stated otherwise, which is experimentally realizable 
e.g. by a mixture of isotopes of $^{40}$K and $^{6}$Li \cite{wille2008,cetina2016}. 
The mixture is further confined in a one-dimensional species selective external potential, a scenario that can be 
experimentally achieved using optical trapping \cite{grimm2000,cetina2016}. 
Indeed, the different isotopes exhibit distinct polarizations and therefore experience different potentials. 
In particular, the majority species (Fermi sea) is trapped in a double-well composed of a harmonic oscillator with frequency $\omega_A$ and a centered Gaussian barrier 
characterized by a height $h$ and width $w$.
The impurity atoms experience a harmonic oscillator of frequency $\omega_B=0.6\omega_A$ [see Fig.~\ref{fig:setup} (a)]. 
The many-body Hamiltonian of this system reads 
\begin{equation}
\begin{split} 
H&=\sum\limits_{\sigma=A,B} \sum\limits_{i=1}^{N_\sigma}\left[  -\frac{\hbar^2}{2M_\sigma}\left( \frac{\partial}{\partial x_i^\sigma}\right)^2
+\frac{1}{2}M_\sigma\omega_\sigma^2(x_i^\sigma)^2 \right] 
\\&+\sum\limits_{i=1}^{N_A}\frac{h}{w \sqrt{2\pi}} e^{-\frac{(x_i^A)^2 }{2w^2}} + g_{AB} \sum\limits_{i=1}^{N_A} \sum \limits_{j=1}^{N_B}\delta(x_i^{A}-x_j^{B}).
\label{eq:hamilt} 
\end{split}
\end{equation} 
The interspecies interactions are modeled by contact interactions scaling with the effective one-dimensional coupling strength $g_{AB}$. 
The latter is an adequate approximation since we operate within the ultracold regime where $s$-wave scattering constitutes the predominant interaction process. 
Note also that $s$-wave scattering is forbidden for spin-polarized fermions \cite{pethick2008,pitaevskii2003,lewenstein2012} and hence intraspecies interactions are 
neglected. 
Moreover, the effective interspecies one-dimensional coupling strength \cite{Olshanii1998} possesses the form   
${g_{AB}} =\frac{{2{\hbar ^2}{a^s_{AB}}}}{{\mu a_ \bot ^2}}{\left( {1 - {\left|{\zeta (1/2)} \right|{a^s_{AB}}}/{{\sqrt 2 {a_ \bot }}}} \right)^{ -
1}}$. 
Here, $\mu=\frac{M_AM_B}{M_A+M_B}$ refers to the reduced mass, $\zeta$ is the Riemann zeta function and ${a^s_{AB}}$ denotes the three-dimensional $s$-wave 
interspecies scattering length. 
Also, ${a_ \bot } = \sqrt{\hbar /{\mu{\omega _ \bot }}}$ is the transversal length scale with ${{\omega _ \bot }}$ being the transversal confinement frequency. 
The interspecies interaction strength $g_{AB}$ can be experimentally adjusted either via ${a^s_{AB}}$ with the aid of Feshbach resonances \cite{kohler2006,chin2010} 
or by means of ${{\omega _ \bot }}$ utilizing the corresponding confinement-induced resonances \cite{Olshanii1998, kim2006}. 

In this work we employ a longitudinal trapping frequency $\omega_A=0.1\approx 2\pi \times 20 Hz$. 
To restrict the dynamics to one-dimension one can e.g. use a typical for one-dimensional experiments transversal confinement frequency 
being of the order of $\omega_{\perp}\approx 2\pi \times 200 Hz$ \cite{bersano2018three,serwane2011,wenz2013,katsimiga2020observation}. 
In this way, it is possible to adequately neglect the dynamics in the perpendicular directions which might be non-negligible when higher-lying excitations participate in the dynamics. 
Indeed, if the aspect ratio of the longitudinal and the perpendicular confinement frequencies is not sufficiently large higher-lying excitations will 
disturb the one-dimensional description. 
This phenomenon is not encountered in the present investigation. 
Additionally, we remark that since few-body systems exhibit low-densities the contribution of incoherent processes e.g. three-body recombination are drastically reduced, a fact 
that leads to increased coherence times.  
In this sense, it is reasonable to consider a coherent evolution of our system. 

For our purposes the many-body Hamiltonian (\ref{eq:hamilt}) is rescaled in units of $\hbar  \omega_{\perp}$. 
As a result the length, time, and interaction strength scales are expressed in units of $\sqrt{\frac{\hbar}{M_{A} \omega_{\perp}}}$, 
$\omega_{\perp}^{-1}$ and $\sqrt{\frac{\hbar^3 \omega_{\perp}}{M_{A}}}$ respectively. 
The corresponding amplitude of the Gaussian barrier $h$ and its width $w$ are provided in terms of $\sqrt{ \frac{ \hbar^3 \omega_{\perp} }{ M_{A} }}$ and $\sqrt{\frac{\hbar}{M_{A} \omega_{\perp}}}$. 
Furthermore, for computational convenience the spatial extension of the system is restricted to a finite one by imposing hard-wall boundary conditions 
at $x_\pm=\pm40$. 
This choice does not affect our results since for all numerical simulations the dynamics of e.g. the $\sigma$-species single-particle densities takes 
place within the spatial interval $-15<x<15$. 

The system is initialized in its interspecies interacting ground state of the composite external potential described by the Hamiltonian (\ref{eq:hamilt}), 
see also Fig. \ref{fig:setup} (a). 
Numerically, in order to find the ground state of the system we employ the imaginary propagation or the improved relaxation method within ML-MCTDHX \cite{cao2017unified}. 
Note that throughout this work we consider a height $h=8$ and width $w=1$ of the Gaussian barrier for the fermionic sea, thus having 30 doublets 
below the maximum of the barrier. 
Depending on $g_{AB}$, the emergent ground state of the mixture is characterized by two symmetrically placed fragments of the Fermi sea around $x=0$ while the impurity atoms 
reside at their trap minimum $x=0$, see also Fig. \ref{fig:setup} (a). 
Most importantly, the two species remain spatially separated (immiscible) independently of the value of $g_{AB}$. 

To induce the nonequilibrium dynamics of the FF mixture at $t=0$ we suddenly ramp-down the Gaussian barrier of the Fermi sea from $h=8$ to $h=0$ and let 
the system evolve in time [see Fig.~\ref{fig:setup} (b)]. 
Quenching the barrier height of the Fermi sea to zero triggers a counterflow between its initially separated fragments favoring an overall breathing motion of this species 
and a spatial overlap between the different species [Fig. \ref{fig:setup} (b)]. 
Consequently, the impurity atoms couple with the quench-induced excitations of the fermionic sea and form quasiparticles known as Fermi polarons \cite{Mistakidis2019,tylutki2017,Massignan2014,schmidt2018universal}. 
The aim of the present work is to shed light on the complex motion of a single [Sec. \ref{sec:one_impurity}] and two impurities [Sec. \ref{sec:two_impurities}] coupled to the 
fermionic sea for different interspecies interaction strengths within the interval $g_{AB} \in [0,4]$. 
Additionally, we shall examine the dependence on the mass ratio between the species and expose the emergent correlation effects with a special focus on 
impurity-impurity induced correlations mediated by the Fermi sea in the case of two fermionic impurities.

\subsection{Many-body approach} \label{sec:ml-x}	

To simulate the quench dynamics of the FF mixture we solve the many-body Schr{\"o}dinger equation within ML-MCTDHX \cite{cao2017unified,cao2013multi,kronke2013non}. 
This method rests on the expansion of the MB wavefunction in terms of a time-dependent and variationally optimized basis, enabling us 
to take into account both the inter- and intraspecies correlations. 
Indeed, the FF mixture is a bipartite composite system whose Hilbert space is $\mathcal{H}^{AB}=\mathcal{H}^A\otimes\mathcal{H}^B$,  
where $\mathcal{H}^{\sigma}$ denotes the individual $\sigma$-species Hilbert space. 
In order to incorporate interspecies correlations we employ $D$ distinct species functions, $\Psi^{\sigma}_k (\vec x^{\sigma};t)$, for 
each species consisting of $N_{\sigma}$ fermions. 
Note that $\vec x^{\sigma}=\left( x^{\sigma}_1, \dots, x^{\sigma}_{N_{\sigma}} \right)$ denote the spatial $\sigma=A,B$ species coordinates. 
Consequently, the many-body wavefunction, $\Psi_{MB}$, acquires the form of a truncated Schmidt decomposition \cite{horodecki2009} of 
rank $D$ 
\begin{equation}
\Psi_{MB}(\vec x^A,\vec x^B;t) = \sum_{k=1}^D \sqrt{ \lambda_k(t) }~ \Psi^A_k (\vec x^A;t) \Psi^B_k (\vec x^B;t).    
\label{eq:WF}
\end{equation} 
In this expression, $\{\Psi_k^{\sigma}\}$ forms an orthonormal $N_{\sigma}$-body wavefunction set in a subspace 
of $\mathcal{H}^{\sigma}$. 
Additionally, the Schmidt coefficients $\lambda_k(t)$ in decreasing order are often referred to as natural species populations 
of the $k$-th species function $\Psi^{\sigma}_k$ of the $\sigma$-species. 
Most importantly, the system is called entangled or interspecies correlated \cite{roncaglia2014,horodecki2009} if more than a single coefficients $\lambda_k(t)$ are significantly populated. 
In this case, the many-body state [Eq. (\ref{eq:WF})] is a complex superposition and not a direct product of two states stemming from $\mathcal{H}^A$ and $\mathcal{H}^B$. 

As a next step, in order to take into account the intraspecies correlations into our many-body ansatz we further express each of the 
species functions $\Psi^{\sigma}_k (\vec x^{\sigma};t)$ with respect to the determinants of different $d_{\sigma}$ time-dependent 
fermionic single-particle functions (SPFs), $\varphi_1,\dots,\varphi_{d_{\sigma}}$. 
Namely 
\begin{equation}
\begin{split}
&\Psi_k^{\sigma}(\vec x^{\sigma};t) = \sum_{\substack{l_1,\dots,l_{d_{\sigma}} \in \{0,1\} \\ \rm{with} \sum l_i=N_\sigma}} C_{k,(l_1,
	\dots,l_{d_{\sigma}})}(t)\\& \sum_{i=1}^{N_{\sigma}!} {\rm sign}(\mathcal{P}_i) \mathcal{P}_i
 \bigg[ \prod_{\substack{j\in\{1,\dots,d_{\sigma}\} \\ {\rm with}~ l_j=1}} \varphi_j(x_{K(j)};t)\bigg].  
\label{eq:SPFs}
\end{split}
\end{equation} 
Here, $C_{k,(l_1,\dots,l_{m_{\sigma}})}(t)$ are the time-dependent expansion coefficients of a certain determinant while $\mathcal{P}$ 
is the permutation operator exchanging the particle positions $x_{s}$, $s=1,\dots,N_{\sigma}$ within the SPFs. 
$\rm{sign}(\mathcal{P}_i)$ provides the sign of the corresponding permutation and 
$K(j)\equiv \sum_{\nu=1}^{j}l_{\nu}$, with $l_{\nu}$ being the occupation of the $\nu$th SPF. 
The eigenfunctions and eigenvalues of the $\sigma$-species single-particle reduced density matrix 
$\rho_\sigma^{(1)}(x,x';t)=\langle\Psi_{MB}(t)|\hat{\Psi}^{\sigma \dagger}(x)\hat{\Psi}^\sigma(x')|\Psi_{MB}(t)\rangle$ 
are the so-called natural orbitals $\phi^{\sigma}_i(x;t)$ and natural populations $n^{\sigma}_i(t)$ respectively. 
Here, $\hat{\Psi}^{\sigma}(x)$ [$\hat{\Psi}^{\sigma \dagger}(x)$] is the fermionic field operator that annihilates [creates] a 
$\sigma$-species fermion located at position $x$. 
Let us note in passing that the diagonal of the $\sigma$-species single-particle reduced density matrix corresponds to the 
$\sigma$-species one-body density i.e. $\rho_\sigma^{(1)}(x,x'=x;t)=\rho_\sigma^{(1)}(x,;t)$ which is an experimentally 
tractable quantity via {\it in-situ} imaging \cite{fukuhara2013quantum}. 
The $\sigma$-species is said to be intraspecies correlated when more than $N_\sigma$ natural populations are macroscopically 
occupied namely $0<n_i^{\sigma}(t)<1$ with $N_{\sigma}<i<d_{\sigma}$. 
However, in the case that only $N_{\sigma}$ natural populations are contributing such that $\sum_{i=1}^{N_{\sigma}}n_i^{\sigma}(t)=N_{\sigma}$ the many-body wavefunction 
ansatz reduces to the corresponding HF one \cite{pethick2008,pitaevskii2003,giorgini2008}, i.e.  
\begin{equation}
\begin{split}
&\Psi_{HF}(\vec x^{A},\vec x^{B};t) =
	\\&\prod_{\sigma=A,B}\sum_{i=1}^{N_{\sigma}!} {\rm sign}(\mathcal{P}_i) \mathcal{P}_i
\left[\varphi_1(x_1^{\sigma};t) \cdots \varphi_{N_{\sigma}}(x_{N_{\sigma}}^{\sigma};t) \right].  
\label{eq:HF}
\end{split}
\end{equation} 
Indeed in the limit of $D=1$ and $d_{\sigma}=N_{\sigma}$ from Eqs. (\ref{eq:WF}) and (\ref{eq:SPFs}) we retrieve the HF ansatz 
i.e. $\Psi_{MB}(\vec x^A,\vec x^B;t) \to \Psi_{HF}(\vec x^{A},\vec x^{B};t)$. 
To calculate the ML-MCTDHX equations of motion \cite{cao2017unified,kohler2019dynamical} for the FF mixture we employ the Dirac-Frenkel variational principle 
\cite{frenkel1932,dirac1930} for the many-body ansatz as introduced in Eqs.~(\ref{eq:WF}), (\ref{eq:SPFs}).   
In doing so, we obtain a set of $D^2$ linear differential equations of motion for the coefficients $\lambda_k(t)$ coupled to 
$D$[${d_A}\choose{N_A}$+$ {d_B}\choose{N_B}$] non-linear integro-differential equations for the species functions and $d_A+d_B$  
integro-differential equations for the SPFs. 
These constitute the so-called ML-MCTDHF equations of motion \cite{cao2017unified,Koutentakis2019}. 

\subsection{Main observables of interest} \label{sec:observable-of-interest} 

To unravel the role of two-body intra- and interspecies correlations during the nonequilibrium dynamics of the FF mixture 
in a spatially resolved manner, we resort to the second order coherence function \cite{sakmann2008,naraschewski1999,Mistakidis2018}
\begin{align}
G^{(2)}_{\sigma\sigma\prime}(x,x^\prime;t)=\frac{\rho_{\sigma\sigma\prime}^{(2)}(x,x^\prime;t)}{\rho_\sigma^{(1)}(x;t)\rho_{\sigma\prime}^{(1)}(x^\prime;t)}. \label{two_body_cor}
\end{align}
In this expression, $\rho^{(2)}(x,x';t)=\langle\Psi(t)|\hat{\Psi}^{\sigma\dagger}(x')\hat{\Psi}^{\dagger\sigma^\prime}(x)\hat{\Psi}^{\sigma^\prime}
(x)\hat{\Psi}^\sigma(x')|\Psi(t)\rangle$ is the diagonal two-body reduced density matrix. 
This quantity provides the probability of measuring simultaneously two particles of 
species $\sigma$ and $\sigma^\prime$ at positions $x$ and $x^\prime$ respectively. 
Regarding the same or different species, namely $\sigma=\sigma'$ or $\sigma \not=\sigma'$, $|G^{(2)}_{\sigma \sigma'}(x,x^\prime;t)|$ 
unveils the existence of intra- or interspecies two-body correlations correspondingly. 
The many-body state is referred to as two-body correlated [anti-correlated] when 
$G^{(2)}_{\sigma\sigma\prime}(x,x^\prime;t)>1$ [$G^{(2)}_{\sigma\sigma\prime}(x,x^\prime;t)<1$] is satisfied while if 
$G^{(2)}_{\sigma\sigma\prime}(x,x^\prime;t)=1$ holds it is termed fully second order coherent \cite{naraschewski1999,Mistakidis2018,Siegl2018,erdmann2019phase}. 
Furthermore, $G^{(2)}_{\sigma\sigma\prime}(x,x^\prime;t)$ can be experimentally assessed via {\it in-situ} density-density fluctuation 
measurements \cite{tavares2017,nguyen2019}.   

To reveal the emergence of interspecies correlations or entanglement between the two species of the FF 
mixture \cite{Koutentakis2019,cao2017unified,fasshauer2016} we calculate the eigenvalues $\lambda_k$ of the species reduced density matrix
$\rho^{(N_{\sigma})} (\vec{x}^{\sigma}, \vec{x}'^{\sigma};t)=\int d^{N_{\sigma'}} x^{\sigma'} \Psi^*_{MB}(\vec{x}^{\sigma}, 
\vec{x}^{\sigma'};t) \Psi_{MB}(\vec{x}'^{\sigma},\vec{x}^{\sigma'};t)$. 
Here, $\vec{x}^{\sigma}=(x^{\sigma}_1, \cdots, x^{\sigma}_{N_{\sigma}})$, and of course $\sigma\neq \sigma'$ [see also Eq. (\ref{eq:WF})]. 
Indeed, in the case that multiple eigenvalues of $\rho^{(N_{\sigma})}$ possess a macroscopic population the system is said to be species entangled 
or interspecies correlated, otherwise it is termed non-entangled. 
To quantify the degree of the system's entanglement we employ the so-called von-Neumann entropy \cite{yu2009,catani2009}  
\begin{align}
S(t)=-\sum\limits_{k=1}^D \lambda_{k}(t)\ln[\lambda_k(t)]. 
\label{eq:entropy}
\end{align} 
This quantity vanishes in the, non-entangled, HF limit, namely $S(t)=0$ since $\lambda_1(t)=1$. 
However, for a beyond HF state where more than a single $\lambda_k$ is nonzero $S(t)\neq0$. 

To unveil the degree of the intraspecies correlations or fragmentation of each species we extract the eigenvalues of the $\sigma$-species one-body reduced 
density matrix, $\rho^{(1)}_{\sigma}(x,x';t)$, being the so-called $\sigma$-species natural populations $n^{\sigma}_i(t)$. 
As discussed above [Sec. \ref{sec:ml-x}] it can be easily shown that if $\sum_i^{N{\sigma}}n_i^{\sigma}(t)=N_{\sigma}$ with $n_{i>N_{\sigma}}^{\sigma}(t)=0$, we can retrieve 
the corresponding HF wavefunction i.e. $\Psi_{MB}(\vec x^A,\vec x^B;t) \to \Psi_{HF}(\vec x^A,\vec x^B;t)$. 
Indeed, the system's wavefunction deviates from the HF state only when more than $N_{\sigma}$ natural orbitals are occupied \cite{pethick2008,erdmann2019phase,erdmann2018}. 
Therefore, in order to identify the degree of the $\sigma$-species fragmentation we measure the deviation  
\begin{equation}
F_{\sigma}(t) = N_{\sigma} - \sum\limits_{i=1}^{N_{\sigma}}n_i^\sigma(t).
\label{eq:fragmentation}
\end{equation} 
This quantity offers an indicator for the occupation of the $d_{\sigma} - N_{\sigma}$ natural orbitals and therefore $F_{\sigma}(t)>0$ signifies the deviation 
from a HF state. 

To estimate the breathing motion, namely the expansion and contraction dynamics \cite{abraham2014,koutentakis2017,Siegl2018} of the $\sigma$-species fermionic 
cloud we inspect the corresponding position variance 
\begin{equation}
\begin{split} 
\braket{X ^2_{\sigma}(t)}=\bra{\Psi_{MB}(t)}&(\hat{x}^{\sigma})^2\ket{\Psi_{MB}(t)}\\&- \bra{\Psi_{MB}(t)}\hat{x}^{\sigma}\ket{\Psi_{MB}(t)}^2.
\label{eq:variance}
\end{split}
\end{equation}
In this expression, $\hat{x}^{\sigma}=\int_{\mathcal{R}}dx~x_{\sigma}~\hat{\Psi}^{\sigma \dagger}(x)\hat{\Psi}^{\sigma}(x)$, and 
$(\hat{x}^{\sigma})^{2}=\int_{\mathcal{R}}dx~(x^{\sigma})^2~\hat{\Psi}^{\sigma \dagger}(x)\hat{\Psi}^{\sigma}(x)$ refer to one-body operators.  
Also, $\mathcal{R}$ denotes the spatial region of integration being in our case $\mathcal{R}\in [-40, 40]$. 
$\braket{X ^2_{\sigma}(t)}$ probes the spatial size of the $\sigma$-species cloud at each time instant of the evolution \cite{fukuhara2013quantum} 
and it is experimentally accessible \cite{ronzheimer2013}.

\section{Quench Dynamics of a Single Impurity} \label{sec:one_impurity} 

We assume a heavy impurity, $N_B=1$, repulsively interacting with a one-dimensional Fermi sea of $N_A=6$ fermions. 
Recall that in one-dimension a Fermi sea with $N_A>5$ atoms approaches the behavior of a many-body fermionic environment as it has been demonstrated 
in Refs. \cite{wenz2013,Mistakidis2019}. 
The mass imbalance is fixed to $M_B=6M_A$ unless it is stated otherwise. 
The system is trapped in a species selective optical potential and it is initially prepared into its ground state described by the Hamiltonian of Eq.~(\ref{eq:hamilt}). 
In particular, the Fermi sea is confined in a double-well potential characterized by a frequency $\omega_A=0.1$, barrier height $h=8$ and width $w=1$ whilst 
the impurity is trapped in a harmonic oscillator potential of frequency $\omega_B = 0.6\omega_A$ \cite{cetina2016}. 
We remark that the ground state configuration of the system can be described as a superposition of different eigenstates of the corresponding external potential of each species consisting of spin-polarized fermions. 
More specifically, for the majority species the first six (or otherwise the three two-fold degenerate) lowest-lying single-particle eigenstates of the double-well potential predominantly contribute to the 
initial state, while the energetically lowest harmonic oscillator eigenstate is dominantly populated for the impurity (results not shown). 
Of course, the above-mentioned states provide the dominant contributions to the initial state, while higher-lying states possess a comparatively much smaller contribution which increases 
for a larger interspecies interaction strength. 

This setup enforces an initial state of negligible spatial overlap between the species, immiscible components \cite{erdmann2019phase,Mistakidis2018}, independently of the value 
of the interspecies coupling $g_{AB}$ [Fig. \ref{fig:setup} (a)]. 
Specifically, the impurity is located around its trap center while the atoms of the Fermi sea are symmetrically placed with respect to $x=0$ in the left 
and right sides of their double-well, e.g. see Fig. \ref{fig:one_imp:ob_dens} (d). 
Note that such a ground state configuration is reminiscent of the one for the same setup with $h=0$ and strong interspecies repulsion, where an interspecies spatial separation takes place \cite{erdmann2019phase}. 
To induce the dynamics we ramp-down at $t=0$ the potential barrier of the Fermi sea from $h=8$ to $h=0$ and monitor the time-evolution of each species for 
a wide range of weak and strong interspecies repulsions lying in the interval $g_{AB} \in [0,4]$. 
Moreover, the emergent dynamical response is showcased both within the ML-MCTDHX many-body approach and the commonly used HF theoretical framework.

\subsection{One-body density evolution and position variance} \label{sec:one_imp:one_body_density}

\begin{figure}[ht] 
\includegraphics[width=0.47\textwidth]{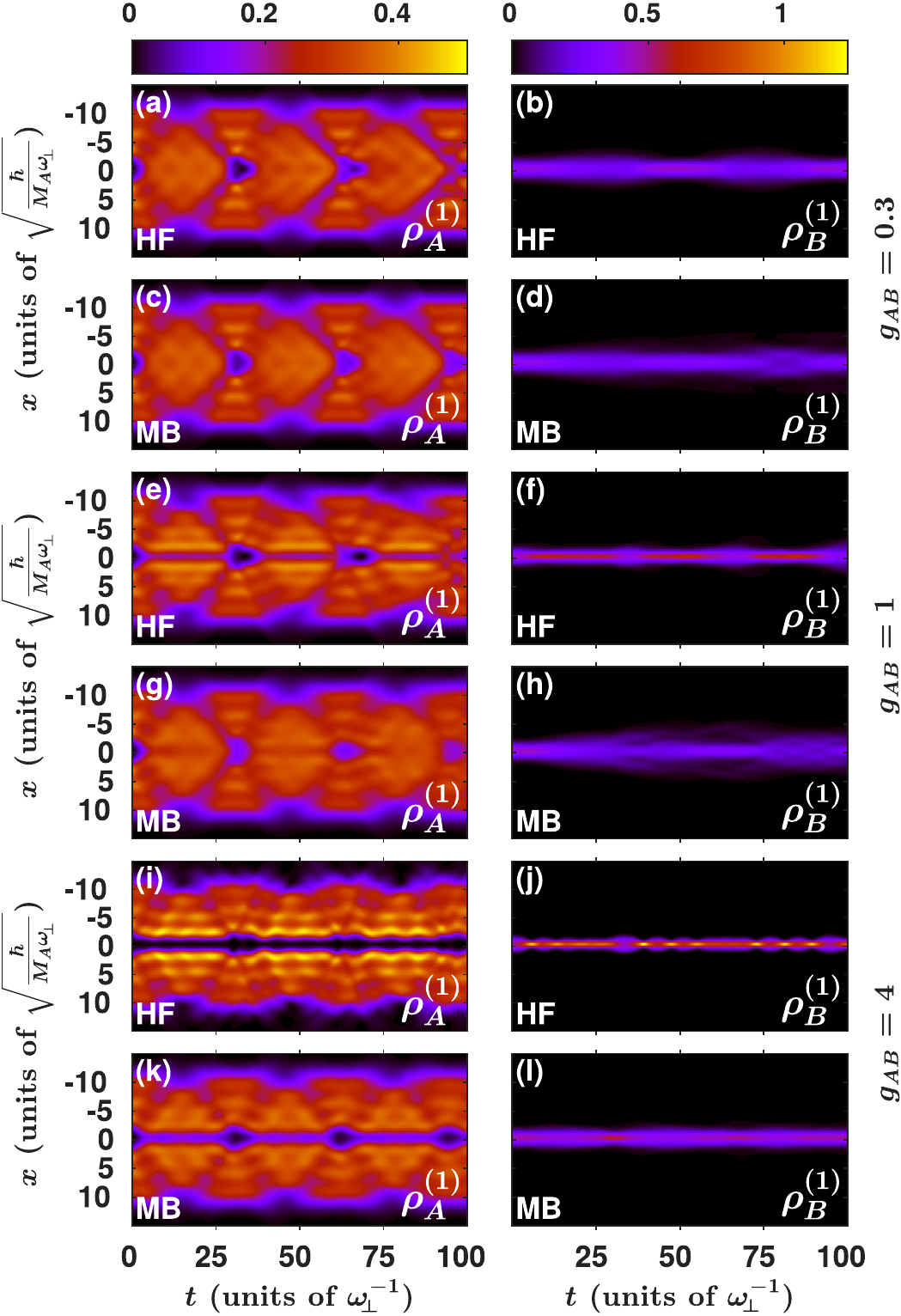}
\caption{Time-evolution of the $\sigma$-species one-body density $\rho^{(1)}_\sigma(x;t)$ of the FF mixture, after switching-off the potential barrier of the Fermi sea, 
for (a)-(d) weak ($g_{AB}=0.3$), (e)-(h) intermediate ($g_{AB}=1.0$) and (i)-(l) strong ($g_{AB}=4.0$) interspecies repulsions. 
The dynamics for each interaction strength is shown within (a), (b), (e), (f), (i), (j) the HF and 
(c), (d), (g), (h), (k), (l) the many-body approach.
The system consists of a Fermi sea (left panels) with $N_A = 6$ atoms and a single impurity $N_B=1$ (right panels) possessing a mass-imbalance of $M_B=6 M_A$. }
\label{fig:one_imp:ob_dens_dev}
\end{figure}

\begin{figure}[ht]
\includegraphics[width=0.45\textwidth]{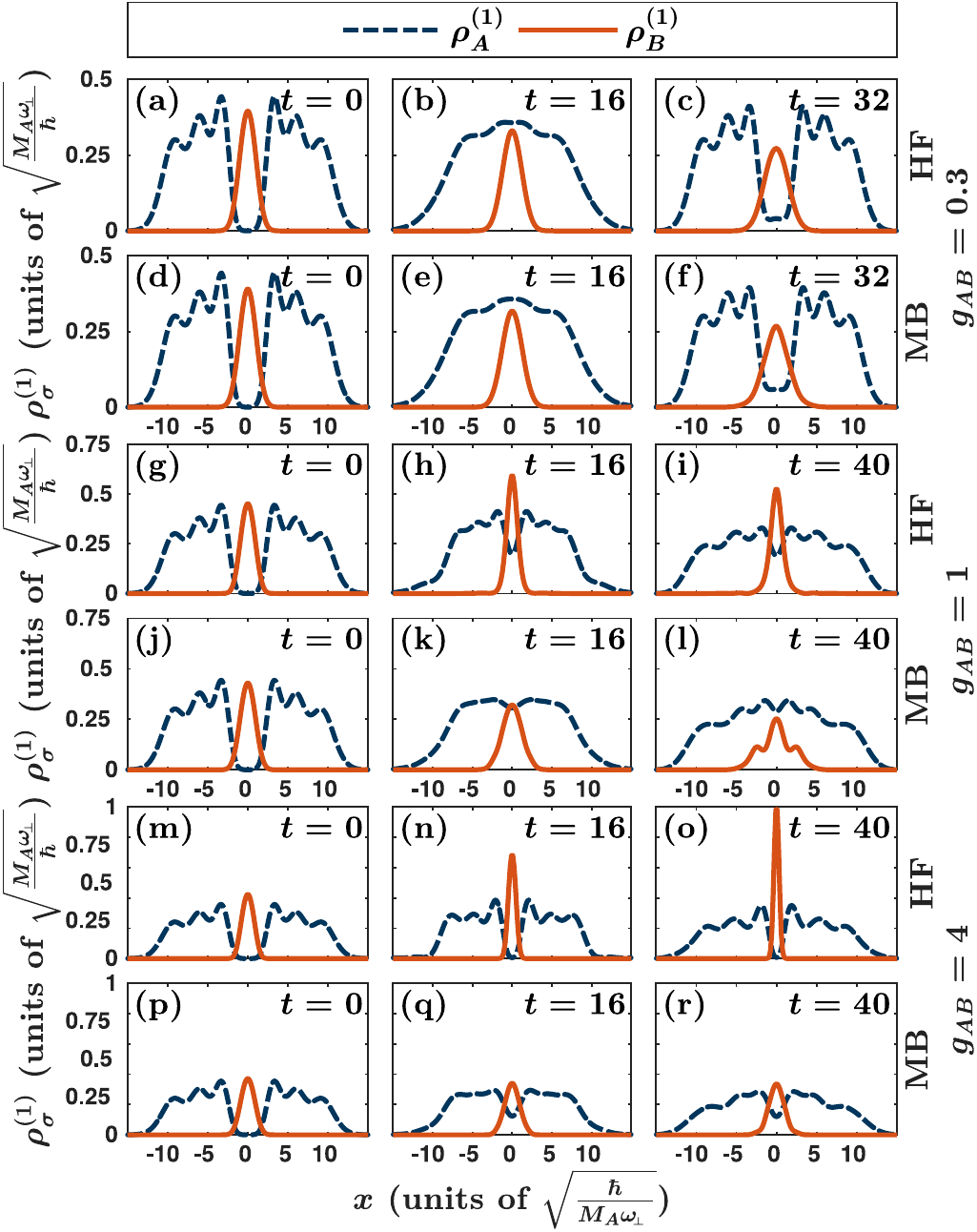}
\caption{Instantaneous profiles of the $\sigma$-species one-body density $\rho^{(1)}_\sigma(x;t)$ of the FF mixture for (a)-(f) weak ($g_{AB}=0.3$), (g)-(l) intermediate ($g_{AB}=1.0$) 
and (m)-(r) strong ($g_{AB}=4.0$) interspecies interaction strengths. 
The densities are provided within (a)-(c), (g)-(i), (m)-(o) the HF and (d)-(f), (j)-(l), (p)-(r) the many-body method. 
The mixture consists of a Fermi sea with $N_A=6$ and a heavy single impurity $N_B=1$ with $M_B=6M_A$. 
The fermionic bath and the impurity are initially trapped in double-well and a harmonic oscillator respectively. 
The dynamics is triggered by ramping-down the central potential barrier of the double-well.}
\label{fig:one_imp:ob_dens}
\end{figure}

\begin{figure}[ht]
\includegraphics[width=0.45\textwidth]{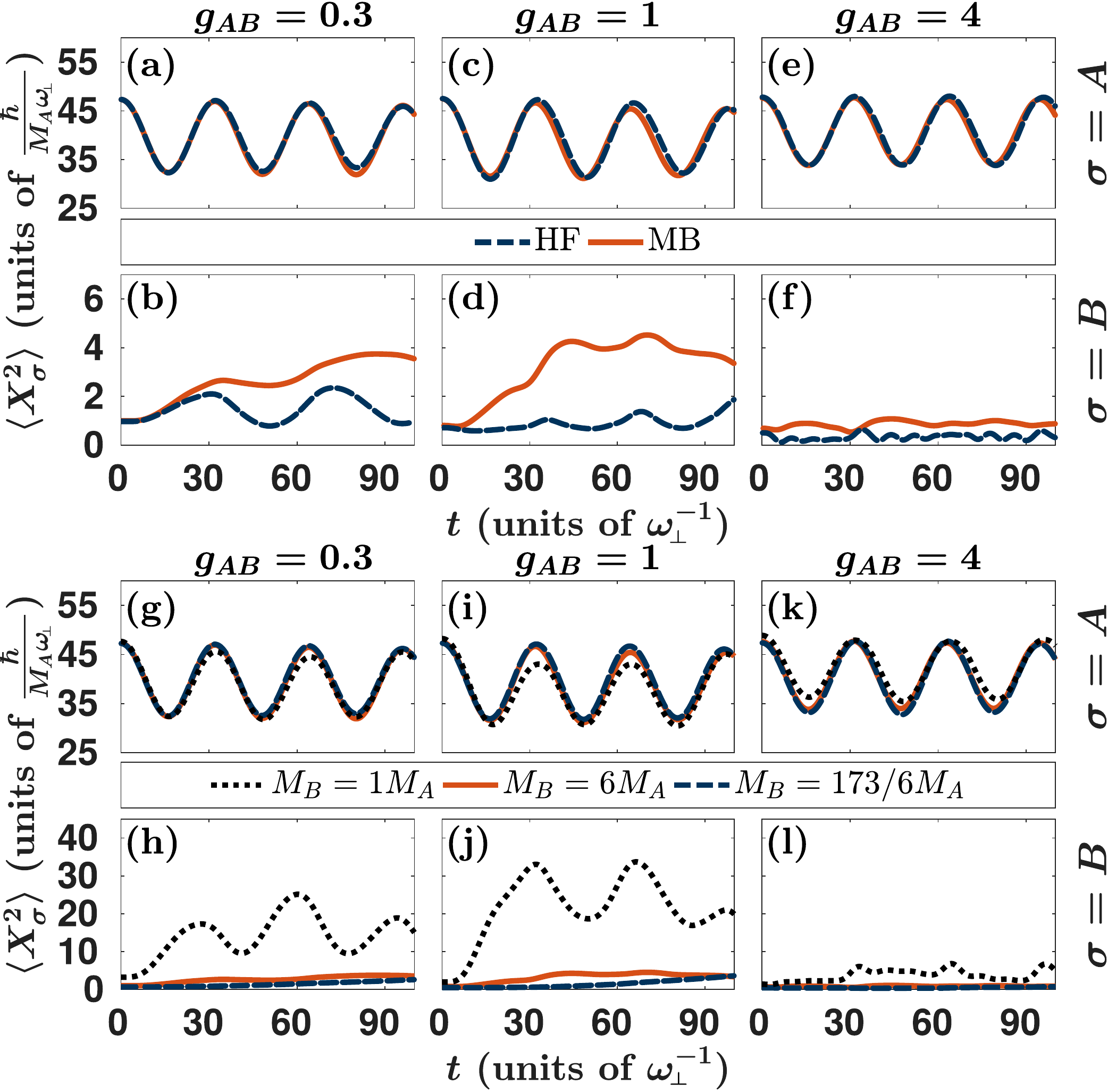}
\caption{Evolution of the variance of the one-body density $\braket{X^2_\sigma(t)}$ of (a), (c), (e) the Fermi sea ($\sigma=A$) and 
(b), (d), (f) the impurity ($\sigma=B$) for different interspecies repulsions $g_{AB}$ (see legends) and obtained within HF  
and the many-body approaches (see legends). 
The fermionic bath consists of $N_A=6$ particles and the single more massive impurity $N_B=1$ such that $M_B=6M_A$. 
$\braket{X^2_\sigma(t)}$ within the many-body method for (g), (i), (k) the Fermi sea ($\sigma=A$) and (h), (j), (l) the impurity ($\sigma=B$) 
for specific values of $g_{AB}$ (see legends) and different masses $M_B$ of the impurity (see legend).}
\label{fig:one_imp:variance}
\end{figure}

To inspect the quench-induced dynamics of the FF mixture we first invoke the time-evolution of the $\sigma$-species one-body density [Fig. \ref{fig:one_imp:ob_dens_dev}] and the 
position variance $\braket{X^2_\sigma(t)}$ [Fig. \ref{fig:one_imp:variance}] following a quench of the height of the potential barrier from $h=8$ to $h=0$ for 
different interspecies interaction strengths. 
We observe that almost independently of the interspecies repulsion [see e.g. Figs. \ref{fig:one_imp:ob_dens} (d), (j)] and both in the many-body and the HF 
approaches [Figs. \ref{fig:one_imp:ob_dens} (a), (d)] 
the one-body density of the Fermi sea, due to the presence of the potential barrier, is initially segregated into two fragments. 
In particular, each fragment exhibits three local density maxima indicating that three fermions are located in the left and the other three 
at the right side of the double-well \cite{erdmann2019phase}. 
Note that a larger $g_{AB}$ essentially causes a slightly increasing phase separation between the two species \cite{pecak2016two,erdmann2019phase}, 
compare Fig. \ref{fig:one_imp:ob_dens} (d) with Fig. \ref{fig:one_imp:ob_dens} (p). 

The quench triggers a counterflow dynamics of these fragments and $\rho^{(1)}_A(x;t)$ performs an overall breathing motion of frequency $\omega_{A}^{br} \approx 0.2 = 2 \omega_A$ for 
every $g_{AB}$ \cite{abraham2012,abraham2014} and in both approaches [Figs. \ref{fig:one_imp:variance} (a), (c), (e)]. 
This value of the breathing frequency of the Fermi sea, i.e. $\omega_A^{br}=2\omega_A$, is in accordance with the corresponding theoretical prediction \cite{bauch2010quantum,bauch2009quantum,abraham2012quantum}. 
Note that the instantaneous density patterns building upon $\rho^{(1)}_A(x;t)$ depend crucially on both the value of $g_{AB}$ and the correlations (see the discussion below). 
This breathing motion of $\rho^{(1)}_A(x;t)$ is directly captured by the oscillatory motion of the corresponding position variance $\braket{X^2_A(t)}$ which is found to be almost the same in both the 
HF and the many-body approaches for every $g_{AB}$. 
We remark that only some minor deviations occur in the shape of $\braket{X^2_A(t)}$ between the HF and the many-body calculations [hardly discernible 
in Figs. \ref{fig:one_imp:variance} (a), (c), (e)]. 
These negligible deviations of $\braket{X^2_A(t)}$ for fixed $g_{AB}$ can in part be explained by the fact that the postquench ground state density of the Fermi sea is almost the 
same in the HF and the many-body case (not presented for brevity). 

Focusing on weak repulsions, e.g. $g_{AB}=0.3$, and within the HF framework we observe that after the quench the initially separated density fragments of the Fermi sea [Fig. \ref{fig:one_imp:ob_dens} (a)] 
come close and collide around $x=0$ forming a peak at the same location and two local density minima symmetrically placed with respect to $x=0$ [Fig. \ref{fig:one_imp:ob_dens} (b)]. 
Subsequently $\rho^{(1)}_A(x;t)$ splits again into two counter propagating density fragments, 
each of them exhibiting three local maxima, traveling towards the edges of the harmonic trap and having a more 
shallow density dip at $x=0$ compared to the one at $t=0$, see Figs. \ref{fig:one_imp:ob_dens} (a), (c). 
This motion of $\rho^{(1)}_A(x;t)$ is periodically repeated in the course of the evolution, thus reflecting the counterflow dynamics of the 
Fermi sea [Fig. \ref{fig:one_imp:ob_dens_dev} (a)]. 
On the other hand, the impurity which is not directly affected by the quench is implicitly perturbed by the Fermi sea due to the finite $g_{AB}$. 
More precisely, the impurity resides at the trap center during the entire time-evolution [Fig. \ref{fig:one_imp:ob_dens_dev} (b)] and its cloud shows a periodic expansion and contraction as also captured 
by $\braket{X^2_B(t)}$ [Fig. \ref{fig:one_imp:variance} (b)] with frequency $\omega_B^{br}=0.13$. 
As a result of the combined motion of the two species the system undergoes a periodic mixing and demixing dynamics, i.e. it oscillates between 
miscibility and immiscibility, see Figs. \ref{fig:one_imp:ob_dens_dev} (a), (b). 

Inspecting the corresponding one-body densities of the many-body evolution, depicted in Figs. \ref{fig:one_imp:ob_dens_dev} (c), (d) and Figs. \ref{fig:one_imp:ob_dens} (d)-(f), we deduce 
that a qualitatively similar phenomenology to the HF case takes place for both components. 
However $\rho^{(1)}_B(x;t)$ appears to be slightly wider within the many-body approach [Fig. \ref{fig:one_imp:ob_dens_dev} (d)] as compared to the HF case [Fig. \ref{fig:one_imp:ob_dens_dev} (b)] 
and in particular $\rho^{(1)}_B(x;t)$ exhibits an overall expansion tendency which is not present in the HF evolution. 
This overall expansion of $\rho^{(1)}_B(x;t)$ during the evolution is also represented by the increasing tendency of the amplitude of its position variance $\braket{X^2_B(t)}$ 
as illustrated in Fig. \ref{fig:one_imp:variance} (b). 
Additionally, the density dip of $\rho^{(1)}_A(x;t)$ formed after the collision of the fragments at $x=0$ is shallower in the many-body [Fig. \ref{fig:one_imp:ob_dens_dev} (c)] compared 
to the HF case [Fig. \ref{fig:one_imp:ob_dens_dev} (a)]. 
The latter results to a relatively larger spatial overlap between the components in the course of time, and thus to a larger degree of miscibility, 
within the many-body approach \cite{mistakidis2019correlated,Mistakidis2018,erdmann2019phase}. 

Increasing the interspecies repulsion, for instance to $g_{AB}=1$, and referring to the HF evolution [Figs. \ref{fig:one_imp:ob_dens_dev} (e), (f)] it becomes evident that the dynamical spatial 
separation between the species is enhanced compared to $g_{AB}=0.3$, see also the instantaneous density profiles in Figs. \ref{fig:one_imp:ob_dens} (g)-(i). 
Moreover, the density fragments building upon $\rho^{(1)}_A(x;t)$ and accompanied by a dip around $x=0$ persist during the entire time-evolution. 
Note that the dip appearing in the vicinity of $x=0$ is shallower within the time-interval of contraction of $\rho^{(1)}_A(x;t)$ whilst it is deeper during the expansion 
of $\rho^{(1)}_A(x;t)$ [Fig. \ref{fig:one_imp:ob_dens_dev} (e)]. 
Consequently the impurity's cloud is effectively trapped by the density dip of $\rho^{(1)}_A(x;t)$ and shows a rather localized shape forming a density peak which is centered at $x=0$. 
The above-mentioned behavior of each of the species becomes more pronounced for stronger repulsive values of $g_{AB}$, as it can be readily seen by monitoring the dynamics of each 
$\rho^{(1)}_{\sigma}(x;t)$ presented in Figs. \ref{fig:one_imp:ob_dens_dev} (i), (j) for $g_{AB}=4$. 
Indeed, the impurity possesses a highly localized density distribution which lies around the density dip of $\rho^{(1)}_A(x;t)$ for all times and as a consequence gives rise to a complete 
dynamical phase separation between the species, see Figs. \ref{fig:one_imp:ob_dens_dev} (i), (j). 
Note also here that the density dip of $\rho^{(1)}_A(x;t)$, with $\rho^{(1)}_A(x\approx0;t)\approx0$, is more pronounced and the emergent density fragments are deeper 
[Figs. \ref{fig:one_imp:ob_dens} (m)-(o)] compared to the $g_{AB}=1$ case [Figs. \ref{fig:one_imp:ob_dens} (g)-(i)]. 
Moreover, $\rho^{(1)}_B(x;t)$ performs a weak amplitude but multifrequency breathing motion identified by the irregular oscillatory behavior of its 
variance $\braket{X^2_B(t)}$ shown in Fig. \ref{fig:one_imp:variance} (d) and Fig. \ref{fig:one_imp:variance} (f) for $g_{AB}=1$ and $g_{AB}=4$ respectively. 
Evidently for increasing $g_{AB}$ a larger number of frequencies participate in the dynamics of $\braket{X^2_B(t)}$, compare Fig. \ref{fig:one_imp:variance} (d) with 
Fig. \ref{fig:one_imp:variance} (f). 
Indeed, for $g_{AB}=1$ the predominant frequencies of $\braket{X^2_B(t)}$ are $\omega_1\approx0.19$, $\omega_2\approx0.44$ whilst for $g_{AB}=4$ they correspond to 
$\omega_1\approx0.19$, $\omega_2\approx0.38$, $\omega_3\approx0.56$ and $\omega_4\approx0.75$.  

Turning to the correlated dynamics for strong interactions we can infer that an overall similar dynamics to the HF case takes place but significant alterations occur for 
the instantaneous pattern formation of both species. 
Referring to the Fermi sea it can be deduced that both the density dips and the fragments of $\rho^{(1)}_A(x;t)$ are shallower from the HF case, compare Fig. \ref{fig:one_imp:ob_dens_dev} (g) with 
Fig. \ref{fig:one_imp:ob_dens_dev} (e) for $g_{AB}=1$ and Fig. \ref{fig:one_imp:ob_dens_dev} (i) to Fig. \ref{fig:one_imp:ob_dens_dev} (k) at $g_{AB}=4$. 
The aforementioned differences are also directly imprinted in the instantaneous density profiles of $\rho^{(1)}_A(x;t)$ within the many-body [Figs. \ref{fig:one_imp:ob_dens} (k)-(l), (q)-(r)] and the 
HF approaches [Fig. \ref{fig:one_imp:ob_dens} (h)-(i), (n)-(o)]. 
On the other hand, the density of the impurity is more spread within the many-body than the HF approach, contrast in particular Fig. \ref{fig:one_imp:ob_dens_dev} (h) with 
Fig. \ref{fig:one_imp:ob_dens_dev} (f) where this phenomenon is enhanced as well as Fig. \ref{fig:one_imp:ob_dens_dev} (l) to Fig. \ref{fig:one_imp:ob_dens_dev} (j). 
This spreading behavior of $\rho^{(1)}_B(x;t)$ together with the fact that the density dip of $\rho^{(1)}_A(x;t)$ is shallower in the many-body case results to a smaller degree of 
phase separation between the species when correlations are taken into account. 
For instance, inspecting $\rho^{(1)}_{\sigma}(x;t)$ for $g_{AB}=4$ we can easily deduce that the almost perfect phase separation emerging in the HF approach throughout the dynamics 
[Figs. \ref{fig:one_imp:ob_dens_dev} (i), (j)] is altered to periodic oscillations between miscibility and immiscibility among the species in the many-body evolution [Figs. \ref{fig:one_imp:ob_dens_dev} (k), (l)]. 
Additionally, in the presence of correlations the impurity cloud shows an expansion tendency in the course of the evolution as it can be identified by the gradually increasing 
amplitude of $\braket{X^2_B(t)}$ [Figs. \ref{fig:one_imp:variance} (d), (f)], a behavior that is absent within the HF framework. 
Notice that this expanding behavior of $\rho^{(1)}_B(x;t)$ is more pronounced for $g_{AB}=1$ [see Fig. \ref{fig:one_imp:ob_dens_dev} (h) and Fig. \ref{fig:one_imp:variance} (d)] than at 
$g_{AB}=4$ [Fig. \ref{fig:one_imp:ob_dens_dev} (l) and Fig. \ref{fig:one_imp:variance} (f)] where the strong interspecies repulsion tends to suppress the expansion of the impurity cloud \cite{mistakidis2019correlated}.

\subsection{Effect of mass-imbalance on the dynamics} \label{sec:one_imp:mass_imbalance} 

To investigate the impact of the impurity mass on the nonequilibrium dynamics of the FF mixture we next inspect the behavior of the $\sigma$-species position variance 
$\braket{X^2_{\sigma}(t)}$, serving as an indicator of each species dynamical response, for fixed $g_{AB}$ but different masses of the impurity particle. 
Specifically, we consider a mass-balanced mixture i.e. $M_A=M_B$ with $\omega_A=\omega_B$ corresponding to two distinct hyperfine states of $^{6}$Li \cite{serwane2011,wenz2013} 
and a highly mass-imbalanced system where $M_B = 173/6 M_A$ and $\omega_B=0.0125\omega_A$ which is realized by the species $^{173}$Yb and $^{6}$Li \cite{hara2011quantum,khramov2014ultracold}. 
The dynamics of the FF mixture is induced by a sudden change of the height of the potential barrier from $h=8$ to $h=0$ as in Sec. \ref{sec:one_imp:one_body_density}. 
Figure \ref{fig:one_imp:variance} presents $\braket{X^2_{\sigma}(t)}$ for the above-mentioned systems and for specific interspecies interaction strengths, namely $g_{AB}=0.3,1$ and $4$. 
As it can be seen, the dynamical response of the Fermi sea is almost insensitive to the mass of the impurity. 
Indeed for fixed $g_{AB}$ the variance $\braket{X^2_{A}(t)}$ performs an oscillatory motion, reflecting the breathing dynamics of the fermionic bath, whose amplitude and frequency are essentially 
unaffected by the mass of the impurity, see Figs. \ref{fig:one_imp:variance} (g), (i) and (k). 
To be more precise, the amplitude of $\braket{X^2_{A}(t)}$ is slightly smaller for the mass-balanced than the mass-imbalanced case for every $g_{AB}$, a result that is more pronounced 
at $g_{AB}=1$ [Fig. \ref{fig:one_imp:variance} (i)]. 
This behavior can be attributed to the fact that a light impurity perturbs its bath to a smaller extent than a heavier one \cite{Siegl2018,mistakidis2019dissipative}. 
Note also that the postquench ground state of the Fermi sea at a specific $g_{AB}$ is almost unaltered between the $M_B = 6 M_A$ and the $M_B = (173/6) M_A$ cases. 
However, small differences are evident when comparing the mass-imbalanced and the mass-balanced scenaria especially for increasing $g_{AB}$ (not shown here). 

On the contrary, the dynamical response of the impurity for a fixed $g_{AB}$ depends crucially on its mass as illustrated in Figs. \ref{fig:one_imp:variance} (h), (j) and (l). 
Evidently, for a specific $g_{AB}$ the variance of the impurity $\braket{X^2_{B}(t)}$ shows an ``irregular'' oscillatory pattern with an overall increasing amplitude indicating its tendency 
to disperse within the fermionic bath. 
Most importantly, the amplitude of $\braket{X^2_{B}(t)}$ is drastically larger in the case of a light than a heavy impurity independently of the value of $g_{AB}$, a result that is 
particularly pronounced at intermediate interactions e.g. $g_{AB}=1$ [Fig. \ref{fig:one_imp:variance} (j)]. 
However, comparing $\braket{X^2_{B}(t)}$ for the heavy impurities we observe that it acquires smaller values for a reduced mass but its overall shape is not 
significantly affected [Figs. \ref{fig:one_imp:variance} (h), (j) and (l)]. 
Notice also here that for the mass balanced case the degree of miscibility between the species is enhanced compared to the heavy impurity case (results not shown for brevity), a behavior that is 
already known to occur in both the ground state properties \cite{pecak2016two} as well as during the nonequilibrium dynamics \cite{erdmann2018,erdmann2019phase} of FF mixtures.

\subsection{Quantifying the degree of inter- and intraspecies correlations} \label{sec:one_imp:entropy_and_fragmentation} 

To estimate the importance of the interspecies entanglement and intraspecies correlations (or fragmentation) of the many-body state describing the nonequilibrium dynamics 
of the FF mixture we next employ the von-Neumann entropy $S(t)$ [Eq. (\ref{eq:entropy})] and the $\sigma$-species fragmentation measure $F_{\sigma}(t)$ introduced 
in Eq. (\ref{eq:fragmentation}). 
Recall that the system is termed non-entangled in the case of $S(t)=0$, while it is called entangled or interspecies correlated in the case of $S(t)\neq0$. 
Furthermore, the presence of intraspecies correlations is designated by $F_{\sigma}(t)>0$ (see also Sec. \ref{sec:observable-of-interest}). 
Let us remark that in the case of a single impurity atom, as the one considered herein, it follows from the Schmidt decomposition of Eq. (\ref{eq:WF}) that $\lambda_k(t)\equiv n_k^B(t)$. 
Therefore the degree of fragmentation of the impurity here coincides with the degree of interspecies correlations \cite{erdmann2019phase,Mistakidis2019,cao2017unified}.  
As a result, below we examine only $S(t)$ and $F_{A}(t)$ because $S(t)$ provides information about the degree of fragmentation of the impurity. 
It is also important to mention that since the fermions of the same species are assumed here to be spin-polarized, and therefore non-interacting, the 
manifestation of intraspecies correlations during the dynamics is caused due to the occurrence of interspecies interactions and correlations \cite{pitaevskii2003,pethick2008,lewenstein2012}. 
\begin{figure}[ht]
\includegraphics[width=0.45\textwidth]{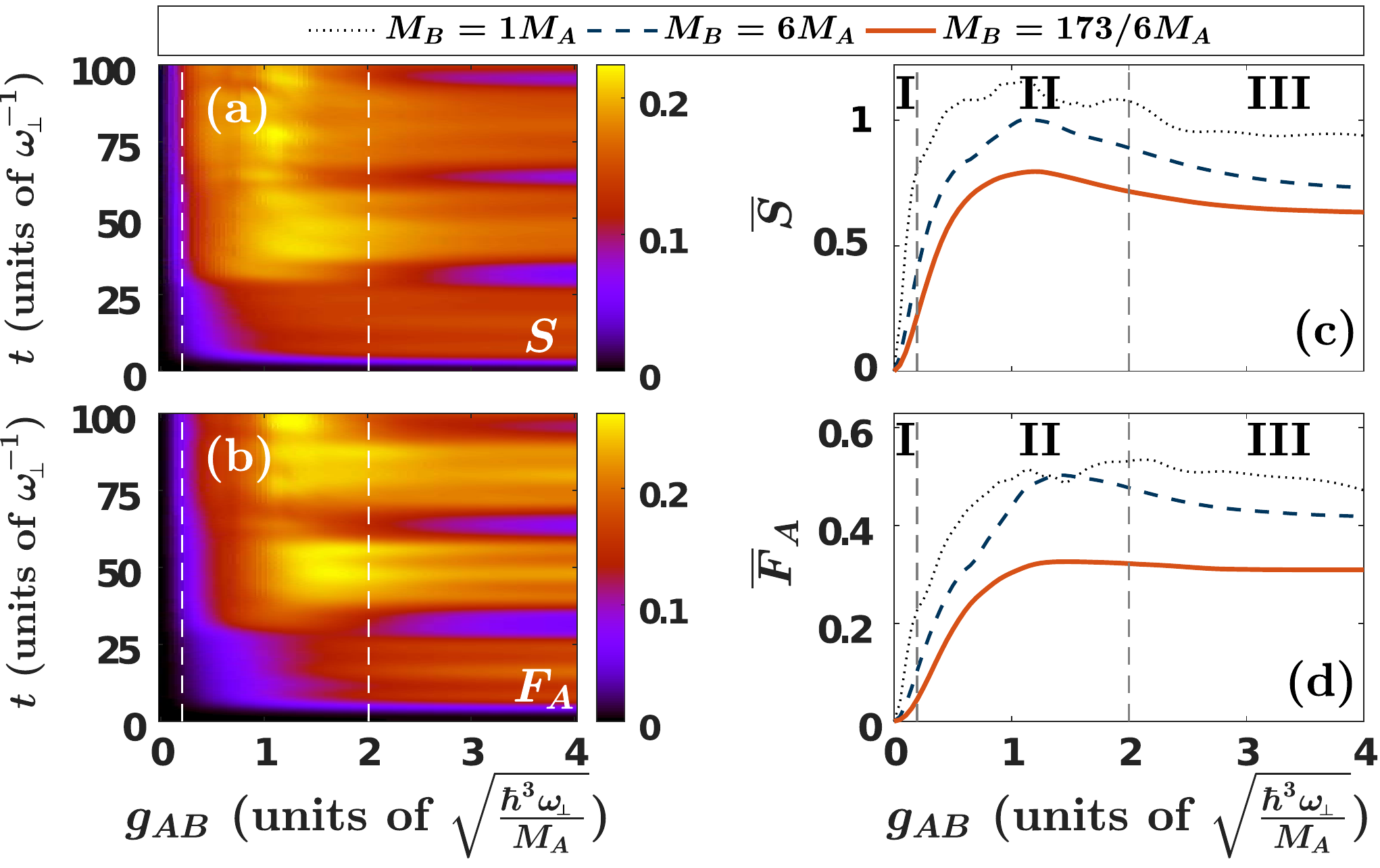}
\caption{Time-evolution of (a) the von-Neumann entropy $S(t)$ and (b) the deviation from unity of the first six natural populations of the Fermi sea $F_A(t)$ for increasing interspecies repulsion $g_{AB}$.  
The system consists of $N_A=6$ fermions and a single impurity $N_B=1$ with a mass-imbalance of $M_B=6M_A$. 
Time-averaged (c) von-Neumann entropy i.e. $\bar{S}(t)$ and (d) $\bar{F}_A(t)$ for three different masses of the impurity (see legend). 
The vertical dashed gray lines indicate the boundaries of the three interaction regimes where the correlation measures show a distinct behavior. 
In all cases the dynamics is induced by a sudden quench at $t=0$ of the potential of the Fermi sea from a double-well to a harmonic oscillator.}
\label{fig:one_imp:entropy_and_fragmentation}
\end{figure}

The time-evolution of $S(t)$ and $F_A(t)$ following a ramp-down of the barrier height of the double-well of the Fermi sea in the case of $M_B=6M_A$ is presented 
in Figs. \ref{fig:one_imp:entropy_and_fragmentation} (a) and (b) respectively for a varying interspecies repulsion $g_{AB}$. 
Also in order to illustrate the average amount of correlations participating in the dynamics we depict in Figs. \ref{fig:one_imp:entropy_and_fragmentation} (c), (d) 
the time-integrated value over the considered evolution time of the above-mentioned measures namely $\bar{S}=(1/T)\int_0^T dt S(t)$ and $\bar{F}_A=(1/T)\int_0^T dt F_A(t)$. 
Since the initial state of the system exhibits an approximately perfect phase separation, see e.g. Fig. \ref{fig:one_imp:ob_dens} (d), it holds that 
$S(t=0)=0$ and $F_{A}(t=0)=0$ reflecting the vanishing inter- and intraspecies correlations at $t=0$ for arbitrary 
$g_{AB}$ [Figs. \ref{fig:one_imp:entropy_and_fragmentation} (a), (b)]. 
As time evolves we observe the build up of both inter- and intraspecies correlations identified by the non-zero values of $S(t)$ and $F_A(t)$ respectively \cite{wlodzynski2020geometry,erdmann2019phase}. 
In particular, the behavior of the correlation measures can be classified into three different interaction regimes marked as I, II and III 
in Fig. \ref{fig:one_imp:entropy_and_fragmentation}. 

For weak interactions i.e. $0<g_{AB}<0.2$ (region I) the amount of entanglement and intraspecies correlations of the Fermi sea is relatively small 
since both $S(t)$ and $F_{A}(t)$ are suppressed. 
This statement is also supported by the fact that the average values $\bar{S}<0.2$ and $\bar{F}_A<0.1$ in this region [Figs. \ref{fig:one_imp:entropy_and_fragmentation} (c), (d)] 
and exhibit a systematically increasing tendency for larger $g_{AB}$. 
Interestingly, for intermediate interactions $0.2<g_{AB}<2$ we enter region II where $S(t)$ and $F_{A}(t)$ increase during 
the evolution and subsequently tend to saturate to a certain finite value for $t>80$ [Figs. \ref{fig:one_imp:entropy_and_fragmentation} (a), (b)]. 
The non-negligible effect of correlations for $0.2<g_{AB}<2$ is also testified by the noticeable increasement of both $\bar{S}$ and $\bar{F}_A$ which acquire their 
maximum values around $g_{AB}\approx1.2$. 
The aforementioned behavior e.g. of $\bar{S}(t)$, $\bar{F}_A(t)$ signifies that the underlying many-body state is strongly entangled and 
intraspecies correlated in this region. 
Entering the strongly interacting regime III characterized by $g_{AB}>2$ both $S(t)$ and $F_A(t)$ possess smaller values than in region II and most importantly 
exhibit an oscillatory behavior taking values in the interval $[0.05, 0.25]$. 
As a consequence $\bar{S}$ and $\bar{F}_A$ become smaller compared to region II. 
We remark that the oscillations of $S(t)$ and $F_A(t)$ occur due to the oscillatory behavior of the system between miscibility and immiscibility observed in the evolution 
of the $\sigma$-species single-particle densities [Figs. \ref{fig:one_imp:ob_dens_dev} (k), (l)]. 
Specifically within the time-intervals of miscibility (immiscibility) $S(t)$ and $F_A(t)$ possess their maximum (minimum) value, compare Figs. \ref{fig:one_imp:ob_dens_dev} (k), (l) with 
Figs. \ref{fig:one_imp:entropy_and_fragmentation} (a), (b). 

To unveil the effect of the mass of the impurity on the average amount of interparticle correlations we inspect the behavior $\bar{S}$ and $\bar{F}_A$ as a function of $g_{AB}$ for 
different masses of the impurity in Figs. \ref{fig:one_imp:entropy_and_fragmentation} (c), (d). 
As it can be readily seen the overall behavior of both $\bar{S}$ and $\bar{F}_A$ with respect to $g_{AB}$ is not affected by the mass of the impurity. 
Namely $\bar{S}$ and $\bar{F}_A$ increase within region I, maximize in region II and then tend to saturate in region III possessing also a smaller value than in region II. 
However, as shown in Figs. \ref{fig:one_imp:entropy_and_fragmentation} (c), (d) both inter- and intraspecies correlations exhibit a hierarchy in terms of the mass of the impurity, 
i.e. there is a smaller amount of correlations for a heavier impurity. 
This observation indicates that mass-balanced setups are more prone to correlation effects than mass-imbalanced ones, a behavior that has already been observed 
in the nonequilibrium dynamics of different setups \cite{erdmann2019phase,Siegl2018} and it is related to the mobility of the impurity \cite{schmidt2018universal}.

\subsection{Two-body correlation dynamics} \label{sec:one_imp:correlations}

\begin{figure}[ht] 
\includegraphics[width=0.45\textwidth]{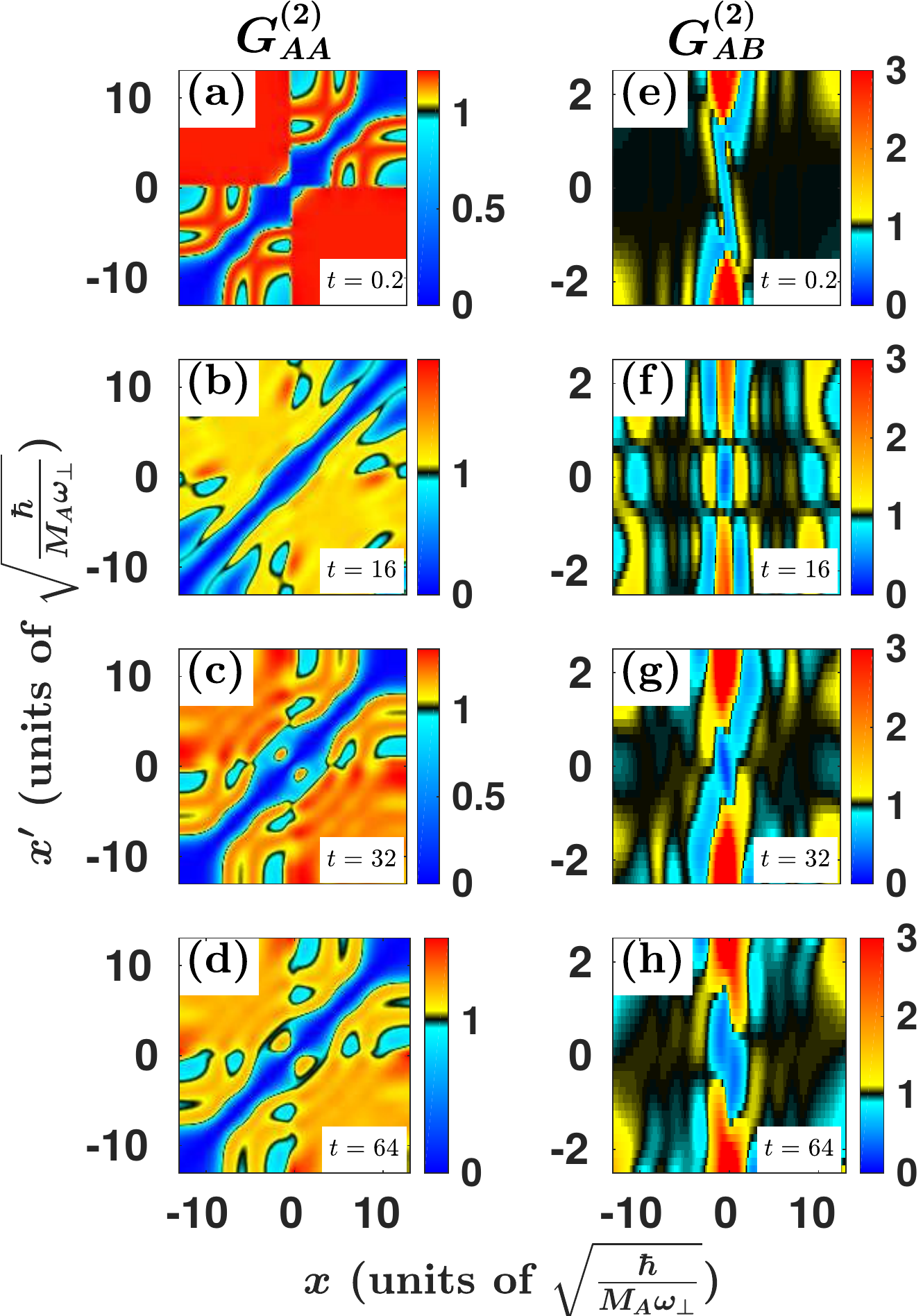}
\caption{Instantaneous two-body (a)-(d) intraspecies $G^{(2)}_{AA}(x,x^\prime;t)$ and (e)-(f) interspecies $G^{(2)}_{AB}(x,x^\prime;t)$ correlations 
for strong interaction strength $g_{AB}=4$ at different time-instants of the evolution (see legend). 
The mixture consists of $N_A=6$ fermions and a heavy single impurity $N_B=1$ with $M_B=6M_A$. 
The Fermi sea is initially trapped in a double-well and the impurity in a harmonic trap. 
The dynamics is induced by quenching the height of the central potential barrier of the double-well to zero.}
\label{fig:one_imp:2b_correlations}
\end{figure}

To reveal the underlying two-body correlation properties of the FF mixture \cite{pkecak2019intercomponent}, we study the corresponding two-body intra- and interspecies correlation functions, 
$G_{\sigma\sigma'}^{(2)}(x,x',t)$ [Eq. (\ref{two_body_cor})], in the course of the many-body quench dynamics [Fig. \ref{fig:one_imp:2b_correlations}]. 
We remind that for two particles of species $\sigma$ and $\sigma^\prime$ located at $x$ and $x^\prime$ respectively if $G^{(2)}_{\sigma\sigma^\prime}(x,x^\prime;t) > 1$ 
[$G^{(2)}_{\sigma\sigma^\prime}(x,x^\prime;t) < 1$] holds it indicates the emergence of two-body correlations [anti-correlations]. 
However in case that $G_{\sigma\sigma^\prime}(x,x^\prime;t) = 1$ is satisfied the two particles are uncorrelated. 
Note also that for the setup under consideration involving only a single impurity the intraspecies two-body correlation function for the $B$-species is by definition zero. 
Therefore, below we investigate the time-evolution of $G_{AA}^{(2)}(x,x',t)$ [Figs. \ref{fig:one_imp:2b_correlations} (a)-(d)] and $G_{AB}^{(2)}(x,x',t)$ [Figs. \ref{fig:one_imp:2b_correlations} (e)-(h)] 
following a ramp-down of the potential barrier height of the Fermi sea at strong interspecies repulsions $g_{AB}=4$.

Regarding the Fermi sea we observe that the diagonal $G_{AA}^{(2)}(x,x'=x,t)<1$, see Figs. \ref{fig:one_imp:2b_correlations} (a)-(d), throughout the evolution implying that two fermions are not likely 
to reside at the same location as a consequence of the Pauli exclusion principle. 
Recall that the single-particle density of the fermionic bath $\rho^{(1)}_A(x;t)$ is splitted into two fragments, created by the initial double-well potential and the presence of strong 
repulsions. 
Each of these fragments exhibits at most three local maxima, see Fig. \ref{fig:one_imp:ob_dens_dev} (k) and Figs. \ref{fig:one_imp:ob_dens} (p)-(r). 
Therefore the fact that $G_{AA}^{(2)}(x,x'=x,t)<1$ indicates also that only one fermion can populate each of these local density maxima. 
Most importantly, two-body correlations occur between different spatial regions of the Fermi sea since $G_{AA}^{(2)}(x,x'\neq x,t)>1$ holds predominantly during the evolution 
[Figs. \ref{fig:one_imp:2b_correlations} (a)-(d)]. 
This behavior of $G_{AA}^{(2)}(x,x'\neq x,t)$ signifies that two fermions of the bath either reside in the same fragment but e.g. distinct local density maxima or they are 
simply located at different density fragments. 

Turning to the interspecies two-body correlations it is evident by the structures building upon $G_{AB}^{(2)}(x,x',t)$ [Figs. \ref{fig:one_imp:2b_correlations} (e)-(h)] that a phase separation between 
the impurity and the fermionic bath occurs during the evolution. 
Indeed an anti-correlated behavior takes place between one fermion of the bath and the impurity in the vicinity of $x=0$ as it can inferred 
by $G_{AB}^{(2)}(x=0,x'=0,t)<1$ for all evolution times. 
However, the impurity is strongly correlated with a fermion residing at the density notch of the fermionic bath, see e.g. 
$G_{AB}^{(2)}(-0.1<x<0.1,-2<x'<-0.2,t)>1$ and $G_{AB}^{(2)}(-0.1<x<0.1,0.2<x'<2,t)>1$ in Figs. \ref{fig:one_imp:2b_correlations} (e)-(h). 
The remaining spatial regions lying beyond the spatial extension of the impurity cloud are predominantly two-body uncorrelated 
i.e. $G_{AB}^{(2)}(x>2,-2<x'<2,t)\approx 1$ and $G_{AB}^{(2)}(x<-2,-2<x'<2,t)\approx 1$. 
Notice also here that on the single-particle density level we observe that the impurity and the Fermi sea are miscible (immiscible) in the time intervals of contraction (expansion) of the 
cloud of the Fermi sea as shown in Figs. \ref{fig:one_imp:ob_dens_dev} (k), (l). 
Interestingly, on the two-body level we can deduce that a phase separation between the impurity and the Fermi sea occurs throughout the dynamics. 
For instance, at $t=16$ where miscibility is inferred between the corresponding single-particle densities,  see Fig. \ref{fig:one_imp:ob_dens} (q), an apparent immiscibility occurs 
on the two-body level, see e.g. $G_{AB}^{(2)}(x=0,x'=0,t=16)<1$ in Fig. \ref{fig:one_imp:2b_correlations} (f). 
Concluding we can deduce that for the many-body dynamics the phase separation between the species is evident only on the two-body level and not by simply observing the corresponding 
$\sigma$-species single-particle densities, a result that suggests a tendency to an anti-ferromagnetic like state of the system \cite{erdmann2019phase,Koutentakis2019}.

\section{Quench dynamics of two fermionic impurities} \label{sec:two_impurities} 

Having discussed the nonequilibrium dynamics of a single impurity immersed in a Fermi sea in Sec. \ref{sec:one_impurity} we now analyze the dynamics of the same setup but including 
two fermionic impurities, i.e. $N_B = 2$. 
All other system parameters are identical to the previous Sec. \ref{sec:one_impurity}, namely $N_A=6$ fermions and $M_B=6M_A$ unless it is stated otherwise. 
Moreover, the Fermi sea is initially trapped in a double-well potential of frequency $\omega_A=0.1$, barrier height $h=8$ and width $w=1$ while 
the impurities reside in a harmonic oscillator with $\omega_B = 0.6\omega_A$. 
The mixture is initialized in its ground state described by the Hamiltonian of Eq.~(\ref{eq:hamilt}). 
As in Sec. \ref{sec:one_impurity} the species selective external potential induces initially a phase separation between the species almost independently 
of the value of $g_{AB}$ or the level of correlations i.e. HF and many-body approaches. 
In particular, the single-particle density of the impurities corresponds to a two-humped Gaussian located at the trap center while 
for the Fermi sea it is segregated into two fragments each of them having three local density maxima [Fig. \ref{fig:two_imp:ob_dens} (d)]. 
Notice that an increasing repulsion $g_{AB}$ gives rise to a slightly larger degree of phase separation between the species, see for instance Fig. \ref{fig:two_imp:ob_dens} (d) and 
Fig. \ref{fig:two_imp:ob_dens} (p). 
The dynamics of the system is triggered by switching-off the potential barrier of the Fermi sea  at $t=0$ from $h=8$ to $h=0$. 

\subsection{Dynamics of the density and the variance} \label{sec:two_imp:one_body_density}

\begin{figure}[ht] 
\includegraphics[width=0.45\textwidth]{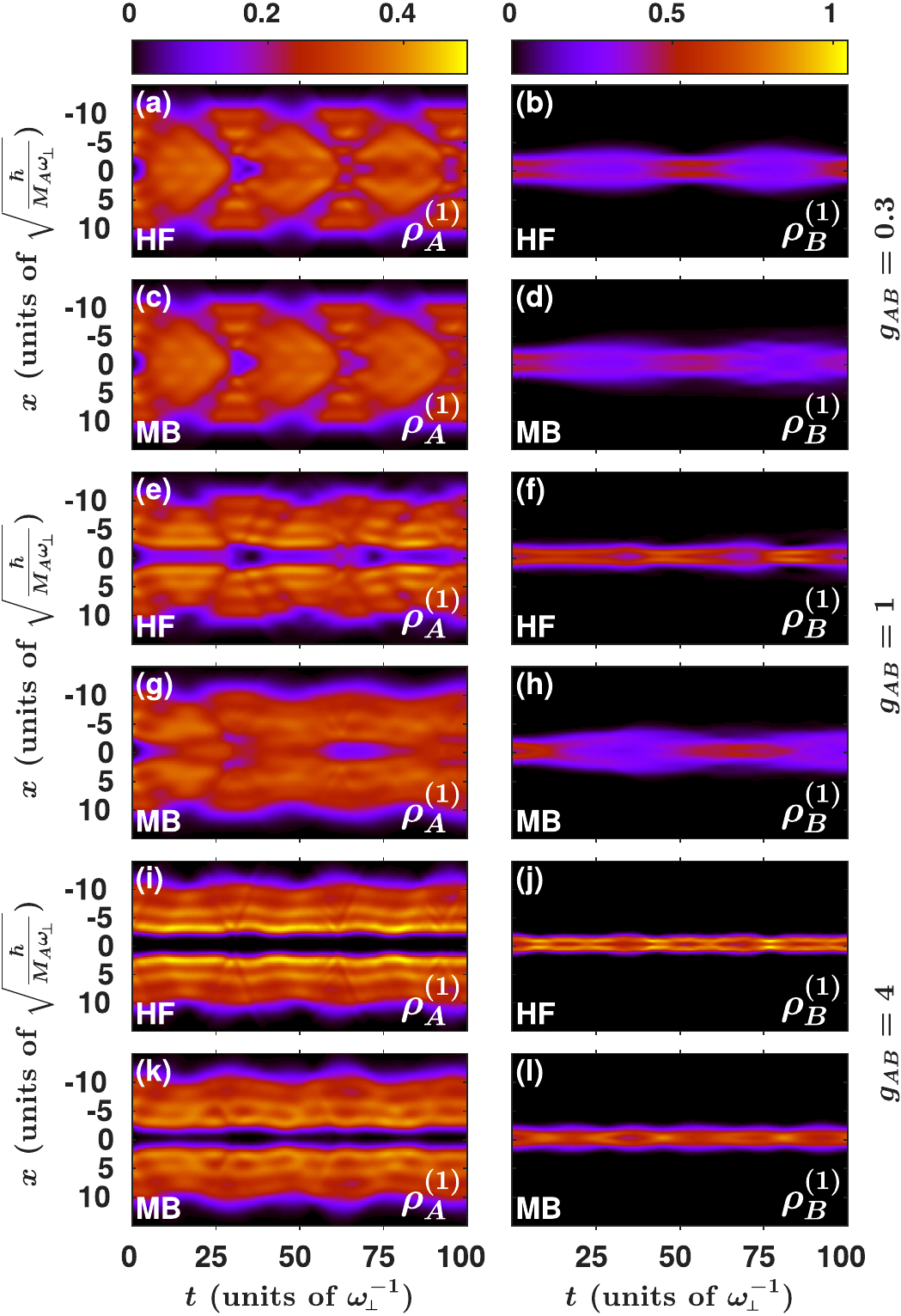}
\caption{Evolution of the $\sigma$-species one-body density $\rho^{(1)}_\sigma(x;t)$ of the FF mixture, following a sudden ramp-down of the potential barrier of the Fermi sea, 
for (a)-(d) weak ($g_{AB}=0.3$), (e)-(h) intermediate ($g_{AB}=1.0$) and (i)-(l) strong ($g_{AB}=4.0$) interspecies repulsions. 
In all cases the dynamics is presented in (a), (b), (e), (f), (i), (j) the HF and (c), (d), (g), (h), (k), (l) the many-body approach. 
The system consists of a Fermi sea (left panels) with $N_A = 6$ atoms and two impurities $N_B=2$ (right panels) exhibiting a mass-imbalance of $M_B=6 M_A$.} 
\label{fig:two_imp:ob_dens_dev}
\end{figure}

\begin{figure}[ht] 
\includegraphics[width=0.45\textwidth]{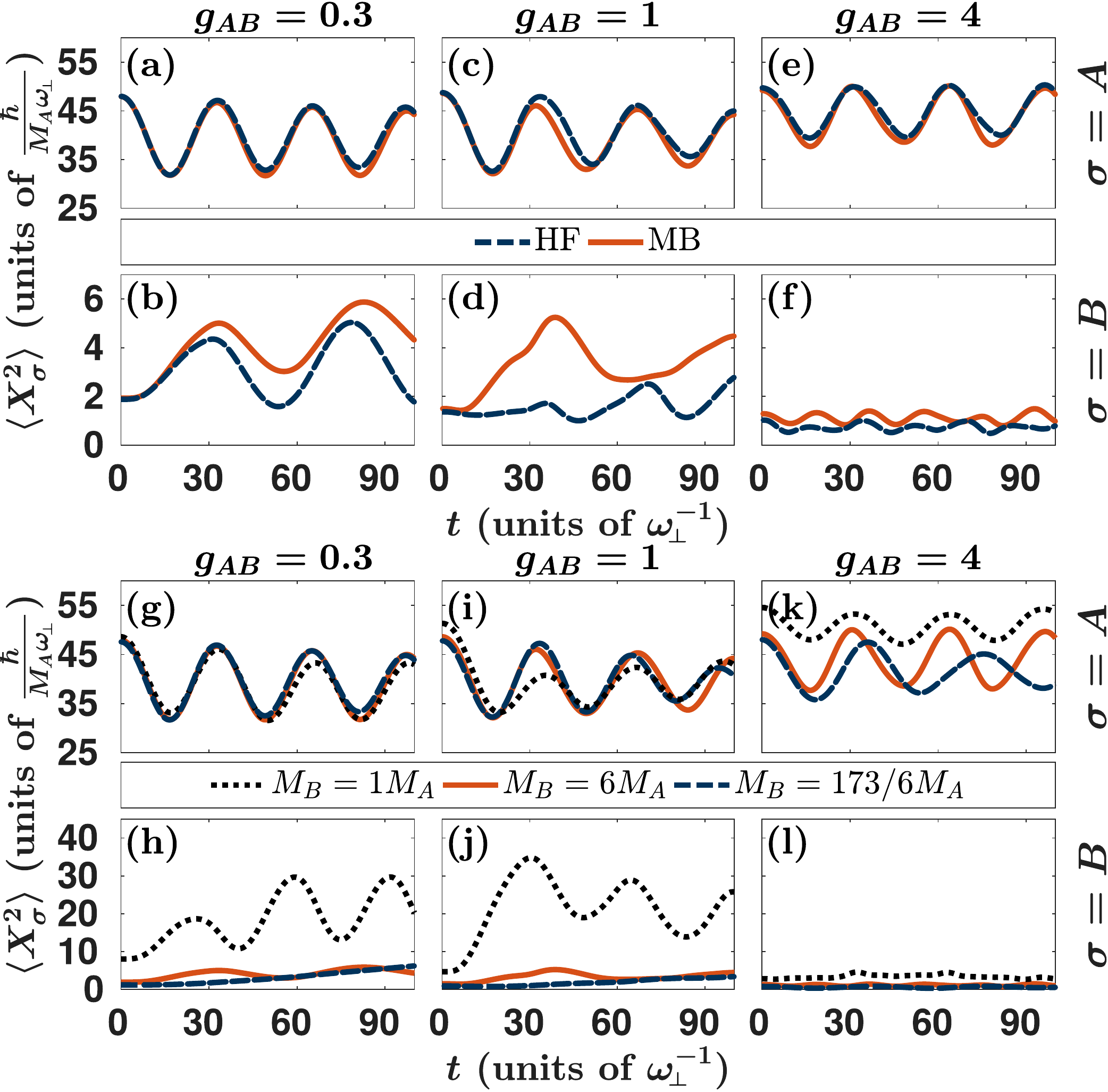}
\caption{Temporal-evolution of the variance of the one-body density $\braket{X^2_\sigma(t)}$ of (a), (c), (e) the Fermi sea ($\sigma=A$) and 
(b), (d), (f) the two spin-polarized impurities ($\sigma=B$) for varying interaction strength $g_{AB}$ (see legends). 
The dynamics of $\braket{X^2_\sigma(t)}$ is showcased both in the HF and the many-body approaches (see legends). 
The Fermi sea consists of $N_A=6$ atoms and the two heavy impurities $N_B=2$ possess $M_B=6M_A$. 
(g), (i), (k) $\braket{X^2_A(t)}$ and (h), (j), (l) $\braket{X^2_B(t)}$ in the many-body approach for distinct values of $g_{AB}$ 
and masses $M_B$ of the impurities (see legend).}
\label{fig:two_imp:variance}
\end{figure}

To monitor the quench dynamics of the FF mixture we inspect the $\sigma$-species single-particle density [Fig. \ref{fig:two_imp:ob_dens_dev}] and the underlying 
variance $\braket{X^2_\sigma(t)}$ [Fig. \ref{fig:two_imp:variance}] after the quench for varying interspecies repulsion. 
Quenching the potential barrier height of the Fermi sea from $h=8$ to $h=0$ induces a counterflow of the initial density fragments and as a 
consequence an overall breathing motion of $\rho^{(1)}_A(x;t)$ takes place with frequency $\omega_{A}^{br} \approx 0.2 = 2 \omega_A$. 
Interestingly, this collective motion is found to be almost insensitive of $g_{AB}$ \cite{erdmann2019phase,Mistakidis2018} and the presence of correlations as identified by 
the oscillatory motion of $\braket{X^2_A(t)}$ illustrated in Figs. \ref{fig:two_imp:variance} (a), (c), (e). 
Indeed, only small deviations are seen in $\braket{X^2_A(t)}$ between the HF and the many-body methods being more pronounced for intermediate interactions, e.g. $g_{AB}=1$ 
in Fig. \ref{fig:two_imp:variance} (c). 
We note that this similar behavior of $\braket{X^2_A(t)}$ for the HF and the many-body approach is in part attributed to the fact that for a certain $g_{AB}$ the postquench 
ground state density of the majority species remains to a large extend unchanged. 

\begin{figure}[ht]
\includegraphics[width=0.45\textwidth]{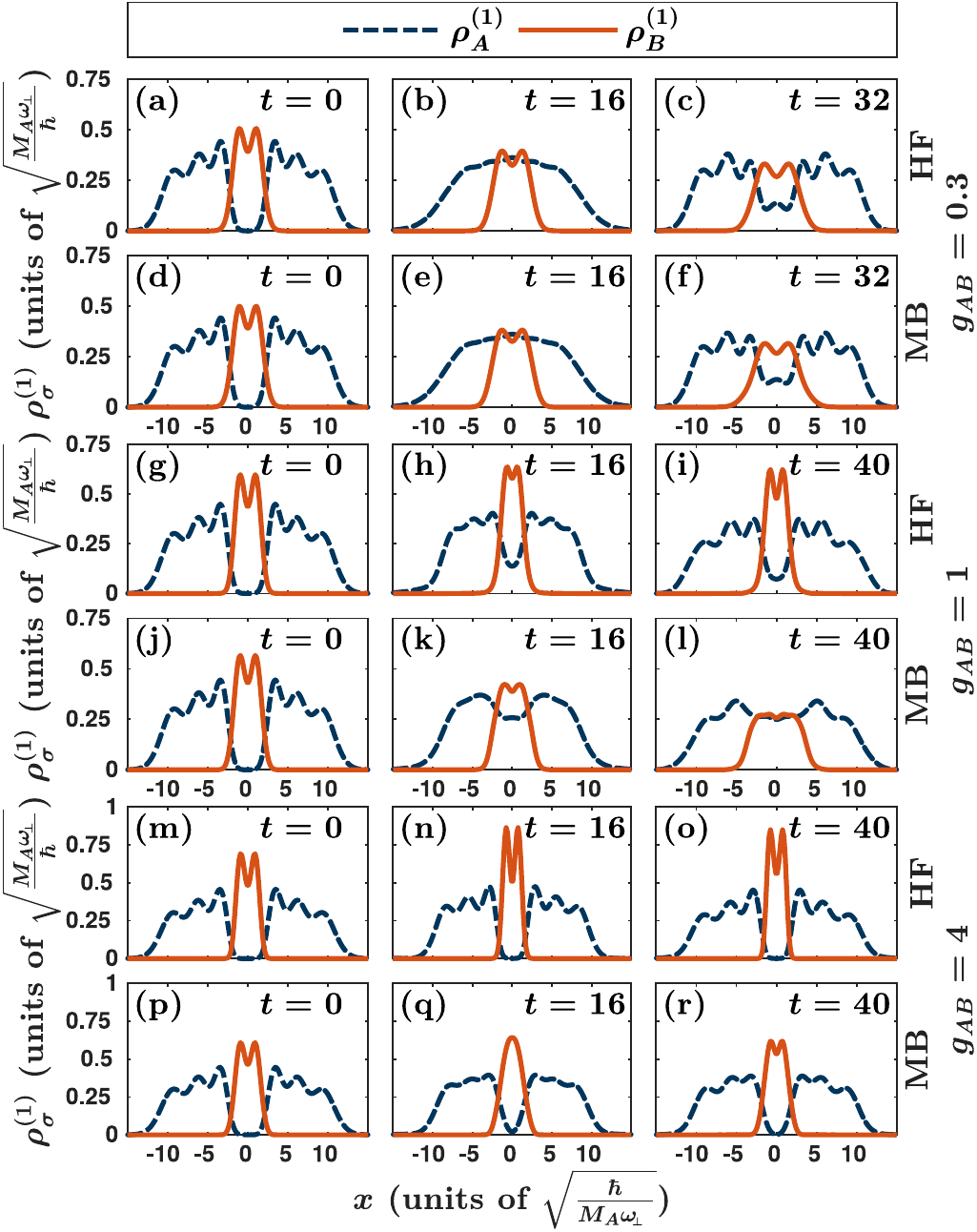}
\caption{One-body density profiles of the $\sigma$-species $\rho^{(1)}_\sigma(x;t)$ of the fermionic mixture for (a)-(f) weak ($g_{AB}=0.3$), (g)-(l) intermediate ($g_{AB}=1.0$) 
and (m)-(r) strong ($g_{AB}=4.0$) interspecies repulsions at distinct time-instants of the evolution (see legend). 
The corresponding densities are obtained within (a)-(c), (g)-(i), (m)-(o) the HF and (d)-(f), (j)-(l), (p)-(r) the many-body approach. 
The fermionic mixture consists of a Fermi sea with $N_A=6$ atoms initially confined in a double-well and two harmonically trapped heavy spin-polarized 
impurities $N_B=2$ with $M_B=6M_A$. 
To trigger the dynamics we switch-off at $t=0$ the central potential barrier of the double-well.}
\label{fig:two_imp:ob_dens}
\end{figure} 

For weak interspecies repulsions, such as $g_{AB}=0.3$ shown in Figs. \ref{fig:two_imp:ob_dens_dev} (a), (b), the initial density fragments of the Fermi sea in the 
HF approach [Fig. \ref{fig:two_imp:ob_dens} (a)], 
following the quench, collide at the postquench trap center i.e. $x=0$ forming a wider distribution [Fig. \ref{fig:two_imp:ob_dens} (b)]. 
The latter consequently breaks into two density fragments each of them possessing three local maxima and moving to the corresponding edge of the harmonic trap where the first period of the 
breathing motion is completed. 
A similar to the above-described motion is repeated for later evolution times at every breathing period Fig. \ref{fig:two_imp:ob_dens_dev} (a). 
Notice also that the density dip of the splitted $\rho^{(1)}_A(x;t)$ at $x=0$ is shallower during the dynamics than at $t=0$, see Figs. \ref{fig:two_imp:ob_dens} (a), (c). 
As a result of the motion of the Fermi sea, the impurities are perturbed performing also a breathing dynamics around the trap center [Fig. \ref{fig:two_imp:ob_dens_dev} (b)] 
while their one-body density exhibits a two-hump structure \cite{erdmann2019phase}, see Figs. \ref{fig:two_imp:ob_dens} (a)-(c). 
Indeed, $\braket{X^2_B(t)}$ [Fig. \ref{fig:two_imp:variance} (a)] shows an oscillatory behavior which reflects the expansion and contraction of the impurities cloud 
[Fig. \ref{fig:two_imp:ob_dens_dev} (b)] with a dominant frequency $\omega_B^{br}\approx0.13$. 
Also, the mixture remains miscible, at least on the one-body level, for the entire time-evolution due to the spatially overlapping single-particle densities of the two species, 
see Figs. \ref{fig:two_imp:ob_dens_dev} (a), (b). 
The same overall dynamical response occurs for both species also within the many-body framework, see Figs. \ref{fig:two_imp:ob_dens_dev} (c), (d). 
The most noticeable difference compared to the HF evolution is that $\rho^{(1)}_B(x;t)$ shows an overall expansion tendency as identified by the 
increasing amplitude of $\braket{X^2_B(t)}$ which simultaneously oscillates as depicted in Fig. \ref{fig:two_imp:variance} (b). 
Moreover, the density maxima appearing in the fragmented structure of $\rho^{(1)}_A(x;t)$ are slightly shallower within the many-body [Fig. \ref{fig:two_imp:ob_dens} (f)] as compared 
to the HF approach [Fig. \ref{fig:two_imp:ob_dens} (c)]. 

For larger interspecies repulsions, e.g. $g_{AB}=1$, we observe a tendency for phase separation between the species [Figs. \ref{fig:two_imp:ob_dens_dev} (e), (f)] in the HF framework, 
a behavior that takes place already in the case of single impurity [Figs. \ref{fig:one_imp:ob_dens_dev} (e), (f)]. 
This tendency for spatial phase separation is also clearly visualized by the density snapshots presented in Figs. \ref{fig:two_imp:ob_dens} (g)-(h). 
Here, in contrast to the weakly interacting case, the fragmented structure of $\rho^{(1)}_A(x;t)$ persists throughout the evolution. 
Accordingly, the density dip of $\rho^{(1)}_A(x;t)$ around $x=0$ effectively traps the impurities whose density exhibits a localized two-hump distribution 
in the vicinity of the trap center, see Figs. \ref{fig:two_imp:ob_dens} (g), (h). 
In particular, the impurities perform an ``irregular'' contraction and expansion dynamics as captured by the multifrequency and non-constant amplitude oscillations 
of $\braket{X^2_B(t)}$ [Fig. \ref{fig:two_imp:variance} (d)]. 
The dominant frequencies participating in the dynamics of $\braket{X^2_B(t)}$ are $\omega_1\approx0.19$ and $\omega_2\approx0.31$.   
Turning to the many-body evolution, the postquench pattern formation of each species is drastically altered from its HF counterpart and the degree of species separation is 
suppressed, see Figs. \ref{fig:two_imp:ob_dens_dev} (g), (h). 
More specifically, the local density maxima building upon $\rho^{(1)}_A(x;t)$ either completely disappear [Fig. \ref{fig:two_imp:ob_dens} (k)] or their number reduces 
[Fig. \ref{fig:two_imp:ob_dens} (l)] while being much shallower than in the HF approach. 
Also the density dip of $\rho^{(1)}_A(x;t)$ around $x=0$ becomes more shallow and wider in the many-body dynamics, compare Fig. \ref{fig:two_imp:ob_dens} (h) with Fig. \ref{fig:two_imp:ob_dens} (k), or 
exhibits two local minima at the location of the humps of $\rho^{(1)}_B(x;t)$ [Fig. \ref{fig:two_imp:ob_dens} (l)]. 
As a consequence of the wider density dip of $\rho^{(1)}_A(x;t)$ which acts as an effective trap for $\rho^{(1)}_B(x;t)$, the latter appears to be wider from the HF case. 
Moreover, $\rho^{(1)}_B(x;t)$ exhibits a predominant expansion tendency until $t\approx35$ as captured by the increasing amplitude of $\braket{X^2_B(t)}$ while later on 
contracts and expands, see Fig. \ref{fig:two_imp:variance} (d) with oscillation frequencies $\omega_1\approx0.13$ and $\omega_2\approx0.25$. 

Entering the strongly interacting regime, $g_{AB}=4$, a complete phase separation between the species occurs within the HF approach, see Figs. \ref{fig:two_imp:ob_dens_dev} (i), (j). 
The single-particle density of the Fermi sea is segregated having two fragments, each of them exhibiting three local density maxima, while remaining 
fully dipped in the vicinity of $x=0$ throughout the evolution [Figs. \ref{fig:two_imp:ob_dens} (m)-(o)]. 
In turn, the impurities density distribution $\rho^{(1)}_B(x;t)$ is quite localized being effectively trapped within the density dip of 
$\rho^{(1)}_A(x;t)$ and shows a two-hump structure. 
Moreover, $\rho^{(1)}_B(x;t)$ contracts and expands in the course of the evolution, a behavior that can be identified by the multifrequency small amplitude oscillations 
of $\braket{X^2_B(t)}$ [Fig. \ref{fig:two_imp:variance} (f)]. 
Here, the dominant participating frequencies correspond to $\omega_1\approx0.19$, $\omega_2\approx0.31$ and $\omega_3\approx 0.63$. 
Within the many-body approach a similar to the above-described dynamical response occurs for each species, see Figs. \ref{fig:two_imp:ob_dens_dev} (k), (l). 
However, the local density maxima in each fragment of $\rho^{(1)}_A(x;t)$ appear to be more shallow than their HF counterparts and the density dip of $\rho^{(1)}_A(x;t)$ is slightly wider, 
see Figs. \ref{fig:two_imp:ob_dens} (q), (r). 
Additionally, the width of $\rho^{(1)}_B(x;t)$ is somewhat larger as compared to the HF case [Fig. \ref{fig:two_imp:ob_dens_dev} (l) and Fig. \ref{fig:two_imp:ob_dens} (q), (r)] 
and the two-hump structure is lost at certain time-intervals of the contraction of $\rho^{(1)}_B(x;t)$, see e.g. Fig. \ref{fig:two_imp:ob_dens} (q). 
Indeed, also here the impurities cloud undergoes a breathing motion with a slightly increased amplitude compared to the HF case as can be seen from the multifrequency 
oscillatory behavior of $\braket{X^2_B(t)}$ illustrated in Fig. \ref{fig:two_imp:variance} (f). 
Note that the predominant frequencies of these oscillations are $\omega_1\approx 0.19$, $\omega_2\approx0.38$ and $\omega_3\approx0.6$. 
Finally, there is a small spatial overlap between $\rho^{(1)}_A(x;t)$ and $\rho^{(1)}_B(x;t)$ in the many-body [Fig. \ref{fig:two_imp:ob_dens} (q)] as compared to the 
HF case [Fig. \ref{fig:two_imp:ob_dens} (n)], leading in turn to a smaller degree of phase separation in the former case.

\subsection{Impact of the impurities mass} \label{sec:two_imp:masses}

Next, we study the influence of the impurities mass on the overall expansion of each species by invoking the $\sigma$-species position variance 
$\braket{X^2_{\sigma}(t)}$ for different values of $g_{AB}$, shown in Figs. \ref{fig:two_imp:variance} (g)-(l). 
To perform this comparison, as in Sec. \ref{sec:one_imp:mass_imbalance}, we assume a mass-balanced and a mass-imbalanced mixture where $M_A=M_B$, $\omega_A=\omega_B$ and 
$M_B=(173/6)M_A$, $\omega_B=0.0125\omega_A$ respectively. 
Evidently for $g_{AB}=0.3$ the dynamical response of the Fermi sea as captured via $\braket{X^2_{A}(t)}$ is to a great extent insensitive to impurities mass [Fig. \ref{fig:two_imp:variance} (g)]. 
However for intermediate repulsions, $g_{AB}=1$, $\braket{X^2_{A}(t)}$ is almost identical between the cases $M_B=6M_A$ and $M_B=(173/6)M_A$ performing decaying amplitude oscillations while 
for a light impurity the oscillation amplitude of $\braket{X^2_{A}(t)}$ becomes slightly smaller [Fig. \ref{fig:two_imp:variance} (i)]. 
The effect of the impurities mass becomes more prominent at strong repulsions, such as $g_{AB}=4$, where both the oscillation amplitude and the frequency of $\braket{X^2_{A}(t)}$ depend crucially 
on $M_B$ [Fig. \ref{fig:two_imp:variance} (k)]. 
Namely, for heavier impurities the decay of the oscillation amplitude of $\braket{X^2_{A}(t)}$ is more pronounced and the corresponding frequency becomes smaller. 
Additionally, the oscillation amplitude of $\braket{X^2_{A}(t)}$ is reduced in the case of a light impurity since it affects less its bath compared to a heavier one. 
It should be stressed at this point that the postquench ground state of the Fermi sea is the same for different mass ratios for small $g_{AB}$, while alterations 
come into play (especially between the mass-balanced and the mass-imbalanced scenaria) for increasing $g_{AB}$. 

Turning to the impurities, we observe that the dynamical behavior of $\braket{X^2_{B}(t)}$ is quite insensitive for heavy impurities independently of $g_{AB}$, 
see Figs. \ref{fig:two_imp:variance} (h), (j), (l). 
Notice that for $M_B=6M_A$ the oscillation amplitude of $\braket{X^2_{B}(t)}$ is slightly larger than for $M_B=(173/6)M_A$, especially for weak and intermediate 
repulsions i.e. $g_{AB}=0.3,1$, indicating that 
the expansion of heavier impurities is reduced. 
This enhanced expansion tendency of the impurities becomes even more pronounced in the case of lighter ones, and in particular for weak and intermediate interspecies 
repulsions [Figs. \ref{fig:two_imp:variance} (h), (j)]. 
Here, $\braket{X^2_{B}(t)}$ shows an overall increasing tendency while oscillating further suggesting that the mobility of impurities is larger for a decreasing mass \cite{schmidt2018universal}.

\subsection{Degree of correlations} \label{sec:two_imp:entropy_and_fragmentation}

In order to reveal the presence of inter- and intraspecies correlations in the course of the evolution of the FF mixture we resort to the corresponding von-Neumann entropy 
$S(t)$ described by Eq. (\ref{eq:entropy}) and the deviation from unity $F_{\sigma}(t)$ of the first $N_{\sigma}$ natural populations, see also Eq. (\ref{eq:fragmentation}). 
As already argued in Sec. (\ref{sec:observable-of-interest}), the many-body state of the mixture is termed interspecies correlated only if $S(t)\neq0$ \cite{wlodzynski2020geometry,erdmann2019phase} 
while it is said to be intraspecies correlated when $F_{\sigma}(t)>0$ \cite{erdmann2019phase,erdmann2018}. 
We also remind at this point that we consider spin-polarized fermions in each species and therefore the existence of intraspecies correlations is induced by the presence 
of interspecies ones \cite{erdmann2019phase,mistakidis2019correlated,pkecak2019intercomponent}. 

\begin{figure}[ht] 
\includegraphics[width=0.45\textwidth]{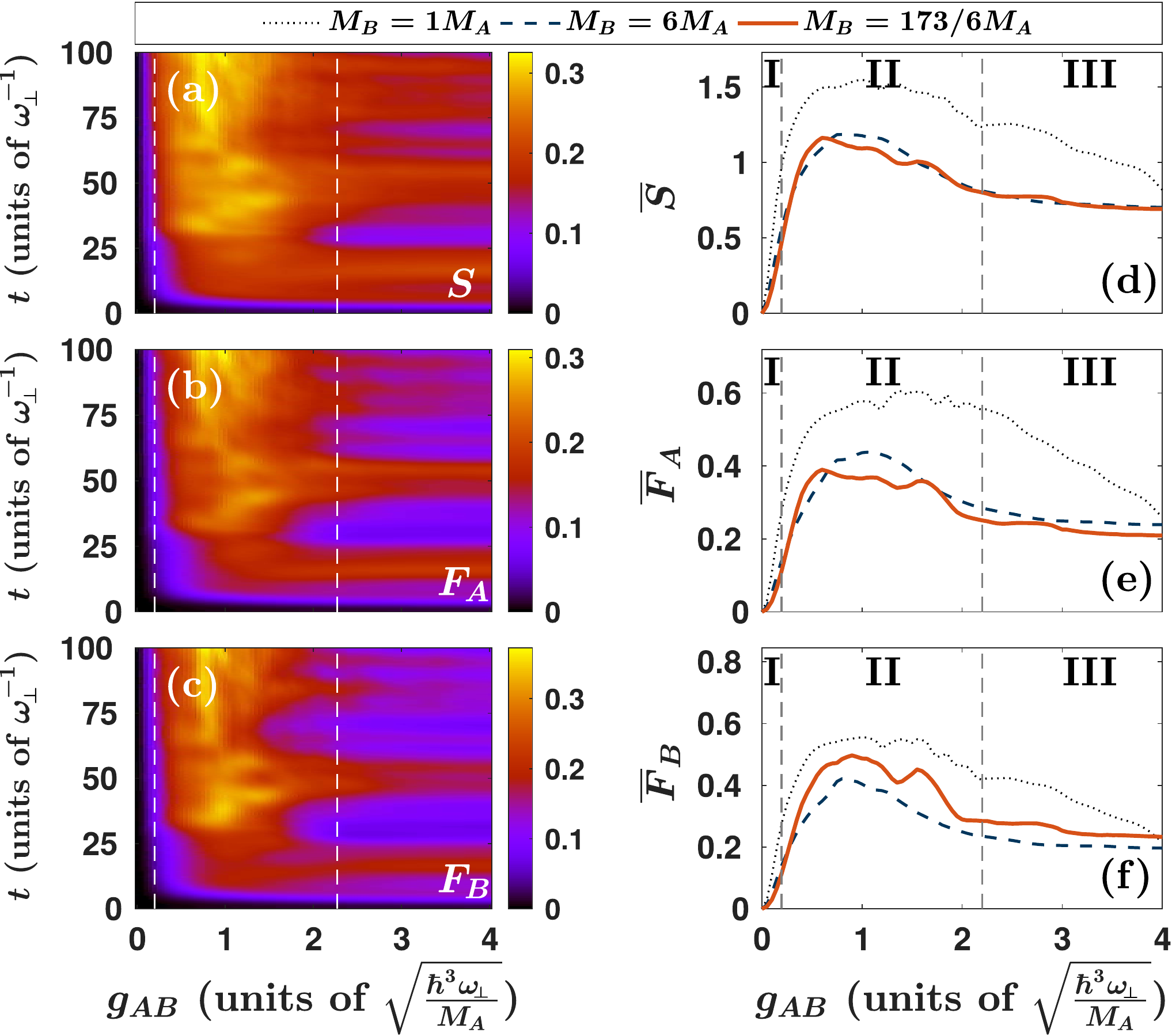}
\caption{Dynamics of (a) the von-Neumann entropy $S(t)$, and the deviation from unity of (b) the first six natural populations of the Fermi sea $F_A(t)$ and (c) the first two natural 
populations of the impurities $F_B(t)$ for different repulsions $g_{AB}$. 
The system comprises of $N_A=6$ fermions and two spin-polarized impurities $N_B=2$ having a mass-imbalance of $M_B=6M_A$. 
Time-averaged (d) von-Neumann entropy i.e. $\bar{S}(t)$, (e) $\bar{F}_A(t)$ and (d) $\bar{F}_B(t)$ for three different masses of the impurities (see legend). 
The vertical dashed gray lines mark the three interaction regimes at which the correlation measures exhibit a distinct behavior. 
In all cases the dynamics is induced at $t=0$ by a sudden ramp-down of the potential barrier of the Fermi sea.}
\label{fig:two_imp:entropy_and_fragmentation}
\end{figure}

Figure \ref{fig:two_imp:entropy_and_fragmentation} depicts $S(t)$ [Fig. \ref{fig:two_imp:entropy_and_fragmentation} (a)], $F_A(t)$ [Fig. \ref{fig:two_imp:entropy_and_fragmentation} (b)] 
and $F_B(t)$ [Fig. \ref{fig:two_imp:entropy_and_fragmentation} (c)] together with their time-average values i.e. $\bar{S}$, $\bar{F}_A$ and $\bar{F}_B$ 
[Figs. \ref{fig:two_imp:entropy_and_fragmentation} (e)-(f)] for varying $g_{AB}$ after quenching the barrier height of the Fermi sea from $h=8$ to $h=0$ when $M_B=6M_A$. 
We remark that an overall similar phenomenology to the single impurity case [Fig. \ref{fig:one_imp:entropy_and_fragmentation}] occurs for the correlation measures $S(t)$, 
$F_A(t)$, $F_B(t)$ and their time-averaged counterparts $\bar{S}$, $\bar{F}_A$ and $\bar{F}_B$ respectively. 
However, as can be deduced by comparing Fig. \ref{fig:one_imp:entropy_and_fragmentation} and Fig. \ref{fig:two_imp:entropy_and_fragmentation} the degree of correlations is enhanced and their 
increase with respect to $g_{AB}$ is steeper in the case of two impurities. 

In particular, the mixture is initially phase separated and therefore the degree of both inter- and intraspecies correlations vanishes for every $g_{AB}$ namely 
$S(t=0)=0$, $F_{A}(t=0)=0$ and $F_{B}(t=0)=0$, see Figs. \ref{fig:two_imp:entropy_and_fragmentation} (a), (b), (c). 
Moreover, during the evolution we can infer the development of inter- and intraspecies correlations followed by the consecutive increase of $S(t)$, $F_A(t)$ and $F_B(t)$ whose 
response can be divided into three distinct interaction regimes indicated by I, II and III in Fig. \ref{fig:two_imp:entropy_and_fragmentation}. 
More specifically, within region I characterized by $0<g_{AB}<0.2$ the degree of inter- and intraspecies correlations is small since 
$\bar{S}<0.3$, $\bar{F}_A<0.1$ and $\bar{F}_B<0.05$ [Figs. \ref{fig:two_imp:entropy_and_fragmentation} (d), (e), (f)]. 
However, for increasing interactions $0.2<g_{AB}<2.2$ (region II) the interparticle correlations are significantly enhanced which can be identified 
via the macroscopic values of $S(t)$, $F_{A}(t)$ and $F_{B}(t)$ as time evolves, see Figs. \ref{fig:two_imp:entropy_and_fragmentation} (a), (b), (c). 
Indeed, the impact of correlations is maximized in this region II since also $\bar{S}$, $\bar{F}_A$ and $\bar{F}_B$ show an increasing tendency for a larger $g_{AB}$ and 
reach their maxima in the vicinity of $g_{AB}\approx0.8$. 
For stronger interactions, namely $g_{AB}>2$ (region III), the degree of inter- and intraspecies correlations is reduced compared to region II but it 
still remains non-negligible. 
The decreasing role of correlations in this region is testified by the oscillating behavior of $S(t)$, $F_{A}(t)$ and $F_{B}(t)$ between the values 0.05 and 0.25 
as well as the smaller average values of $\bar{S}$, $\bar{F}_A$ and $\bar{F}_B$ than in region II. 
This reduced tendency of interparticle correlations can be attributed to the emergent phase separation between the species for strong interactions, 
see e.g. Figs. \ref{fig:two_imp:ob_dens_dev} (k), (l). 

Furthermore, the time-averaged correlation measures for different impurities mass are illustrated in Figs. \ref{fig:two_imp:entropy_and_fragmentation} (d), (e), (f) 
with respect to $g_{AB}$. 
Evidently, the overall shape of $\bar{S}$, $\bar{F}_A$ and $\bar{F}_B$ for increasing $g_{AB}$ remains qualitatively similar for varying impurities mass but the 
corresponding values of these measures depend on $M_B$. 
Indeed, we observe that $\bar{S}$, $\bar{F}_A$ and $\bar{F}_B$ are larger for the light impurities case while they do not significantly change between the $M_B=6M_A$ and 
$M_B=(173/6)M_A$ cases. 
It is worth mentioning here that the corresponding clear hierarchy of inter- and intraspecies correlations in terms of $M_B$ appearing for a single 
impurity [Fig. \ref{fig:two_imp:entropy_and_fragmentation}] is somewhat distorted for two impurities. 
However, the fact that mass-balanced mixtures become more correlated in the course of the evolution than mass-imbalanced ones holds independently of the number 
of impurities.

\subsection{Two-body correlations and induced impurity-impurity interactions}

\begin{figure}[ht] 
\includegraphics[width=0.45\textwidth]{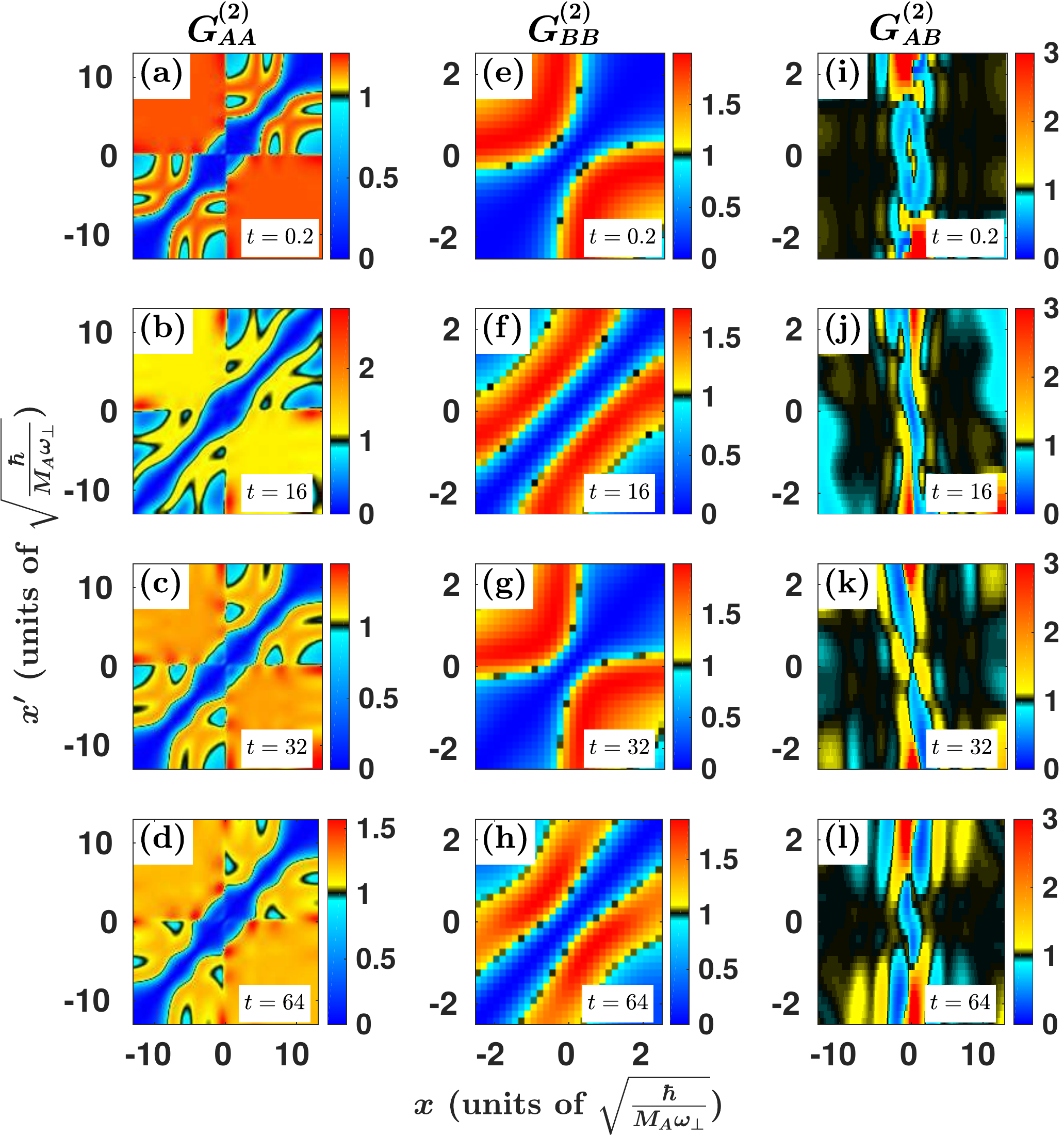}
\caption{Snapshots of the two-body (a)-(d) intraspecies bath $G^{(2)}_{AA}(x,x^\prime;t)$, (e)-(h) impurity-impurity $G^{(2)}_{BB}(x,x^\prime;t)$ and (i)-(l) 
interspecies $G^{(2)}_{AB}(x,x^\prime;t)$ correlations at strong interactions $g_{BI}=4$ at different times of the evolution (see legend). 
The Fermi sea comprises of $N_A=6$ atoms trapped in a double-well and the heavy single impurity $N_B=1$ with $M_B=6M_A$ is confined in a harmonic trap.  
To induce the dynamics we follow a quench of the height of the central potential barrier of the double-well to zero.} 
\label{fig:two_imp:correlations2}
\end{figure}

To expose the intra- and interspecies two-body correlation mechanisms that participate in the nonequilibrium dynamics, we next monitor the two-body correlation functions 
$G_{\sigma\sigma'}^{(2)}(x,x',t)$ introduced in Eq. (\ref{two_body_cor}). 
Recall that the case of $G^{(2)}_{\sigma\sigma^\prime}(x,x^\prime;t) > 1$ [$G^{(2)}_{\sigma\sigma^\prime}(x,x^\prime;t) < 1$] signifies the occurrence of 
two-body correlations [anti-correlations] between two particles of species $\sigma$ and $\sigma^\prime$ respectively while for $G_{\sigma\sigma^\prime}(x,x^\prime;t)=1$ 
the two particles are uncorrelated. 
The emergent time-evolution of the two-body correlation functions within and between the species after a quench of the potential barrier height of the Fermi sea from $h=8$ to $h=0$ 
in the case of $M_B=6M_A$ and for $g_{AB}=4$ is shown in Fig. \ref{fig:two_imp:correlations2}. 

Focusing on the dynamics of $G_{AA}^{(2)}(x,x',t)$ [Figs. \ref{fig:two_imp:correlations2} (a)-(d)] we observe that similar to the single impurity case, see also Figs. \ref{fig:one_imp:2b_correlations} (a)-(d), 
correlation patterns are formed. 
First, due to the Pauli exclusion principle two fermions can not reside at the same location and therefore two-body anti-correlations occur in the diagonal i.e. $G_{AA}^{(2)}(x,x'=x,t)<1$ 
[Figs. \ref{fig:two_imp:correlations2} (a)-(d)] for the entire evolution. 
However, two-body correlations develop between different spatial regions indicated by the fact that the off-diagonal elements satisfy $G_{AA}^{(2)}(x,x'\neq x,t)>1$. 
Also since $\rho^{(1)}_A(x;t)$ shows two fragments, with each of them having at most three local maxima [Figs. \ref{fig:two_imp:ob_dens} (p)-(r)], the fact 
that $G_{AA}^{(2)}(x,x'\neq x,t)>1$ implies that two fermions of the Fermi sea populate either the same fragment but different local density maxima or they  
reside at different fragments. 

Turning to the impurities correlation function i.e. $G_{BB}^{(2)}(x,x^\prime;t)$ [Figs. \ref{fig:two_imp:correlations2} (e)-(h)] we can infer that it mainly alternates between 
two different patterns and most importantly it suggests the emergence of attractive impurity-impurity induced interactions which are of course mediated by the Fermi sea. 
The correlation hole appearing in the diagonal of $G_{BB}^{(2)}(x,x^\prime=x;t)$ throughout the evolution stems from the Pauli's exclusion principle that prevents 
two spin-polarized fermions to reside at the same position. 
Recall that $\rho^{(1)}_B(x;t)$ performs a breathing motion exhibiting a two-hump structure during its expansion and a Gaussian-like distribution when it contracts, see also 
Figs. \ref{fig:two_imp:ob_dens_dev} (k), (l) and Figs. \ref{fig:two_imp:ob_dens} (p)-(r). 
The aforementioned correlation patterns building upon $G_{BB}^{(2)}(x,x^\prime;t)$ are in turn related to this breathing motion. 
Indeed when $\rho^{(1)}_B(x;t)$ expands the two fermions are likely to be close to the trap center with one of them residing around $0<x<2$ and the other at $-2<x'<0$ 
as can be deduced from the fact that $G_{BB}^{(2)}(x,x^\prime;t)>1$ [Figs. \ref{fig:two_imp:correlations2} (e), (g)]. 
In this case each fermion is located on the side of either the left ($x<0$) or the right ($x>0$) hump of $\rho^{(1)}_B(x;t)$. 
However, during the contraction of $\rho^{(1)}_B(x;t)$ the two fermions tend to approach each other [Figs. \ref{fig:two_imp:correlations2} (f), (h)] since short distance 
spatial regions show two-body correlations, see e.g. $G_{BB}^{(2)}(1<x<2,-1<x^\prime<0;t=16)>1$, and longer distance regions become two-body 
anti-correlated, see e.g. $G_{BB}^{(2)}(-2<x<-1,0.1<x^\prime<2;t=16)<1$. 
This latter behavior of $G_{BB}^{(2)}(x,x^\prime;t)$ manifests an attractive tendency between the impurities and it is suggestive of the emergence of attractive 
induced interactions between the impurities \cite{mistakidis2019many,mistakidis2019induced}. 
\begin{figure}[ht] 
\includegraphics[width=0.45\textwidth]{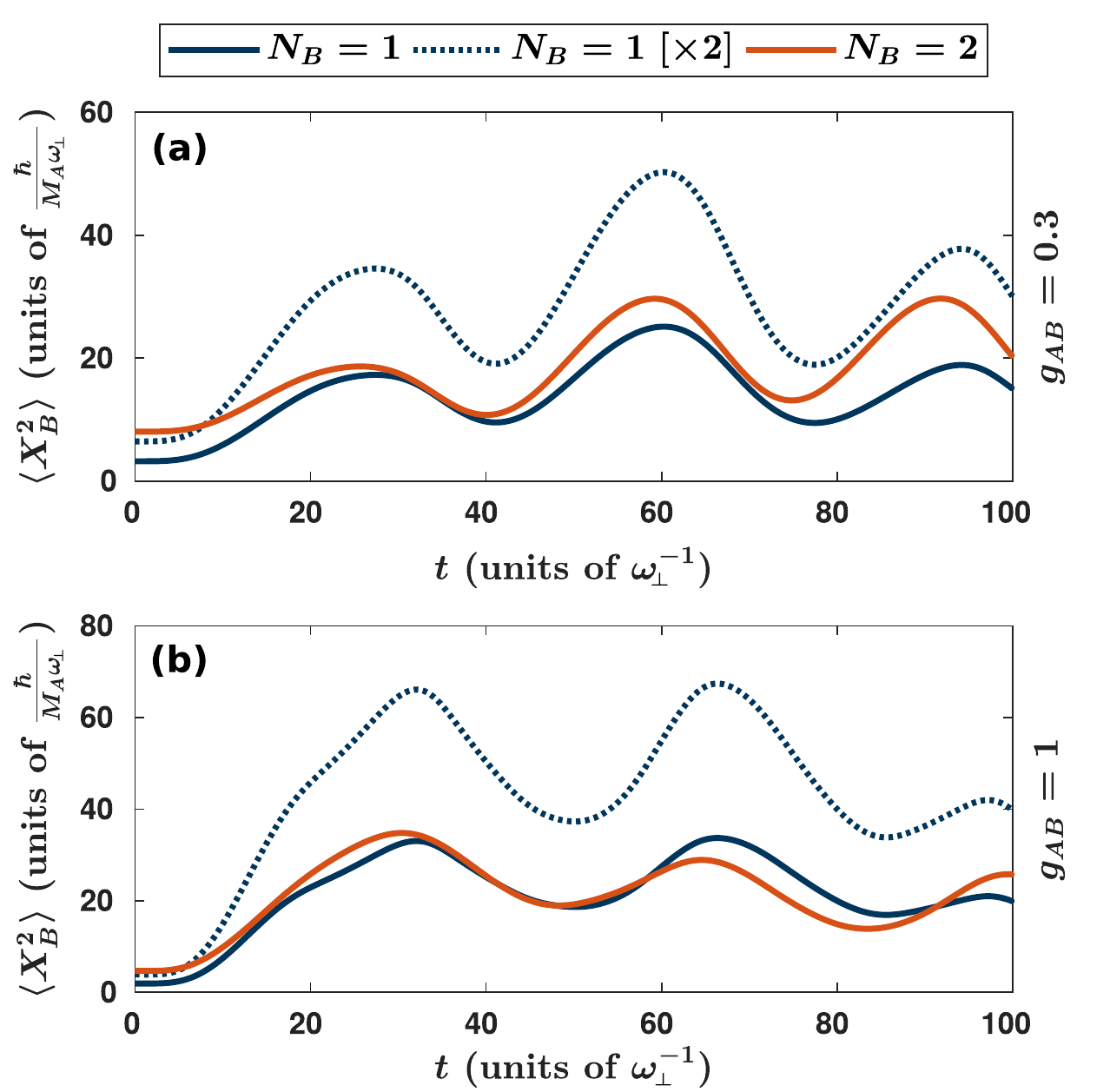}
\caption{Dynamics of the variance of the impurities one-body density $\braket{X^2_B(t)}$ within the many-body approach for (a) $g_{AB}=0.3$ and 
(b) $g_{AB}=1$ in the case of a single impurity, two uncorrelated impurities and two correlated fermionic impurities (see legend). 
The Fermi sea consists of $N_A=6$ atoms in a double-well and either $N_B=1$ or $N_B=2$ harmonically confined fermionic impurities with $M_B=M_A$. 
The dynamics is triggered by quenching at $t=0$ the height of the central potential barrier of the double-well from $h=8$ to $h=0$.} 
\label{fig:variances_compare}
\end{figure}

On the other hand, the two-body interspecies correlation function $G_{AB}^{(2)}(x,x',t)$ [Figs. \ref{fig:two_imp:correlations2} (i)-(l)] reveals from a two-body perspective the phase separated 
behavior between the impurities and the Fermi sea, already discussed on the single-particle level, see also Figs. \ref{fig:two_imp:ob_dens_dev} (k), (l). 
For instance, two-body anti-correlations occur between a fermion of the bath and one of the impurities around the trap center $x=0$ since 
$G_{AB}^{(2)}(x=0,x'=0,t)<1$. 
Additionally, two-body correlations emerge among an impurity atom and a fermion of the bath located in either the left ($x<0$) or the right ($x>0$) vicinity of the density notch 
of the Fermi sea, see e.g. $G_{AB}^{(2)}(0<x<0.1,-2<x'<-0.8,t)>1$ and $G_{AB}^{(2)}(-0.1<x<0,0.5<x'<2,t)>1$ in Figs. \ref{fig:two_imp:correlations2} (i)-(l). 
Notice that regions beyond the spatial extension of $\rho^{(1)}_B(x;t)$ exhibit in essence a two-body uncorrelated behavior 
e.g. $G_{AB}^{(2)}(x>2,-2<x'<2,t)\approx 1$ and $G_{AB}^{(2)}(x<-2,-2<x'<2,t)\approx 1$. 

\subsection{Comparing the spatial size of a single and two fermionic impurities}

Having identified signatures of attractive impurity-impurity induced interactions mediated by the Fermi sea in the spatiotemporal evolution of the impurities 
two-body correlation function $G_{BB}^{(2)}(x,x';t)$ [Figs. \ref{fig:two_imp:correlations2} (e)-(h)] we subsequently compare the behavior of the variance of a single 
and two impurities $\braket{X^2_B(t)}$ [Eq. (\ref{eq:variance})]. 
Note that $\braket{X^2_B(t)}$ provides an estimate of the spatial size of the impurities cloud during the evolution and the variance $\braket{X^2_B(t)}$ of 
two explicitly uncorrelated impurities is essentially provided by $\braket{X^2_B(t)}$ of a single one multiplied by a factor of two. 

In particular, we aim to expose the existence of impurity-impurity correlations by constrasting $\braket{X^2_B(t)}$ in the case of two explicitly 
uncorrelated impurities with the corresponding variance of two fermionic impurities. 
Recall that such a comparison has already been used in the case of bosonic impurities in order to illustrate the presence of impurity-impurity correlations \cite{mistakidis2019induced}. 
As a case example for this comparison we invoke a mass balanced FF mixture consisting of a single or two impurities and showcase $\braket{X^2_B(t)}$ for a single, 
two uncorrelated and two fermionic impurities in Fig. \ref{fig:variances_compare} for different interspecies interactions strengths $g_{AB}$. 
As already discussed in Secs. \ref{sec:one_impurity} and \ref{sec:two_impurities} $\braket{X^2_B(t)}$ for both a single and two impurities and independently of the value of 
$g_{AB}$ exhibits an ``irregular'' oscillatory behavior with an overall increasing tendency. 
Most importantly, closely inspecting Fig. \ref{fig:variances_compare} we can infer that $\braket{X^2_B(t)}$ of two uncorrelated impurities acquires larger values 
than $\braket{X^2_B(t)}$ referring to two fermionic impurities. 
This behavior indicates the involvement of induced correlations between the fermionic impurities mediated by the Fermi sea, thus further supporting the attraction 
tendency between the impurities observed in their two-body correlation function $G_{BB}^{(2)}(x,x';t)$. 
We remark a similar phenomenology occurs also for heavier impurities and other interspecies repulsions (not shown here for brevity).

\section{Conclusions} \label{sec:conclusions} 

We have investigated the nonequilibrium correlated quantum dynamics of a single and two heavy fermionic impurities immersed in a one-dimensional Fermi sea. 
The latter is initially trapped in a double-well while the impurities reside in a harmonic oscillator. 
The mixture is prepared in its ground state and the dynamics is triggered by ramping down the central potential barrier of the double well, thus inducing a counterflow 
of the Fermi sea which in turn perturbs the impurities. 
The emergent dynamics is studied in detail on both the one- and two-body level for a wide range of repulsive interspecies interactions and selected mass ratios. 
To infer the crucial role of correlations we directly compare and contrast the predictions of the Hartree-Fock and the many-body approach. 
A multitude of interesting phenomena is revealed such as phase separation processes leading to a mixing demixing dynamics, breathing of a strongly correlated Fermi sea 
and impurity-impurity induced interactions. 

Focusing on the single impurity case and weak repulsions we show that in the HF approach the system exhibits a periodic mixing-demixing dynamics. 
Both clouds perform a breathing motion with the impurity residing at the trap center and the Fermi sea splitting into two counter propagating density fragments during expansion which recombine 
at the contraction points. 
A similar dynamical evolution of both components takes place also in the many-body case but the impurity shows an overall expansion tendency which is absent in the HF evolution and in turn results 
in a relatively larger spatial overlap between the components in the course of time as compared to the HF case. 
Entering the intermediate and strongly repulsive regime of interactions and both in the HF and the many-body evolution we reveal that the dynamical spatial separation 
between the species is enhanced especially for stronger repulsions. 
Here, the impurity cloud possesses a rather localized shape and performs a weak amplitude and multifrequency breathing motion. 
However in the presence of correlations the density of the impurity is more spread leading to a smaller degree of phase separation between the species 
while the structures building upon the density of the Fermi sea become shallower compared to the HF case. 
Moreover, we unveil that the dynamical response of the Fermi sea is almost insensitive to the mass of the impurity while 
the expansion tendency of a lighter impurity is more pronounced indicating its tendency to disperse within the fermionic bath independently of the interspecies repulsion. 

For two fermionic impurities we observe a similar to the above-described dynamical behavior of each species at weak, intermediate and strong repulsions 
both in the HF and the many-body framework. 
Namely, the quench protocol enforces a counterflow of the Fermi sea while performing an overall breathing motion. 
Also the structures building upon the density of the Fermi sea appear to be shallower in the many-body case as compared to the HF framework. 
Consequently, the impurities exhibit a breathing motion in the HF case and an additional expansion trend within the many-body approach. 
For strong interactions, a phase separation between the species occurs in the course of time which is found to be more prominent in the HF as compared 
to the many-body evolution. 
The degree of phase separation at a fixed interaction is found to be larger for two fermionic impurities as compared to a single one. 
Additionally, we reveal that the impurities mass affects noticeably the response of both species contrary to the single impurity case. 
Indeed, the expansion amplitude of the Fermi sea is reduced for light impurities while it exhibits a decaying behavior for heavier ones. 
Regarding the impurities we showcase that their expansion tendency reduces for a larger mass. 

To further expose the correlated nature of the dynamics we inspect the time-evolution of the two-body intra- and interspecies correlation function. 
In all cases it is shown that the Fermi sea is strongly correlated on the two-body level and a phase separation becomes evident in the interspecies 
correlation function. 
Most importantly, signatures of attractive impurity-impurity induced interactions mediated by the Fermi sea are found in the two-body correlation of the impurities. 
The existence of such attractive induced interactions is further supported by comparing the spatial size of the two and one impurity atoms. 
Furthermore, examining the time-evolution of the von-Neumann entropy and the population eigenvalues of the single-particle functions appearing in the many-body ansatz 
we estimate the degree of both inter- and intraspecies correlations. 
Interestingly, it is found that the amount of correlations maximizes for intermediate repulsions and reduces for fixed interspecies interaction but a heavier impurity. 

There is a variety of promising research directions that are of interest for future studies. 
A straightforward extension is to unravel the quench dynamics of the Fermi-Fermi mixture for attractive interactions and/or for finite temperature \cite{tajima2018many,field2020fate} in order 
to infer whether signatures of induced impurity-impurity interactions survive and in which regimes. 
Another interesting prospect would be to examine e.g. a lattice trapped Fermi-Fermi mixture \cite{cao2017a} consisting of a fermionic bath and impurity atoms and 
unravel its stationary properties but most importantly the interaction quench dynamics. 
In this context the simulation of the corresponding radiofrequency spectrum \cite{Mistakidis2019} by employing spinor impurities in order to 
identify polaronic states is worth pursuing. 
Certainly, the generalization of the present findings to higher-dimensional settings is of immediate interest. 

\begin{acknowledgments}
P.S. gratefully acknowledges financial support by the Cluster of Excellence 'CUI: Advanced Imaging of Matter' of the Deutsche Forschungsgemeinschaft 
(DFG) - EXC 2056 - project ID 390715994. 
S. I. M  gratefully acknowledges financial support in the framework of the Lenz-Ising Award of the University of Hamburg.
\end{acknowledgments}

\appendix

\section{Effect of the barrier width of the double-well on the quench dynamics}\label{barrier_width} 

In the present Appendix, we discuss the influence of the barrier width of the double-well on the emergent quench-induced dynamics of both species of the FF mixture. 
We consider a heavy impurity, $N_B=1$, repulsively interacting with the Fermi sea consisting of $N_A=6$ fermions while the mass imbalance of the mixture is $M_B=6M_A$. 
The system is initially prepared in its ground state with the Fermi sea trapped in a double-well possessing a frequency $\omega_A=0.1$ and a fixed ratio $h/w=8$ with $h$, $w$ being the barrier height and width respectively,  
while the impurity resides in a harmonic oscillator of frequency $\omega_B = 0.6\omega_A$ \cite{cetina2016}.
To investigate the impact of the barrier width on the dynamics we consider the cases $w=1/4$ and $w=2$. 
Note that the former (latter) value is much smaller (larger) than the width of the impuritys' one-body density distribution, at $t=0$, being $w_B\approx 1.18$. 
\begin{figure*}[ht] 
\includegraphics[width=0.9\textwidth]{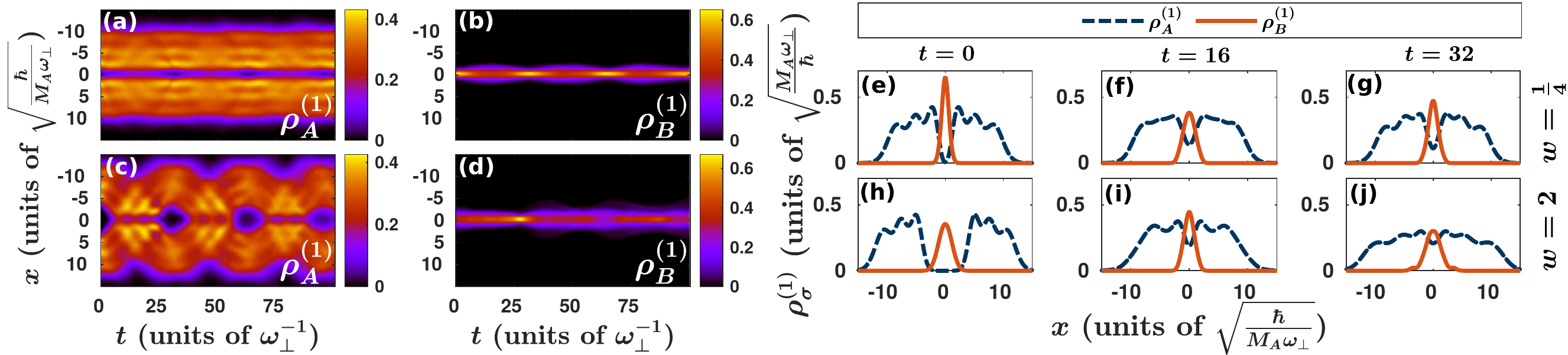}
\caption{(a)-(d) Temporal-evolution of the $\sigma$-species one-body density $\rho^{(1)}_\sigma(x;t)$ of the FF mixture within the many-body approach, following a ramping-down of the potential barrier of the Fermi sea, for $g_{AB}=4.0$.  
The width of the potential barrier corresponds to (a), (b) $w=1/4$ and (c), (d) $w=2$.  
The corresponding instantaneous density profiles of $\rho^{(1)}_\sigma(x;t)$ for (e)-(g) $w=1/4$ and (h)-(j) $w=2$. 
The system consists of a Fermi sea possessing $N_A = 6$ atoms and a single impurity $N_B=1$ with a mass-imbalance of $M_B=6 M_A$.}
\label{fig:widths}
\end{figure*}

Our species selective potential enforces a negligible spatial overlap between the species irrespectively of the interspecies interaction strength. 
In particular, the impurity cloud is distributed around the trap center while the Fermi sea exhibits a symmetric density configuration with respect to $x=0$. 
The density distribution of the Fermi sea is wider for an increasing barrier width $w$ of the double-well by means that $\rho^{(1)}_{A}(x;t)$ is more spatially extended. 
Most importanly, the distance between the density branches residing in the left and right sides of the 
double-well (namely the zero density domain of $\rho^{(1)}_{A}(x;t=0)$ around $x=0$) is larger for $w=2$, see for instance  $\rho^{(1)}_B(x;0)$ for $w=1/4$ and $w=2$ in Figs. \ref{fig:widths} (e), (h) respectively. 
Recall that $w=1/4$ and $w=2$ refer to situations where the barrier of the double-well is much thinner and thicker respectively than the width of the impurity cloud. 
As a consequence, the interspecies spatial separation of the initial state is enhanced for a larger $w$. 
To trigger the dynamics we switch-off, at $t=0$, the potential barrier of the Fermi sea thus quenching from $h/w=8$ to $h/w=0$ and track the time-evolution of each species. 
For simplicity, below, we explicitly focus on the strongly interspecies repulsive case i.e. $g_{AB}=4$ and the single-impurity scenario analyzing the dynamics in the presence of beyond Hartree-Fock correlations. 

To expose the effect of the width $w$ of the barrier of the double-well on the quench dynamics we employ the temporal-evolution of the $\sigma$-species one-body density $\rho^{(1)}_{\sigma}(x;t)$ [Fig. \ref{fig:widths}] for $w=1/4$ and $w=2$. 
As described in the main text, ramping-down the barrier induces a counterflow dynamics of the Fermi sea with $\rho^{(1)}_A(x;t)$ performing an overall breathing motion of frequency $\omega_{A}^{br} \approx 0.2 = 2 \omega_A$ for both $w=1/4$ [Fig. \ref{fig:widths} (a)] and $w=2$ [Fig. \ref{fig:widths} (c)]. 
Of course, the amplitude of this breathing motion is significantly larger for $w=2$ compared to $w=1/4$ since in the former case the distance of the initially segregated density fragments of the Fermi sea is larger and thus their counterflow is performed with an ``effectively'' larger velocity \cite{kiehn2019spontaneous,scott1998formation,weller2008experimental}. 
The latter results in a generically wider $\rho^{(1)}_{A}(x;t)$ for increasing $w$ and a more excited background of the Fermi sea in the course of the evolution, compare for instance the density profiles at $t=16$ and $t=32$ depicted in Figs. \ref{fig:widths} (f), (i) and Figs. \ref{fig:widths} (g), (j) respectively. 

In particular, following the quench the initially, at $t=0$, separated density branches of the Fermi sea [Figs. \ref{fig:widths} (e), (h)] travel towards $x=0$ where they collide and subsequently split again into two counter propagating density fragments irrespectively of $w$. 
Then, the counter propagating fragments move to the edges of the trap and for a specific $w$ exhibit a shallower density dip at time $t$ around $x=0$ compared to the one at $t=0$, see Figs. \ref{fig:widths} (a), (c). 
These density dips building upon $\rho^{(1)}_{A}(x;t)$ are wider and in general shallower [deeper] at the time-intervals of contraction [expansion] of $\rho^{(1)}_{A}(x;t)$ for a larger $w$, compare Figs. \ref{fig:widths} (a) and (c), due to the aforementioned initially larger velocity of the density fragments. 
The above-described dynamical response of $\rho^{(1)}_A(x;t)$ is repeated in a periodic fashion. 
The impurity being indirectly perturbed by the Fermi sea due to the finite $g_{AB}$ resides around the trap center in the course of the evolution. 
Its cloud undergoes an expansion and contraction dynamics with $\rho^{(1)}_B(x;t)$ featuring a predominant expansion tendency independently of the value of $w$ [Figs. \ref{fig:widths} (b), (d)]. 
This overall expansion behavior of $\rho^{(1)}_{B}(x;t)$ is more pronounced for $w=2$ than $w=1/4$. 
Indeed, $\rho^{(1)}_B(x;t)$ is effectively trapped by the density dip of $\rho^{(1)}_A(x;t)$ appearing in the vicinity of $x=0$ with the dip being wider for $w=2$. 
Thus, $\rho^{(1)}_B(x;t)$ becomes more delocalized for $w=2$ compared to $w=1/4$ forming a density peak located at $x=0$. 
We finally remark that the value of $w$ affects in a similar to the above-described way also the dynamics of the FF mixture including two fermionic impurities (not shown here for brevity).

\section{Ingredients and convergence of the many-body simulations}\label{methodology_convergence}

Let us briefly comment on the basic aspects of the deployed numerical method, the Multi-Layer Multi-Configurational Time-Dependent Hartree Method for atomic 
mixtures (ML-MCTDHX) \cite{cao2017unified}, used to simulate the correlated quantum dynamics of the fermionic mixture and subsequently discuss the 
numerical convergence of our results. 
ML-MCTDHX is a variational method for solving the time-dependent many-body Schr{\"o}dinger equation of atomic mixtures with either 
bosonic \cite{mistakidis2019effective,mistakidis2019dissipative,Mistakidis2018} or fermionic \cite{erdmann2019phase,erdmann2018,cao2017a,mistakidis2019correlated} constituents  
including also spin degrees of freedom \cite{Koutentakis2019,mistakidis2019quench,Mistakidis2019}. 
A powerful asset of this method is that it relies on the expansion of the many-body wavefunction with respect to a time-dependent and 
variationally optimized basis. 
The latter allows us to account for the underlying intra- and interspecies correlations of the system under consideration by exploiting a computationally 
feasible basis size and thus choosing the relevant subspace of the Hilbert space at each time instant of the evolution in an efficient manner.

\begin{figure}[ht] 
\includegraphics[width=0.4\textwidth]{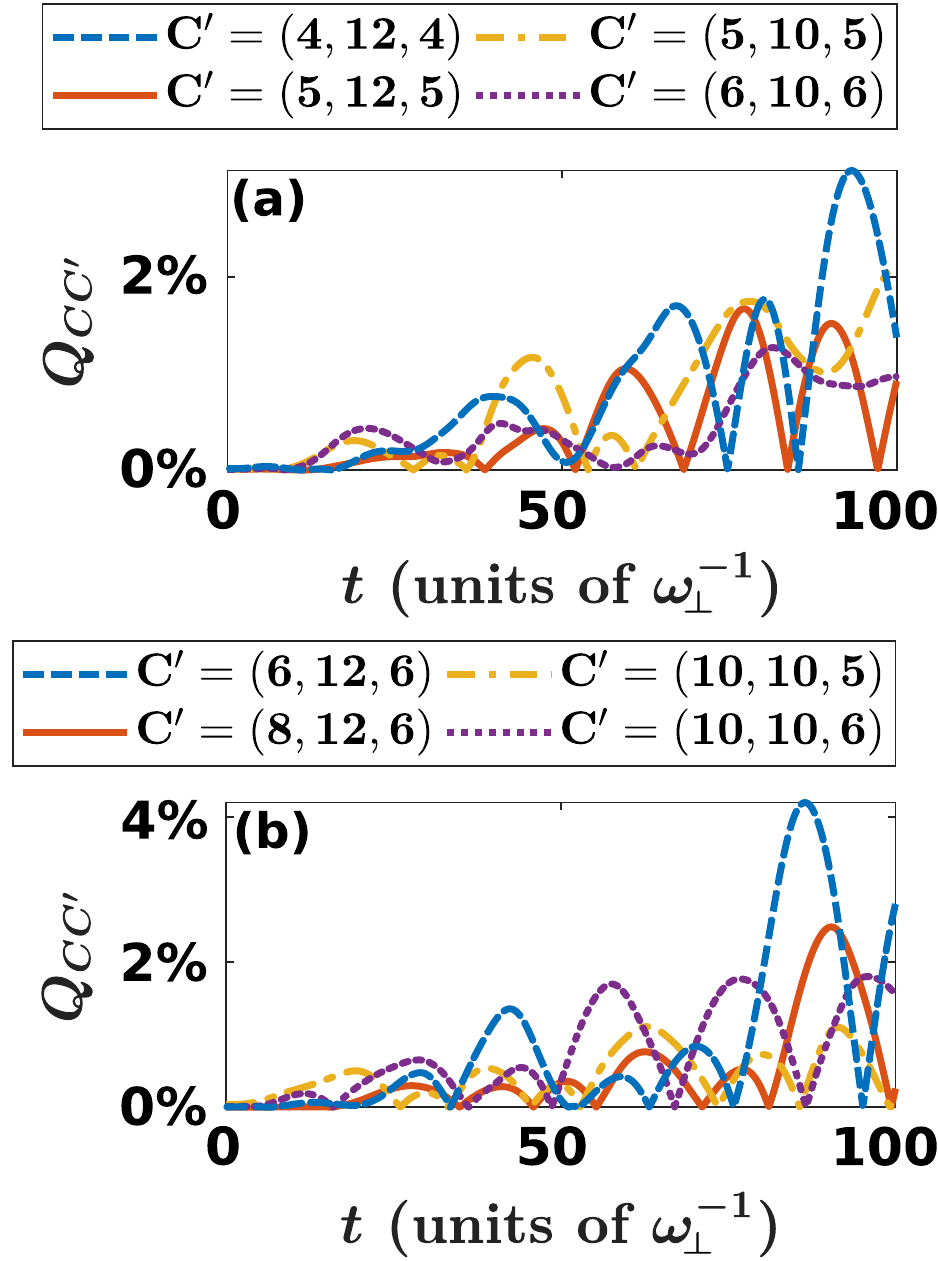}
\caption{Evolution of the relative deviation $Q_{C,C^\prime}$ of $\braket{X^2_A(t)}$ between (a) $C=(6,12,6)$ for $N_B=1$ and (b) $C=(10,12,6)$ for $N_B=2$ 
and other orbital configurations $C^\prime$ (see legends) for $g_{AB}=1$. 
In both cases the Fermi sea consists of $N_A=6$ atoms and it is initially trapped in a double-well while the impurity subsystem is confined in a harmonic trap. 
The dynamics is induced at $t=0$ by ramping-down the potential barrier of the Fermi sea. }
\label{fig:convergence}
\end{figure}

The corresponding degree of Hilbert space truncation is dictated by the employed orbital configuration space denoted in the following by $C=(D,d^A,d^B)$. 
Here, $D=D^A=D^B$ and $d^A$, $d^B$ refer to the number of species and single-particle functions of each species respectively involved in the many-body 
wavefunction ansatz [Eqs. (\ref{eq:WF}), (\ref{eq:SPFs})]. 
Additionally, for our numerical calculations we use a primitive basis based on a sine discrete variable representation which introduces hard-wall 
boundary conditions at both edges of the numerical grid consisting here of 300 grid points. 
Moreover, the hard-wall boundaries are imposed at $x_\pm=\pm40$ and do not impact our results since the spatial extension of the one-body densities of both species 
does not exceed $x_{\pm}=\pm15$. 

To elucidate the numerical convergence of our many-body simulations we assure that all observables of interest become, to a given level of accuracy, almost insensitive 
upon varying the employed orbital configuration space $C=(D,d^A,d^B)$. 
We remark that for many-body calculations, discussed in the main text, we used the orbital configurations $C = (6,12,6)$ and $C = (10,12,6)$ 
for $N_B=1$ and $N_B = 2$ respectively. 
Below, we exemplify the convergence of the center-of-mass variance of the Fermi sea $\braket{X_A^2(t)}$ in the course of the evolution for a different 
number of species and single-particle functions. 
Recall, that the quench protocol acts only on the fermionic bath and the dynamics of the impurities is induced indirectly. 
More precisely, we inspect the absolute deviation of $\braket{X_A^2(t)}_C$ between the $C = (6,12,6)$ for $N_B=1$ [$C = (10,12,6)$ for $N_B=2$] and 
other orbital combinations $C'=(D,d^A,d^B)$ 
\begin{equation}
Q_{C,C^\prime} =\frac{\abs{\braket{ X_A^2}_{C} - \braket{ X_A^2}_{C^\prime}}}{\braket{ X_A^2 }_{C}}. \label{deviation_variance}
\end{equation} 
Figure \ref{fig:convergence} shows $Q_{C,C^\prime}$ after ramping-down the potential barrier of the double-well of the Fermi sea at 
interspecies interaction $g_{AB}=1$ in the case of a single [Fig. \ref{fig:convergence} (a)] and two [Fig. \ref{fig:convergence} (b)] impurities. 
Note that we choose to present here $g_{AB}=1$ since for such intermediate interactions the degree of both inter- and intraspecies correlations are maximized 
[see Fig. \ref{fig:one_imp:entropy_and_fragmentation} and Fig. \ref{fig:two_imp:entropy_and_fragmentation}] and therefore convergence is more difficult to be reached. 
Closely inspecting Fig. \ref{fig:convergence} we can deduce that $\braket{X_A^2(t)}$ is indeed converged for both one and two fermionic impurities. 
Referring to the single impurity case, we observe that $Q_{C,C^\prime}$ between the $C=(6,12,6)$ and $C'=(6,10,6)$ [$C'=(5,12,5)$] orbital configurations is 
below $1\%$ [$1.8\%$] throughout the evolution [Fig. \ref{fig:convergence} (a)]. 
Moreover, turning to the dynamics of two fermionic impurities it can be seen from Fig. \ref{fig:convergence} (b) that $Q_{C,C^\prime}$ when $C=(10,12,6)$ and $C'=(6,12,6)$ becomes at most 
of the order of $4\%$ whilst e.g. for $C=(10,12,6)$ and $C'=(10,10,5)$ it takes a maximum value of $0.9\%$. 
Similar observations can be made by inspecting the corresponding relative deviation of the impurities $\braket{X_B^2(t)}_C$ (not shown here). 
Let us finally mention that the same convergence procedure has been followed for the other interspecies interactions considered in the main text and found to 
exhibit a similar or even better (in some cases) degree of convergence.

\bibliography{bibliography}

\begin{thebibliography}{125}%
\makeatletter
\providecommand \@ifxundefined [1]{%
 \@ifx{#1\undefined}
}%
\providecommand \@ifnum [1]{%
 \ifnum #1\expandafter \@firstoftwo
 \else \expandafter \@secondoftwo
 \fi
}%
\providecommand \@ifx [1]{%
 \ifx #1\expandafter \@firstoftwo
 \else \expandafter \@secondoftwo
 \fi
}%
\providecommand \natexlab [1]{#1}%
\providecommand \enquote  [1]{``#1''}%
\providecommand \bibnamefont  [1]{#1}%
\providecommand \bibfnamefont [1]{#1}%
\providecommand \citenamefont [1]{#1}%
\providecommand \href@noop [0]{\@secondoftwo}%
\providecommand \href [0]{\begingroup \@sanitize@url \@href}%
\providecommand \@href[1]{\@@startlink{#1}\@@href}%
\providecommand \@@href[1]{\endgroup#1\@@endlink}%
\providecommand \@sanitize@url [0]{\catcode `\\12\catcode `\$12\catcode
  `\&12\catcode `\#12\catcode `\^12\catcode `\_12\catcode `\%12\relax}%
\providecommand \@@startlink[1]{}%
\providecommand \@@endlink[0]{}%
\providecommand \url  [0]{\begingroup\@sanitize@url \@url }%
\providecommand \@url [1]{\endgroup\@href {#1}{\urlprefix }}%
\providecommand \urlprefix  [0]{URL }%
\providecommand \Eprint [0]{\href }%
\providecommand \doibase [0]{http://dx.doi.org/}%
\providecommand \selectlanguage [0]{\@gobble}%
\providecommand \bibinfo  [0]{\@secondoftwo}%
\providecommand \bibfield  [0]{\@secondoftwo}%
\providecommand \translation [1]{[#1]}%
\providecommand \BibitemOpen [0]{}%
\providecommand \bibitemStop [0]{}%
\providecommand \bibitemNoStop [0]{.\EOS\space}%
\providecommand \EOS [0]{\spacefactor3000\relax}%
\providecommand \BibitemShut  [1]{\csname bibitem#1\endcsname}%
\let\auto@bib@innerbib\@empty
\bibitem [{\citenamefont {Catani}\ \emph {et~al.}(2008)\citenamefont {Catani},
  \citenamefont {De~Sarlo}, \citenamefont {Barontini}, \citenamefont
  {Minardi},\ and\ \citenamefont {Inguscio}}]{catani2008}%
  \BibitemOpen
  \bibfield  {author} {\bibinfo {author} {\bibfnamefont {J.}~\bibnamefont
  {Catani}}, \bibinfo {author} {\bibfnamefont {L.}~\bibnamefont {De~Sarlo}},
  \bibinfo {author} {\bibfnamefont {G.}~\bibnamefont {Barontini}}, \bibinfo
  {author} {\bibfnamefont {F.}~\bibnamefont {Minardi}}, \ and\ \bibinfo
  {author} {\bibfnamefont {M.}~\bibnamefont {Inguscio}},\ }\href {\doibase
  10.1103/PhysRevA.77.011603} {\bibfield  {journal} {\bibinfo  {journal} {Phys.
  Rev. A}\ }\textbf {\bibinfo {volume} {77}},\ \bibinfo {pages} {011603}
  (\bibinfo {year} {2008})}\BibitemShut {NoStop}%
\bibitem [{\citenamefont {Thalhammer}\ \emph {et~al.}(2008)\citenamefont
  {Thalhammer}, \citenamefont {Barontini}, \citenamefont {De~Sarlo},
  \citenamefont {Catani}, \citenamefont {Minardi},\ and\ \citenamefont
  {Inguscio}}]{thalhammer2008}%
  \BibitemOpen
  \bibfield  {author} {\bibinfo {author} {\bibfnamefont {G.}~\bibnamefont
  {Thalhammer}}, \bibinfo {author} {\bibfnamefont {G.}~\bibnamefont
  {Barontini}}, \bibinfo {author} {\bibfnamefont {L.}~\bibnamefont {De~Sarlo}},
  \bibinfo {author} {\bibfnamefont {J.}~\bibnamefont {Catani}}, \bibinfo
  {author} {\bibfnamefont {F.}~\bibnamefont {Minardi}}, \ and\ \bibinfo
  {author} {\bibfnamefont {M.}~\bibnamefont {Inguscio}},\ }\href {\doibase
  10.1103/PhysRevLett.100.210402} {\bibfield  {journal} {\bibinfo  {journal}
  {Phys. Rev. Lett.}\ }\textbf {\bibinfo {volume} {100}},\ \bibinfo {pages}
  {210402} (\bibinfo {year} {2008})}\BibitemShut {NoStop}%
\bibitem [{\citenamefont {Petrov}(2015)}]{petrov2015}%
  \BibitemOpen
  \bibfield  {author} {\bibinfo {author} {\bibfnamefont {D.~S.}\ \bibnamefont
  {Petrov}},\ }\href {\doibase 10.1103/PhysRevLett.115.155302} {\bibfield
  {journal} {\bibinfo  {journal} {Phys. Rev. Lett.}\ }\textbf {\bibinfo
  {volume} {115}},\ \bibinfo {pages} {155302} (\bibinfo {year}
  {2015})}\BibitemShut {NoStop}%
\bibitem [{\citenamefont {Stan}\ \emph {et~al.}(2004)\citenamefont {Stan},
  \citenamefont {Zwierlein}, \citenamefont {Schunck}, \citenamefont {Raupach},\
  and\ \citenamefont {Ketterle}}]{stan2004}%
  \BibitemOpen
  \bibfield  {author} {\bibinfo {author} {\bibfnamefont {C.~A.}\ \bibnamefont
  {Stan}}, \bibinfo {author} {\bibfnamefont {M.~W.}\ \bibnamefont {Zwierlein}},
  \bibinfo {author} {\bibfnamefont {C.~H.}\ \bibnamefont {Schunck}}, \bibinfo
  {author} {\bibfnamefont {S.~M.~F.}\ \bibnamefont {Raupach}}, \ and\ \bibinfo
  {author} {\bibfnamefont {W.}~\bibnamefont {Ketterle}},\ }\href {\doibase
  10.1103/PhysRevLett.93.143001} {\bibfield  {journal} {\bibinfo  {journal}
  {Physical Review Letters}\ }\textbf {\bibinfo {volume} {93}},\ \bibinfo
  {pages} {143001} (\bibinfo {year} {2004})}\BibitemShut {NoStop}%
\bibitem [{\citenamefont {Ospelkaus}\ \emph {et~al.}(2006)\citenamefont
  {Ospelkaus}, \citenamefont {Ospelkaus}, \citenamefont {Humbert},
  \citenamefont {Sengstock},\ and\ \citenamefont {Bongs}}]{ospelkaus2006}%
  \BibitemOpen
  \bibfield  {author} {\bibinfo {author} {\bibfnamefont {S.}~\bibnamefont
  {Ospelkaus}}, \bibinfo {author} {\bibfnamefont {C.}~\bibnamefont
  {Ospelkaus}}, \bibinfo {author} {\bibfnamefont {L.}~\bibnamefont {Humbert}},
  \bibinfo {author} {\bibfnamefont {K.}~\bibnamefont {Sengstock}}, \ and\
  \bibinfo {author} {\bibfnamefont {K.}~\bibnamefont {Bongs}},\ }\href
  {\doibase 10.1103/PhysRevLett.97.120403} {\bibfield  {journal} {\bibinfo
  {journal} {Phys. Rev. Lett.}\ }\textbf {\bibinfo {volume} {97}},\ \bibinfo
  {pages} {120403} (\bibinfo {year} {2006})}\BibitemShut {NoStop}%
\bibitem [{\citenamefont {Ahufinger}\ \emph {et~al.}(2005)\citenamefont
  {Ahufinger}, \citenamefont {{Sanchez-Palencia}}, \citenamefont {Kantian},
  \citenamefont {Sanpera},\ and\ \citenamefont {Lewenstein}}]{ahufinger2005}%
  \BibitemOpen
  \bibfield  {author} {\bibinfo {author} {\bibfnamefont {V.}~\bibnamefont
  {Ahufinger}}, \bibinfo {author} {\bibfnamefont {L.}~\bibnamefont
  {{Sanchez-Palencia}}}, \bibinfo {author} {\bibfnamefont {A.}~\bibnamefont
  {Kantian}}, \bibinfo {author} {\bibfnamefont {A.}~\bibnamefont {Sanpera}}, \
  and\ \bibinfo {author} {\bibfnamefont {M.}~\bibnamefont {Lewenstein}},\
  }\href {\doibase 10.1103/PhysRevA.72.063616} {\bibfield  {journal} {\bibinfo
  {journal} {Phys. Rev. A}\ }\textbf {\bibinfo {volume} {72}},\ \bibinfo
  {pages} {063616} (\bibinfo {year} {2005})}\BibitemShut {NoStop}%
\bibitem [{\citenamefont {Wille}\ \emph {et~al.}(2008)\citenamefont {Wille},
  \citenamefont {Spiegelhalder}, \citenamefont {Kerner}, \citenamefont {Naik},
  \citenamefont {Trenkwalder}, \citenamefont {Hendl}, \citenamefont {Schreck},
  \citenamefont {Grimm}, \citenamefont {Tiecke}, \citenamefont {Walraven},
  \citenamefont {Kokkelmans}, \citenamefont {Tiesinga},\ and\ \citenamefont
  {Julienne}}]{wille2008}%
  \BibitemOpen
  \bibfield  {author} {\bibinfo {author} {\bibfnamefont {E.}~\bibnamefont
  {Wille}}, \bibinfo {author} {\bibfnamefont {F.~M.}\ \bibnamefont
  {Spiegelhalder}}, \bibinfo {author} {\bibfnamefont {G.}~\bibnamefont
  {Kerner}}, \bibinfo {author} {\bibfnamefont {D.}~\bibnamefont {Naik}},
  \bibinfo {author} {\bibfnamefont {A.}~\bibnamefont {Trenkwalder}}, \bibinfo
  {author} {\bibfnamefont {G.}~\bibnamefont {Hendl}}, \bibinfo {author}
  {\bibfnamefont {F.}~\bibnamefont {Schreck}}, \bibinfo {author} {\bibfnamefont
  {R.}~\bibnamefont {Grimm}}, \bibinfo {author} {\bibfnamefont {T.~G.}\
  \bibnamefont {Tiecke}}, \bibinfo {author} {\bibfnamefont {J.~T.~M.}\
  \bibnamefont {Walraven}}, \bibinfo {author} {\bibfnamefont {S.~J. J. M.~F.}\
  \bibnamefont {Kokkelmans}}, \bibinfo {author} {\bibfnamefont
  {E.}~\bibnamefont {Tiesinga}}, \ and\ \bibinfo {author} {\bibfnamefont
  {P.~S.}\ \bibnamefont {Julienne}},\ }\href {\doibase
  10.1103/PhysRevLett.100.053201} {\bibfield  {journal} {\bibinfo  {journal}
  {Phys. Rev. Lett.}\ }\textbf {\bibinfo {volume} {100}},\ \bibinfo {pages}
  {053201} (\bibinfo {year} {2008})}\BibitemShut {NoStop}%
\bibitem [{\citenamefont {K{\"o}hl}\ \emph {et~al.}(2005)\citenamefont
  {K{\"o}hl}, \citenamefont {Moritz}, \citenamefont {St{\"o}ferle},
  \citenamefont {G{\"u}nter},\ and\ \citenamefont {Esslinger}}]{kohl2005}%
  \BibitemOpen
  \bibfield  {author} {\bibinfo {author} {\bibfnamefont {M.}~\bibnamefont
  {K{\"o}hl}}, \bibinfo {author} {\bibfnamefont {H.}~\bibnamefont {Moritz}},
  \bibinfo {author} {\bibfnamefont {T.}~\bibnamefont {St{\"o}ferle}}, \bibinfo
  {author} {\bibfnamefont {K.}~\bibnamefont {G{\"u}nter}}, \ and\ \bibinfo
  {author} {\bibfnamefont {T.}~\bibnamefont {Esslinger}},\ }\href {\doibase
  10.1103/PhysRevLett.94.080403} {\bibfield  {journal} {\bibinfo  {journal}
  {Phys. Rev. Lett.}\ }\textbf {\bibinfo {volume} {94}},\ \bibinfo {pages}
  {080403} (\bibinfo {year} {2005})}\BibitemShut {NoStop}%
\bibitem [{\citenamefont {Hadzibabic}\ \emph {et~al.}(2002)\citenamefont
  {Hadzibabic}, \citenamefont {Stan}, \citenamefont {Dieckmann}, \citenamefont
  {Gupta}, \citenamefont {Zwierlein}, \citenamefont {G{\"o}rlitz},\ and\
  \citenamefont {Ketterle}}]{hadzibabic2002}%
  \BibitemOpen
  \bibfield  {author} {\bibinfo {author} {\bibfnamefont {Z.}~\bibnamefont
  {Hadzibabic}}, \bibinfo {author} {\bibfnamefont {C.~A.}\ \bibnamefont
  {Stan}}, \bibinfo {author} {\bibfnamefont {K.}~\bibnamefont {Dieckmann}},
  \bibinfo {author} {\bibfnamefont {S.}~\bibnamefont {Gupta}}, \bibinfo
  {author} {\bibfnamefont {M.~W.}\ \bibnamefont {Zwierlein}}, \bibinfo {author}
  {\bibfnamefont {A.}~\bibnamefont {G{\"o}rlitz}}, \ and\ \bibinfo {author}
  {\bibfnamefont {W.}~\bibnamefont {Ketterle}},\ }\href {\doibase
  10.1103/PhysRevLett.88.160401} {\bibfield  {journal} {\bibinfo  {journal}
  {Phys. Rev. Lett.}\ }\textbf {\bibinfo {volume} {88}},\ \bibinfo {pages}
  {160401} (\bibinfo {year} {2002})}\BibitemShut {NoStop}%
\bibitem [{\citenamefont {Chin}\ \emph {et~al.}(2010)\citenamefont {Chin},
  \citenamefont {Grimm}, \citenamefont {Julienne},\ and\ \citenamefont
  {Tiesinga}}]{chin2010}%
  \BibitemOpen
  \bibfield  {author} {\bibinfo {author} {\bibfnamefont {C.}~\bibnamefont
  {Chin}}, \bibinfo {author} {\bibfnamefont {R.}~\bibnamefont {Grimm}},
  \bibinfo {author} {\bibfnamefont {P.}~\bibnamefont {Julienne}}, \ and\
  \bibinfo {author} {\bibfnamefont {E.}~\bibnamefont {Tiesinga}},\ }\href
  {\doibase 10.1103/RevModPhys.82.1225} {\bibfield  {journal} {\bibinfo
  {journal} {Rev. Mod. Phys.}\ }\textbf {\bibinfo {volume} {82}},\ \bibinfo
  {pages} {1225} (\bibinfo {year} {2010})}\BibitemShut {NoStop}%
\bibitem [{\citenamefont {Inouye}\ \emph {et~al.}(1998)\citenamefont {Inouye},
  \citenamefont {Andrews}, \citenamefont {Stenger}, \citenamefont {Miesner},
  \citenamefont {{Stamper-Kurn}},\ and\ \citenamefont {Ketterle}}]{inouye1998}%
  \BibitemOpen
  \bibfield  {author} {\bibinfo {author} {\bibfnamefont {S.}~\bibnamefont
  {Inouye}}, \bibinfo {author} {\bibfnamefont {M.~R.}\ \bibnamefont {Andrews}},
  \bibinfo {author} {\bibfnamefont {J.}~\bibnamefont {Stenger}}, \bibinfo
  {author} {\bibfnamefont {H.-J.}\ \bibnamefont {Miesner}}, \bibinfo {author}
  {\bibfnamefont {D.~M.}\ \bibnamefont {{Stamper-Kurn}}}, \ and\ \bibinfo
  {author} {\bibfnamefont {W.}~\bibnamefont {Ketterle}},\ }\href {\doibase
  10.1038/32354} {\bibfield  {journal} {\bibinfo  {journal} {Nature}\ }\textbf
  {\bibinfo {volume} {392}},\ \bibinfo {pages} {151} (\bibinfo {year}
  {1998})}\BibitemShut {NoStop}%
\bibitem [{\citenamefont {Greiner}\ \emph {et~al.}(2002)\citenamefont
  {Greiner}, \citenamefont {Mandel}, \citenamefont {Esslinger}, \citenamefont
  {H{\"a}nsch},\ and\ \citenamefont {Bloch}}]{greiner2002}%
  \BibitemOpen
  \bibfield  {author} {\bibinfo {author} {\bibfnamefont {M.}~\bibnamefont
  {Greiner}}, \bibinfo {author} {\bibfnamefont {O.}~\bibnamefont {Mandel}},
  \bibinfo {author} {\bibfnamefont {T.}~\bibnamefont {Esslinger}}, \bibinfo
  {author} {\bibfnamefont {T.~W.}\ \bibnamefont {H{\"a}nsch}}, \ and\ \bibinfo
  {author} {\bibfnamefont {I.}~\bibnamefont {Bloch}},\ }\href {\doibase
  10.1038/415039a} {\bibfield  {journal} {\bibinfo  {journal} {Nature}\
  }\textbf {\bibinfo {volume} {415}},\ \bibinfo {pages} {39} (\bibinfo {year}
  {2002})}\BibitemShut {NoStop}%
\bibitem [{\citenamefont {Bloch}\ \emph {et~al.}(2008)\citenamefont {Bloch},
  \citenamefont {Dalibard},\ and\ \citenamefont {Zwerger}}]{bloch2008}%
  \BibitemOpen
  \bibfield  {author} {\bibinfo {author} {\bibfnamefont {I.}~\bibnamefont
  {Bloch}}, \bibinfo {author} {\bibfnamefont {J.}~\bibnamefont {Dalibard}}, \
  and\ \bibinfo {author} {\bibfnamefont {W.}~\bibnamefont {Zwerger}},\ }\href
  {\doibase 10.1103/RevModPhys.80.885} {\bibfield  {journal} {\bibinfo
  {journal} {Rev. Mod. Phys.}\ }\textbf {\bibinfo {volume} {80}},\ \bibinfo
  {pages} {885} (\bibinfo {year} {2008})}\BibitemShut {NoStop}%
\bibitem [{\citenamefont {Wenz}\ \emph {et~al.}(2013)\citenamefont {Wenz},
  \citenamefont {Zurn}, \citenamefont {Murmann}, \citenamefont {Brouzos},
  \citenamefont {Lompe},\ and\ \citenamefont {Jochim}}]{wenz2013}%
  \BibitemOpen
  \bibfield  {author} {\bibinfo {author} {\bibfnamefont {A.~N.}\ \bibnamefont
  {Wenz}}, \bibinfo {author} {\bibfnamefont {G.}~\bibnamefont {Zurn}}, \bibinfo
  {author} {\bibfnamefont {S.}~\bibnamefont {Murmann}}, \bibinfo {author}
  {\bibfnamefont {I.}~\bibnamefont {Brouzos}}, \bibinfo {author} {\bibfnamefont
  {T.}~\bibnamefont {Lompe}}, \ and\ \bibinfo {author} {\bibfnamefont
  {S.}~\bibnamefont {Jochim}},\ }\href {\doibase 10.1126/science.1240516}
  {\bibfield  {journal} {\bibinfo  {journal} {Science}\ }\textbf {\bibinfo
  {volume} {342}},\ \bibinfo {pages} {457} (\bibinfo {year}
  {2013})}\BibitemShut {NoStop}%
\bibitem [{\citenamefont {Serwane}\ \emph {et~al.}(2011)\citenamefont
  {Serwane}, \citenamefont {Zurn}, \citenamefont {Lompe}, \citenamefont
  {Ottenstein}, \citenamefont {Wenz},\ and\ \citenamefont
  {Jochim}}]{serwane2011}%
  \BibitemOpen
  \bibfield  {author} {\bibinfo {author} {\bibfnamefont {F.}~\bibnamefont
  {Serwane}}, \bibinfo {author} {\bibfnamefont {G.}~\bibnamefont {Zurn}},
  \bibinfo {author} {\bibfnamefont {T.}~\bibnamefont {Lompe}}, \bibinfo
  {author} {\bibfnamefont {T.~B.}\ \bibnamefont {Ottenstein}}, \bibinfo
  {author} {\bibfnamefont {A.~N.}\ \bibnamefont {Wenz}}, \ and\ \bibinfo
  {author} {\bibfnamefont {S.}~\bibnamefont {Jochim}},\ }\href {\doibase
  10.1126/science.1201351} {\bibfield  {journal} {\bibinfo  {journal}
  {Science}\ }\textbf {\bibinfo {volume} {332}},\ \bibinfo {pages} {336}
  (\bibinfo {year} {2011})}\BibitemShut {NoStop}%
\bibitem [{\citenamefont {Z{\"u}rn}\ \emph {et~al.}(2012)\citenamefont
  {Z{\"u}rn}, \citenamefont {Serwane}, \citenamefont {Lompe}, \citenamefont
  {Wenz}, \citenamefont {Ries}, \citenamefont {Bohn},\ and\ \citenamefont
  {Jochim}}]{zurn2012}%
  \BibitemOpen
  \bibfield  {author} {\bibinfo {author} {\bibfnamefont {G.}~\bibnamefont
  {Z{\"u}rn}}, \bibinfo {author} {\bibfnamefont {F.}~\bibnamefont {Serwane}},
  \bibinfo {author} {\bibfnamefont {T.}~\bibnamefont {Lompe}}, \bibinfo
  {author} {\bibfnamefont {A.~N.}\ \bibnamefont {Wenz}}, \bibinfo {author}
  {\bibfnamefont {M.~G.}\ \bibnamefont {Ries}}, \bibinfo {author}
  {\bibfnamefont {J.~E.}\ \bibnamefont {Bohn}}, \ and\ \bibinfo {author}
  {\bibfnamefont {S.}~\bibnamefont {Jochim}},\ }\href {\doibase
  10.1103/PhysRevLett.108.075303} {\bibfield  {journal} {\bibinfo  {journal}
  {Phys. Rev. Lett.}\ }\textbf {\bibinfo {volume} {108}},\ \bibinfo {pages}
  {075303} (\bibinfo {year} {2012})}\BibitemShut {NoStop}%
\bibitem [{\citenamefont {Wu}\ \emph {et~al.}(2012)\citenamefont {Wu},
  \citenamefont {Park}, \citenamefont {Ahmadi}, \citenamefont {Will},\ and\
  \citenamefont {Zwierlein}}]{Wu2012}%
  \BibitemOpen
  \bibfield  {author} {\bibinfo {author} {\bibfnamefont {C.-H.}\ \bibnamefont
  {Wu}}, \bibinfo {author} {\bibfnamefont {J.~W.}\ \bibnamefont {Park}},
  \bibinfo {author} {\bibfnamefont {P.}~\bibnamefont {Ahmadi}}, \bibinfo
  {author} {\bibfnamefont {S.}~\bibnamefont {Will}}, \ and\ \bibinfo {author}
  {\bibfnamefont {M.~W.}\ \bibnamefont {Zwierlein}},\ }\href {\doibase
  10.1103/PhysRevLett.109.085301} {\bibfield  {journal} {\bibinfo  {journal}
  {Phys. Rev. Lett.}\ }\textbf {\bibinfo {volume} {109}},\ \bibinfo {pages}
  {085301} (\bibinfo {year} {2012})}\BibitemShut {NoStop}%
\bibitem [{\citenamefont {Heo}\ \emph {et~al.}(2012)\citenamefont {Heo},
  \citenamefont {Wang}, \citenamefont {Christensen}, \citenamefont {Rvachov},
  \citenamefont {Cotta}, \citenamefont {Choi}, \citenamefont {Lee},\ and\
  \citenamefont {Ketterle}}]{Heo2012}%
  \BibitemOpen
  \bibfield  {author} {\bibinfo {author} {\bibfnamefont {M.-S.}\ \bibnamefont
  {Heo}}, \bibinfo {author} {\bibfnamefont {T.~T.}\ \bibnamefont {Wang}},
  \bibinfo {author} {\bibfnamefont {C.~A.}\ \bibnamefont {Christensen}},
  \bibinfo {author} {\bibfnamefont {T.~M.}\ \bibnamefont {Rvachov}}, \bibinfo
  {author} {\bibfnamefont {D.~A.}\ \bibnamefont {Cotta}}, \bibinfo {author}
  {\bibfnamefont {J.-H.}\ \bibnamefont {Choi}}, \bibinfo {author}
  {\bibfnamefont {Y.-R.}\ \bibnamefont {Lee}}, \ and\ \bibinfo {author}
  {\bibfnamefont {W.}~\bibnamefont {Ketterle}},\ }\href {\doibase
  10.1103/PhysRevA.86.021602} {\bibfield  {journal} {\bibinfo  {journal} {Phys.
  Rev. A}\ }\textbf {\bibinfo {volume} {86}},\ \bibinfo {pages} {021602}
  (\bibinfo {year} {2012})}\BibitemShut {NoStop}%
\bibitem [{\citenamefont {Roati}\ \emph {et~al.}(2007)\citenamefont {Roati},
  \citenamefont {Zaccanti}, \citenamefont {D'Errico}, \citenamefont {Catani},
  \citenamefont {Modugno}, \citenamefont {Simoni}, \citenamefont {Inguscio},\
  and\ \citenamefont {Modugno}}]{Roati2007}%
  \BibitemOpen
  \bibfield  {author} {\bibinfo {author} {\bibfnamefont {G.}~\bibnamefont
  {Roati}}, \bibinfo {author} {\bibfnamefont {M.}~\bibnamefont {Zaccanti}},
  \bibinfo {author} {\bibfnamefont {C.}~\bibnamefont {D'Errico}}, \bibinfo
  {author} {\bibfnamefont {J.}~\bibnamefont {Catani}}, \bibinfo {author}
  {\bibfnamefont {M.}~\bibnamefont {Modugno}}, \bibinfo {author} {\bibfnamefont
  {A.}~\bibnamefont {Simoni}}, \bibinfo {author} {\bibfnamefont
  {M.}~\bibnamefont {Inguscio}}, \ and\ \bibinfo {author} {\bibfnamefont
  {G.}~\bibnamefont {Modugno}},\ }\href {\doibase
  10.1103/PhysRevLett.99.010403} {\bibfield  {journal} {\bibinfo  {journal}
  {Phys. Rev. Lett.}\ }\textbf {\bibinfo {volume} {99}},\ \bibinfo {pages}
  {010403} (\bibinfo {year} {2007})}\BibitemShut {NoStop}%
\bibitem [{\citenamefont {Schirotzek}\ \emph {et~al.}(2009)\citenamefont
  {Schirotzek}, \citenamefont {Wu}, \citenamefont {Sommer},\ and\ \citenamefont
  {Zwierlein}}]{Schirotzek2009}%
  \BibitemOpen
  \bibfield  {author} {\bibinfo {author} {\bibfnamefont {A.}~\bibnamefont
  {Schirotzek}}, \bibinfo {author} {\bibfnamefont {C.-H.}\ \bibnamefont {Wu}},
  \bibinfo {author} {\bibfnamefont {A.}~\bibnamefont {Sommer}}, \ and\ \bibinfo
  {author} {\bibfnamefont {M.~W.}\ \bibnamefont {Zwierlein}},\ }\href {\doibase
  10.1103/PhysRevLett.102.230402} {\bibfield  {journal} {\bibinfo  {journal}
  {Phys. Rev. Lett.}\ }\textbf {\bibinfo {volume} {102}},\ \bibinfo {pages}
  {230402} (\bibinfo {year} {2009})}\BibitemShut {NoStop}%
\bibitem [{\citenamefont {Kohstall}\ \emph {et~al.}(2012)\citenamefont
  {Kohstall}, \citenamefont {Zaccanti}, \citenamefont {Jag}, \citenamefont
  {Trenkwalder}, \citenamefont {Massignan}, \citenamefont {Bruun},
  \citenamefont {Schreck},\ and\ \citenamefont
  {Grimm}}]{kohstall2012metastability}%
  \BibitemOpen
  \bibfield  {author} {\bibinfo {author} {\bibfnamefont {C.}~\bibnamefont
  {Kohstall}}, \bibinfo {author} {\bibfnamefont {M.}~\bibnamefont {Zaccanti}},
  \bibinfo {author} {\bibfnamefont {M.}~\bibnamefont {Jag}}, \bibinfo {author}
  {\bibfnamefont {A.}~\bibnamefont {Trenkwalder}}, \bibinfo {author}
  {\bibfnamefont {P.}~\bibnamefont {Massignan}}, \bibinfo {author}
  {\bibfnamefont {G.~M.}\ \bibnamefont {Bruun}}, \bibinfo {author}
  {\bibfnamefont {F.}~\bibnamefont {Schreck}}, \ and\ \bibinfo {author}
  {\bibfnamefont {R.}~\bibnamefont {Grimm}},\ }\href@noop {} {\bibfield
  {journal} {\bibinfo  {journal} {Nature}\ }\textbf {\bibinfo {volume} {485}},\
  \bibinfo {pages} {615} (\bibinfo {year} {2012})}\BibitemShut {NoStop}%
\bibitem [{\citenamefont {Koschorreck}\ \emph {et~al.}(2012)\citenamefont
  {Koschorreck}, \citenamefont {Pertot}, \citenamefont {Vogt}, \citenamefont
  {Fr{\"o}hlich}, \citenamefont {Feld},\ and\ \citenamefont
  {K{\"o}hl}}]{Koschorreck2012}%
  \BibitemOpen
  \bibfield  {author} {\bibinfo {author} {\bibfnamefont {M.}~\bibnamefont
  {Koschorreck}}, \bibinfo {author} {\bibfnamefont {D.}~\bibnamefont {Pertot}},
  \bibinfo {author} {\bibfnamefont {E.}~\bibnamefont {Vogt}}, \bibinfo {author}
  {\bibfnamefont {B.}~\bibnamefont {Fr{\"o}hlich}}, \bibinfo {author}
  {\bibfnamefont {M.}~\bibnamefont {Feld}}, \ and\ \bibinfo {author}
  {\bibfnamefont {M.}~\bibnamefont {K{\"o}hl}},\ }\href@noop {} {\bibfield
  {journal} {\bibinfo  {journal} {Nature}\ }\textbf {\bibinfo {volume} {485}},\
  \bibinfo {pages} {619} (\bibinfo {year} {2012})}\BibitemShut {NoStop}%
\bibitem [{\citenamefont {Zhang}\ \emph {et~al.}(2012)\citenamefont {Zhang},
  \citenamefont {Ong}, \citenamefont {Arakelyan},\ and\ \citenamefont
  {Thomas}}]{zhang2012polaron}%
  \BibitemOpen
  \bibfield  {author} {\bibinfo {author} {\bibfnamefont {Y.}~\bibnamefont
  {Zhang}}, \bibinfo {author} {\bibfnamefont {W.}~\bibnamefont {Ong}}, \bibinfo
  {author} {\bibfnamefont {I.}~\bibnamefont {Arakelyan}}, \ and\ \bibinfo
  {author} {\bibfnamefont {J.}~\bibnamefont {Thomas}},\ }\href@noop {}
  {\bibfield  {journal} {\bibinfo  {journal} {Phys. Rev. Lett.}\ }\textbf
  {\bibinfo {volume} {108}},\ \bibinfo {pages} {235302} (\bibinfo {year}
  {2012})}\BibitemShut {NoStop}%
\bibitem [{\citenamefont {Scazza}\ \emph
  {et~al.}(2017{\natexlab{a}})\citenamefont {Scazza}, \citenamefont
  {Valtolina}, \citenamefont {Massignan}, \citenamefont {Recati}, \citenamefont
  {Amico}, \citenamefont {Burchianti}, \citenamefont {Fort}, \citenamefont
  {Inguscio}, \citenamefont {Zaccanti},\ and\ \citenamefont
  {Roati}}]{scazza2017}%
  \BibitemOpen
  \bibfield  {author} {\bibinfo {author} {\bibfnamefont {F.}~\bibnamefont
  {Scazza}}, \bibinfo {author} {\bibfnamefont {G.}~\bibnamefont {Valtolina}},
  \bibinfo {author} {\bibfnamefont {P.}~\bibnamefont {Massignan}}, \bibinfo
  {author} {\bibfnamefont {A.}~\bibnamefont {Recati}}, \bibinfo {author}
  {\bibfnamefont {A.}~\bibnamefont {Amico}}, \bibinfo {author} {\bibfnamefont
  {A.}~\bibnamefont {Burchianti}}, \bibinfo {author} {\bibfnamefont
  {C.}~\bibnamefont {Fort}}, \bibinfo {author} {\bibfnamefont {M.}~\bibnamefont
  {Inguscio}}, \bibinfo {author} {\bibfnamefont {M.}~\bibnamefont {Zaccanti}},
  \ and\ \bibinfo {author} {\bibfnamefont {G.}~\bibnamefont {Roati}},\ }\href
  {\doibase 10.1103/PhysRevLett.118.083602} {\bibfield  {journal} {\bibinfo
  {journal} {Phys. Rev. Lett.}\ }\textbf {\bibinfo {volume} {118}},\ \bibinfo
  {pages} {083602} (\bibinfo {year} {2017}{\natexlab{a}})}\BibitemShut
  {NoStop}%
\bibitem [{\citenamefont {Landau}(1933)}]{Landau1933}%
  \BibitemOpen
  \bibfield  {author} {\bibinfo {author} {\bibfnamefont {L.~D.}\ \bibnamefont
  {Landau}},\ }\href@noop {} {\bibfield  {journal} {\bibinfo  {journal} {Phys.
  Z. Sowjetunion}\ }\textbf {\bibinfo {volume} {3}},\ \bibinfo {pages} {644}
  (\bibinfo {year} {1933})}\BibitemShut {NoStop}%
\bibitem [{\citenamefont {Fröhlich}(1954)}]{Frohlich1954}%
  \BibitemOpen
  \bibfield  {author} {\bibinfo {author} {\bibfnamefont {H.}~\bibnamefont
  {Fröhlich}},\ }\href@noop {} {\bibfield  {journal} {\bibinfo  {journal}
  {Advances in Physics}\ }\textbf {\bibinfo {volume} {3}},\ \bibinfo {pages}
  {325} (\bibinfo {year} {1954})}\BibitemShut {NoStop}%
\bibitem [{\citenamefont {Schmidt}\ \emph {et~al.}(2018)\citenamefont
  {Schmidt}, \citenamefont {Knap}, \citenamefont {Ivanov}, \citenamefont {You},
  \citenamefont {Cetina},\ and\ \citenamefont {Demler}}]{schmidt2018universal}%
  \BibitemOpen
  \bibfield  {author} {\bibinfo {author} {\bibfnamefont {R.}~\bibnamefont
  {Schmidt}}, \bibinfo {author} {\bibfnamefont {M.}~\bibnamefont {Knap}},
  \bibinfo {author} {\bibfnamefont {D.~A.}\ \bibnamefont {Ivanov}}, \bibinfo
  {author} {\bibfnamefont {J.-S.}\ \bibnamefont {You}}, \bibinfo {author}
  {\bibfnamefont {M.}~\bibnamefont {Cetina}}, \ and\ \bibinfo {author}
  {\bibfnamefont {E.}~\bibnamefont {Demler}},\ }\href@noop {} {\bibfield
  {journal} {\bibinfo  {journal} {Rep. Progr. Phys.}\ }\textbf {\bibinfo
  {volume} {81}},\ \bibinfo {pages} {024401} (\bibinfo {year}
  {2018})}\BibitemShut {NoStop}%
\bibitem [{\citenamefont {Massignan}\ \emph {et~al.}(2014)\citenamefont
  {Massignan}, \citenamefont {Zaccanti},\ and\ \citenamefont
  {Bruun}}]{Massignan2014}%
  \BibitemOpen
  \bibfield  {author} {\bibinfo {author} {\bibfnamefont {P.}~\bibnamefont
  {Massignan}}, \bibinfo {author} {\bibfnamefont {M.}~\bibnamefont {Zaccanti}},
  \ and\ \bibinfo {author} {\bibfnamefont {G.~M.}\ \bibnamefont {Bruun}},\
  }\href {\doibase 10.1088/0034-4885/77/3/034401} {\bibfield  {journal}
  {\bibinfo  {journal} {Rep. Progr. Phys.}\ }\textbf {\bibinfo {volume} {77}},\
  \bibinfo {pages} {034401} (\bibinfo {year} {2014})}\BibitemShut {NoStop}%
\bibitem [{\citenamefont {Khandekar}\ \emph {et~al.}(1988)\citenamefont
  {Khandekar}, \citenamefont {Bhagwat},\ and\ \citenamefont
  {Lawande}}]{Khandekar1988}%
  \BibitemOpen
  \bibfield  {author} {\bibinfo {author} {\bibfnamefont {D.~C.}\ \bibnamefont
  {Khandekar}}, \bibinfo {author} {\bibfnamefont {K.~V.}\ \bibnamefont
  {Bhagwat}}, \ and\ \bibinfo {author} {\bibfnamefont {S.~V.}\ \bibnamefont
  {Lawande}},\ }\href {\doibase 10.1103/PhysRevB.37.3085} {\bibfield  {journal}
  {\bibinfo  {journal} {Phys. Rev. B}\ }\textbf {\bibinfo {volume} {37}},\
  \bibinfo {pages} {3085} (\bibinfo {year} {1988})}\BibitemShut {NoStop}%
\bibitem [{\citenamefont {Feynman}\ \emph {et~al.}(1962)\citenamefont
  {Feynman}, \citenamefont {Hellwarth}, \citenamefont {Iddings},\ and\
  \citenamefont {Platzman}}]{Feynman1962}%
  \BibitemOpen
  \bibfield  {author} {\bibinfo {author} {\bibfnamefont {R.~P.}\ \bibnamefont
  {Feynman}}, \bibinfo {author} {\bibfnamefont {R.~W.}\ \bibnamefont
  {Hellwarth}}, \bibinfo {author} {\bibfnamefont {C.~K.}\ \bibnamefont
  {Iddings}}, \ and\ \bibinfo {author} {\bibfnamefont {P.~M.}\ \bibnamefont
  {Platzman}},\ }\href {\doibase 10.1103/PhysRev.127.1004} {\bibfield
  {journal} {\bibinfo  {journal} {Phys. Rev.}\ }\textbf {\bibinfo {volume}
  {127}},\ \bibinfo {pages} {1004} (\bibinfo {year} {1962})}\BibitemShut
  {NoStop}%
\bibitem [{\citenamefont {J\o{}rgensen}\ \emph {et~al.}(2016)\citenamefont
  {J\o{}rgensen}, \citenamefont {Wacker}, \citenamefont {Skalmstang},
  \citenamefont {Parish}, \citenamefont {Levinsen}, \citenamefont
  {Christensen}, \citenamefont {Bruun},\ and\ \citenamefont {Arlt}}]{Nils2016}%
  \BibitemOpen
  \bibfield  {author} {\bibinfo {author} {\bibfnamefont {N.~B.}\ \bibnamefont
  {J\o{}rgensen}}, \bibinfo {author} {\bibfnamefont {L.}~\bibnamefont
  {Wacker}}, \bibinfo {author} {\bibfnamefont {K.~T.}\ \bibnamefont
  {Skalmstang}}, \bibinfo {author} {\bibfnamefont {M.~M.}\ \bibnamefont
  {Parish}}, \bibinfo {author} {\bibfnamefont {J.}~\bibnamefont {Levinsen}},
  \bibinfo {author} {\bibfnamefont {R.~S.}\ \bibnamefont {Christensen}},
  \bibinfo {author} {\bibfnamefont {G.~M.}\ \bibnamefont {Bruun}}, \ and\
  \bibinfo {author} {\bibfnamefont {J.~J.}\ \bibnamefont {Arlt}},\ }\href
  {\doibase 10.1103/PhysRevLett.117.055302} {\bibfield  {journal} {\bibinfo
  {journal} {Phys. Rev. Lett.}\ }\textbf {\bibinfo {volume} {117}},\ \bibinfo
  {pages} {055302} (\bibinfo {year} {2016})}\BibitemShut {NoStop}%
\bibitem [{\citenamefont {Hu}\ \emph {et~al.}(2016)\citenamefont {Hu},
  \citenamefont {Van~de Graaff}, \citenamefont {Kedar}, \citenamefont {Corson},
  \citenamefont {Cornell},\ and\ \citenamefont {Jin}}]{Hu2016}%
  \BibitemOpen
  \bibfield  {author} {\bibinfo {author} {\bibfnamefont {M.-G.}\ \bibnamefont
  {Hu}}, \bibinfo {author} {\bibfnamefont {M.~J.}\ \bibnamefont {Van~de
  Graaff}}, \bibinfo {author} {\bibfnamefont {D.}~\bibnamefont {Kedar}},
  \bibinfo {author} {\bibfnamefont {J.~P.}\ \bibnamefont {Corson}}, \bibinfo
  {author} {\bibfnamefont {E.~A.}\ \bibnamefont {Cornell}}, \ and\ \bibinfo
  {author} {\bibfnamefont {D.~S.}\ \bibnamefont {Jin}},\ }\href {\doibase
  10.1103/PhysRevLett.117.055301} {\bibfield  {journal} {\bibinfo  {journal}
  {Phys. Rev. Lett.}\ }\textbf {\bibinfo {volume} {117}},\ \bibinfo {pages}
  {055301} (\bibinfo {year} {2016})}\BibitemShut {NoStop}%
\bibitem [{\citenamefont {Catani}\ \emph
  {et~al.}(2009{\natexlab{a}})\citenamefont {Catani}, \citenamefont
  {Barontini}, \citenamefont {Lamporesi}, \citenamefont {Rabatti},
  \citenamefont {Thalhammer}, \citenamefont {Minardi}, \citenamefont
  {Stringari},\ and\ \citenamefont {Inguscio}}]{catani2009entropy}%
  \BibitemOpen
  \bibfield  {author} {\bibinfo {author} {\bibfnamefont {J.}~\bibnamefont
  {Catani}}, \bibinfo {author} {\bibfnamefont {G.}~\bibnamefont {Barontini}},
  \bibinfo {author} {\bibfnamefont {G.}~\bibnamefont {Lamporesi}}, \bibinfo
  {author} {\bibfnamefont {F.}~\bibnamefont {Rabatti}}, \bibinfo {author}
  {\bibfnamefont {G.}~\bibnamefont {Thalhammer}}, \bibinfo {author}
  {\bibfnamefont {F.}~\bibnamefont {Minardi}}, \bibinfo {author} {\bibfnamefont
  {S.}~\bibnamefont {Stringari}}, \ and\ \bibinfo {author} {\bibfnamefont
  {M.}~\bibnamefont {Inguscio}},\ }\href@noop {} {\bibfield  {journal}
  {\bibinfo  {journal} {Phys. Rev. Lett.}\ }\textbf {\bibinfo {volume} {103}},\
  \bibinfo {pages} {140401} (\bibinfo {year} {2009}{\natexlab{a}})}\BibitemShut
  {NoStop}%
\bibitem [{\citenamefont {Fukuhara}\ \emph {et~al.}(2013)\citenamefont
  {Fukuhara}, \citenamefont {Kantian}, \citenamefont {Endres}, \citenamefont
  {Cheneau}, \citenamefont {Schau{\ss}}, \citenamefont {Hild}, \citenamefont
  {Bellem}, \citenamefont {Schollw{\"o}ck}, \citenamefont {Giamarchi},
  \citenamefont {Gross}, \citenamefont {Bloch},\ and\ \citenamefont
  {Kuhr}}]{fukuhara2013quantum}%
  \BibitemOpen
  \bibfield  {author} {\bibinfo {author} {\bibfnamefont {T.}~\bibnamefont
  {Fukuhara}}, \bibinfo {author} {\bibfnamefont {A.}~\bibnamefont {Kantian}},
  \bibinfo {author} {\bibfnamefont {M.}~\bibnamefont {Endres}}, \bibinfo
  {author} {\bibfnamefont {M.}~\bibnamefont {Cheneau}}, \bibinfo {author}
  {\bibfnamefont {P.}~\bibnamefont {Schau{\ss}}}, \bibinfo {author}
  {\bibfnamefont {S.}~\bibnamefont {Hild}}, \bibinfo {author} {\bibfnamefont
  {D.}~\bibnamefont {Bellem}}, \bibinfo {author} {\bibfnamefont
  {U.}~\bibnamefont {Schollw{\"o}ck}}, \bibinfo {author} {\bibfnamefont
  {T.}~\bibnamefont {Giamarchi}}, \bibinfo {author} {\bibfnamefont
  {C.}~\bibnamefont {Gross}}, \bibinfo {author} {\bibfnamefont
  {I.}~\bibnamefont {Bloch}}, \ and\ \bibinfo {author} {\bibfnamefont
  {S.}~\bibnamefont {Kuhr}},\ }\href@noop {} {\bibfield  {journal} {\bibinfo
  {journal} {Nat. Phys.}\ }\textbf {\bibinfo {volume} {9}},\ \bibinfo {pages}
  {235} (\bibinfo {year} {2013})}\BibitemShut {NoStop}%
\bibitem [{\citenamefont {Scazza}\ \emph
  {et~al.}(2017{\natexlab{b}})\citenamefont {Scazza}, \citenamefont
  {Valtolina}, \citenamefont {Massignan}, \citenamefont {Recati}, \citenamefont
  {Amico}, \citenamefont {Burchianti}, \citenamefont {Fort}, \citenamefont
  {Inguscio}, \citenamefont {Zaccanti},\ and\ \citenamefont
  {Roati}}]{scazza2017repulsive}%
  \BibitemOpen
  \bibfield  {author} {\bibinfo {author} {\bibfnamefont {F.}~\bibnamefont
  {Scazza}}, \bibinfo {author} {\bibfnamefont {G.}~\bibnamefont {Valtolina}},
  \bibinfo {author} {\bibfnamefont {P.}~\bibnamefont {Massignan}}, \bibinfo
  {author} {\bibfnamefont {A.}~\bibnamefont {Recati}}, \bibinfo {author}
  {\bibfnamefont {A.}~\bibnamefont {Amico}}, \bibinfo {author} {\bibfnamefont
  {A.}~\bibnamefont {Burchianti}}, \bibinfo {author} {\bibfnamefont
  {C.}~\bibnamefont {Fort}}, \bibinfo {author} {\bibfnamefont {M.}~\bibnamefont
  {Inguscio}}, \bibinfo {author} {\bibfnamefont {M.}~\bibnamefont {Zaccanti}},
  \ and\ \bibinfo {author} {\bibfnamefont {G.}~\bibnamefont {Roati}},\
  }\href@noop {} {\bibfield  {journal} {\bibinfo  {journal} {Phys. Rev. Lett.}\
  }\textbf {\bibinfo {volume} {118}},\ \bibinfo {pages} {083602} (\bibinfo
  {year} {2017}{\natexlab{b}})}\BibitemShut {NoStop}%
\bibitem [{\citenamefont {Cetina}\ \emph {et~al.}(2015)\citenamefont {Cetina},
  \citenamefont {Jag}, \citenamefont {Lous}, \citenamefont {Walraven},
  \citenamefont {Grimm}, \citenamefont {Christensen},\ and\ \citenamefont
  {Bruun}}]{cetina2015decoherence}%
  \BibitemOpen
  \bibfield  {author} {\bibinfo {author} {\bibfnamefont {M.}~\bibnamefont
  {Cetina}}, \bibinfo {author} {\bibfnamefont {M.}~\bibnamefont {Jag}},
  \bibinfo {author} {\bibfnamefont {R.~S.}\ \bibnamefont {Lous}}, \bibinfo
  {author} {\bibfnamefont {J.~T.}\ \bibnamefont {Walraven}}, \bibinfo {author}
  {\bibfnamefont {R.}~\bibnamefont {Grimm}}, \bibinfo {author} {\bibfnamefont
  {R.~S.}\ \bibnamefont {Christensen}}, \ and\ \bibinfo {author} {\bibfnamefont
  {G.~M.}\ \bibnamefont {Bruun}},\ }\href@noop {} {\bibfield  {journal}
  {\bibinfo  {journal} {Phys. Rev. Lett.}\ }\textbf {\bibinfo {volume} {115}},\
  \bibinfo {pages} {135302} (\bibinfo {year} {2015})}\BibitemShut {NoStop}%
\bibitem [{\citenamefont {Cetina}\ \emph
  {et~al.}(2016{\natexlab{a}})\citenamefont {Cetina}, \citenamefont {Jag},
  \citenamefont {Lous}, \citenamefont {Fritsche}, \citenamefont {Walraven},
  \citenamefont {Grimm}, \citenamefont {Levinsen}, \citenamefont {Parish},
  \citenamefont {Schmidt}, \citenamefont {Knap},\ and\ \citenamefont
  {Demler}}]{cetina2016ultrafast}%
  \BibitemOpen
  \bibfield  {author} {\bibinfo {author} {\bibfnamefont {M.}~\bibnamefont
  {Cetina}}, \bibinfo {author} {\bibfnamefont {M.}~\bibnamefont {Jag}},
  \bibinfo {author} {\bibfnamefont {R.~S.}\ \bibnamefont {Lous}}, \bibinfo
  {author} {\bibfnamefont {I.}~\bibnamefont {Fritsche}}, \bibinfo {author}
  {\bibfnamefont {J.~T.}\ \bibnamefont {Walraven}}, \bibinfo {author}
  {\bibfnamefont {R.}~\bibnamefont {Grimm}}, \bibinfo {author} {\bibfnamefont
  {J.}~\bibnamefont {Levinsen}}, \bibinfo {author} {\bibfnamefont {M.~M.}\
  \bibnamefont {Parish}}, \bibinfo {author} {\bibfnamefont {R.}~\bibnamefont
  {Schmidt}}, \bibinfo {author} {\bibfnamefont {M.}~\bibnamefont {Knap}}, \
  and\ \bibinfo {author} {\bibfnamefont {E.}~\bibnamefont {Demler}},\
  }\href@noop {} {\bibfield  {journal} {\bibinfo  {journal} {Science}\ }\textbf
  {\bibinfo {volume} {354}},\ \bibinfo {pages} {96} (\bibinfo {year}
  {2016}{\natexlab{a}})}\BibitemShut {NoStop}%
\bibitem [{\citenamefont {Volosniev}\ and\ \citenamefont
  {Hammer}(2017)}]{volosniev2017analytical}%
  \BibitemOpen
  \bibfield  {author} {\bibinfo {author} {\bibfnamefont {A.~G.}\ \bibnamefont
  {Volosniev}}\ and\ \bibinfo {author} {\bibfnamefont {H.-W.}\ \bibnamefont
  {Hammer}},\ }\href@noop {} {\bibfield  {journal} {\bibinfo  {journal} {Phys.
  Rev. A}\ }\textbf {\bibinfo {volume} {96}},\ \bibinfo {pages} {031601}
  (\bibinfo {year} {2017})}\BibitemShut {NoStop}%
\bibitem [{\citenamefont {Dehkharghani}\ \emph {et~al.}(2018)\citenamefont
  {Dehkharghani}, \citenamefont {Volosniev},\ and\ \citenamefont
  {Zinner}}]{dehkharghani2018coalescence}%
  \BibitemOpen
  \bibfield  {author} {\bibinfo {author} {\bibfnamefont {A.~S.}\ \bibnamefont
  {Dehkharghani}}, \bibinfo {author} {\bibfnamefont {A.~G.}\ \bibnamefont
  {Volosniev}}, \ and\ \bibinfo {author} {\bibfnamefont {N.~T.}\ \bibnamefont
  {Zinner}},\ }\href@noop {} {\bibfield  {journal} {\bibinfo  {journal} {Phys.
  Rev. Lett.}\ }\textbf {\bibinfo {volume} {121}},\ \bibinfo {pages} {080405}
  (\bibinfo {year} {2018})}\BibitemShut {NoStop}%
\bibitem [{\citenamefont {Mistakidis}\ \emph
  {et~al.}(2019{\natexlab{a}})\citenamefont {Mistakidis}, \citenamefont
  {Katsimiga}, \citenamefont {Koutentakis},\ and\ \citenamefont
  {Schmelcher}}]{Mistakidis2019}%
  \BibitemOpen
  \bibfield  {author} {\bibinfo {author} {\bibfnamefont {S.~I.}\ \bibnamefont
  {Mistakidis}}, \bibinfo {author} {\bibfnamefont {G.~C.}\ \bibnamefont
  {Katsimiga}}, \bibinfo {author} {\bibfnamefont {G.~M.}\ \bibnamefont
  {Koutentakis}}, \ and\ \bibinfo {author} {\bibfnamefont {P.}~\bibnamefont
  {Schmelcher}},\ }\href {\doibase 10.1088/1367-2630/ab1045} {\bibfield
  {journal} {\bibinfo  {journal} {New J. Phys.}\ }\textbf {\bibinfo {volume}
  {21}},\ \bibinfo {pages} {043032} (\bibinfo {year}
  {2019}{\natexlab{a}})}\BibitemShut {NoStop}%
\bibitem [{\citenamefont {Ardila}\ and\ \citenamefont
  {Pohl}(2018)}]{Ardila2018}%
  \BibitemOpen
  \bibfield  {author} {\bibinfo {author} {\bibfnamefont {L.~A.~P.}\
  \bibnamefont {Ardila}}\ and\ \bibinfo {author} {\bibfnamefont
  {T.}~\bibnamefont {Pohl}},\ }\href {\doibase 10.1088/1361-6455/aaf35e}
  {\bibfield  {journal} {\bibinfo  {journal} {J. Phys. B: At. Mol. and Opt.
  Phys.}\ }\textbf {\bibinfo {volume} {52}},\ \bibinfo {pages} {015004}
  (\bibinfo {year} {2018})}\BibitemShut {NoStop}%
\bibitem [{\citenamefont {Ardila}\ and\ \citenamefont
  {Giorgini}(2015)}]{ardila2015impurity}%
  \BibitemOpen
  \bibfield  {author} {\bibinfo {author} {\bibfnamefont {L.~A.~P.}\
  \bibnamefont {Ardila}}\ and\ \bibinfo {author} {\bibfnamefont
  {S.}~\bibnamefont {Giorgini}},\ }\href@noop {} {\bibfield  {journal}
  {\bibinfo  {journal} {Phys. Rev. A}\ }\textbf {\bibinfo {volume} {92}},\
  \bibinfo {pages} {033612} (\bibinfo {year} {2015})}\BibitemShut {NoStop}%
\bibitem [{\citenamefont {Ardila}\ \emph {et~al.}(2019)\citenamefont {Ardila},
  \citenamefont {J{\o}rgensen}, \citenamefont {Pohl}, \citenamefont {Giorgini},
  \citenamefont {Bruun},\ and\ \citenamefont {Arlt}}]{ardila2019analyzing}%
  \BibitemOpen
  \bibfield  {author} {\bibinfo {author} {\bibfnamefont {L.~A.~P.}\
  \bibnamefont {Ardila}}, \bibinfo {author} {\bibfnamefont {N.~B.}\
  \bibnamefont {J{\o}rgensen}}, \bibinfo {author} {\bibfnamefont
  {T.}~\bibnamefont {Pohl}}, \bibinfo {author} {\bibfnamefont {S.}~\bibnamefont
  {Giorgini}}, \bibinfo {author} {\bibfnamefont {G.}~\bibnamefont {Bruun}}, \
  and\ \bibinfo {author} {\bibfnamefont {J.}~\bibnamefont {Arlt}},\ }\href@noop
  {} {\bibfield  {journal} {\bibinfo  {journal} {Phys. Rev. A}\ }\textbf
  {\bibinfo {volume} {99}},\ \bibinfo {pages} {063607} (\bibinfo {year}
  {2019})}\BibitemShut {NoStop}%
\bibitem [{\citenamefont {Grusdt}\ \emph {et~al.}(2018)\citenamefont {Grusdt},
  \citenamefont {Seetharam}, \citenamefont {Shchadilova},\ and\ \citenamefont
  {Demler}}]{Grusdt2018}%
  \BibitemOpen
  \bibfield  {author} {\bibinfo {author} {\bibfnamefont {F.}~\bibnamefont
  {Grusdt}}, \bibinfo {author} {\bibfnamefont {K.}~\bibnamefont {Seetharam}},
  \bibinfo {author} {\bibfnamefont {Y.}~\bibnamefont {Shchadilova}}, \ and\
  \bibinfo {author} {\bibfnamefont {E.}~\bibnamefont {Demler}},\ }\href
  {\doibase 10.1103/PhysRevA.97.033612} {\bibfield  {journal} {\bibinfo
  {journal} {Phys. Rev. A}\ }\textbf {\bibinfo {volume} {97}},\ \bibinfo
  {pages} {033612} (\bibinfo {year} {2018})}\BibitemShut {NoStop}%
\bibitem [{\citenamefont {Grusdt}\ \emph
  {et~al.}(2017{\natexlab{a}})\citenamefont {Grusdt}, \citenamefont {Schmidt},
  \citenamefont {Shchadilova},\ and\ \citenamefont {Demler}}]{Grusdt2017}%
  \BibitemOpen
  \bibfield  {author} {\bibinfo {author} {\bibfnamefont {F.}~\bibnamefont
  {Grusdt}}, \bibinfo {author} {\bibfnamefont {R.}~\bibnamefont {Schmidt}},
  \bibinfo {author} {\bibfnamefont {Y.~E.}\ \bibnamefont {Shchadilova}}, \ and\
  \bibinfo {author} {\bibfnamefont {E.}~\bibnamefont {Demler}},\ }\href
  {\doibase 10.1103/PhysRevA.96.013607} {\bibfield  {journal} {\bibinfo
  {journal} {Phys. Rev. A}\ }\textbf {\bibinfo {volume} {96}},\ \bibinfo
  {pages} {013607} (\bibinfo {year} {2017}{\natexlab{a}})}\BibitemShut
  {NoStop}%
\bibitem [{\citenamefont {Grusdt}\ and\ \citenamefont
  {Demler}(2015)}]{grusdt2015new}%
  \BibitemOpen
  \bibfield  {author} {\bibinfo {author} {\bibfnamefont {F.}~\bibnamefont
  {Grusdt}}\ and\ \bibinfo {author} {\bibfnamefont {E.}~\bibnamefont
  {Demler}},\ }\href@noop {} {\bibfield  {journal} {\bibinfo  {journal}
  {Quantum Matter at Ultralow Temperatures}\ }\textbf {\bibinfo {volume}
  {191}},\ \bibinfo {pages} {325} (\bibinfo {year} {2015})}\BibitemShut
  {NoStop}%
\bibitem [{\citenamefont {Tempere}\ \emph {et~al.}(2009)\citenamefont
  {Tempere}, \citenamefont {Casteels}, \citenamefont {Oberthaler},
  \citenamefont {Knoop}, \citenamefont {Timmermans},\ and\ \citenamefont
  {Devreese}}]{Tempere2009}%
  \BibitemOpen
  \bibfield  {author} {\bibinfo {author} {\bibfnamefont {J.}~\bibnamefont
  {Tempere}}, \bibinfo {author} {\bibfnamefont {W.}~\bibnamefont {Casteels}},
  \bibinfo {author} {\bibfnamefont {M.~K.}\ \bibnamefont {Oberthaler}},
  \bibinfo {author} {\bibfnamefont {S.}~\bibnamefont {Knoop}}, \bibinfo
  {author} {\bibfnamefont {E.}~\bibnamefont {Timmermans}}, \ and\ \bibinfo
  {author} {\bibfnamefont {J.~T.}\ \bibnamefont {Devreese}},\ }\href {\doibase
  10.1103/PhysRevB.80.184504} {\bibfield  {journal} {\bibinfo  {journal} {Phys.
  Rev. B}\ }\textbf {\bibinfo {volume} {80}},\ \bibinfo {pages} {184504}
  (\bibinfo {year} {2009})}\BibitemShut {NoStop}%
\bibitem [{\citenamefont {Panochko}\ and\ \citenamefont
  {Pastukhov}(2019)}]{panochko2019two}%
  \BibitemOpen
  \bibfield  {author} {\bibinfo {author} {\bibfnamefont {G.}~\bibnamefont
  {Panochko}}\ and\ \bibinfo {author} {\bibfnamefont {V.}~\bibnamefont
  {Pastukhov}},\ }\href@noop {} {\bibfield  {journal} {\bibinfo  {journal}
  {arXiv:\textbf{1909.01256}}\ } (\bibinfo {year} {2019})}\BibitemShut
  {NoStop}%
\bibitem [{\citenamefont {Mistakidis}\ \emph
  {et~al.}(2019{\natexlab{b}})\citenamefont {Mistakidis}, \citenamefont
  {Volosniev}, \citenamefont {Zinner},\ and\ \citenamefont
  {Schmelcher}}]{mistakidis2019effective}%
  \BibitemOpen
  \bibfield  {author} {\bibinfo {author} {\bibfnamefont {S.~I.}\ \bibnamefont
  {Mistakidis}}, \bibinfo {author} {\bibfnamefont {A.~G.}\ \bibnamefont
  {Volosniev}}, \bibinfo {author} {\bibfnamefont {N.~T.}\ \bibnamefont
  {Zinner}}, \ and\ \bibinfo {author} {\bibfnamefont {P.}~\bibnamefont
  {Schmelcher}},\ }\href@noop {} {\bibfield  {journal} {\bibinfo  {journal}
  {Phys. Rev. A}\ }\textbf {\bibinfo {volume} {100}},\ \bibinfo {pages}
  {013619} (\bibinfo {year} {2019}{\natexlab{b}})}\BibitemShut {NoStop}%
\bibitem [{\citenamefont {Mistakidis}\ \emph
  {et~al.}(2019{\natexlab{c}})\citenamefont {Mistakidis}, \citenamefont
  {Katsimiga}, \citenamefont {Koutentakis}, \citenamefont {Busch},\ and\
  \citenamefont {Schmelcher}}]{mistakidis2019quench}%
  \BibitemOpen
  \bibfield  {author} {\bibinfo {author} {\bibfnamefont {S.~I.}\ \bibnamefont
  {Mistakidis}}, \bibinfo {author} {\bibfnamefont {G.~C.}\ \bibnamefont
  {Katsimiga}}, \bibinfo {author} {\bibfnamefont {G.~M.}\ \bibnamefont
  {Koutentakis}}, \bibinfo {author} {\bibfnamefont {T.}~\bibnamefont {Busch}},
  \ and\ \bibinfo {author} {\bibfnamefont {P.}~\bibnamefont {Schmelcher}},\
  }\href@noop {} {\bibfield  {journal} {\bibinfo  {journal} {Phys. Rev. Lett.}\
  }\textbf {\bibinfo {volume} {122}},\ \bibinfo {pages} {183001} (\bibinfo
  {year} {2019}{\natexlab{c}})}\BibitemShut {NoStop}%
\bibitem [{\citenamefont {Volosniev}\ \emph {et~al.}(2015)\citenamefont
  {Volosniev}, \citenamefont {Hammer},\ and\ \citenamefont
  {Zinner}}]{volosniev2015real}%
  \BibitemOpen
  \bibfield  {author} {\bibinfo {author} {\bibfnamefont {A.~G.}\ \bibnamefont
  {Volosniev}}, \bibinfo {author} {\bibfnamefont {H.-W.}\ \bibnamefont
  {Hammer}}, \ and\ \bibinfo {author} {\bibfnamefont {N.~T.}\ \bibnamefont
  {Zinner}},\ }\href@noop {} {\bibfield  {journal} {\bibinfo  {journal} {Phys.
  Rev. A}\ }\textbf {\bibinfo {volume} {92}},\ \bibinfo {pages} {023623}
  (\bibinfo {year} {2015})}\BibitemShut {NoStop}%
\bibitem [{\citenamefont {Mistakidis}\ \emph
  {et~al.}(2019{\natexlab{d}})\citenamefont {Mistakidis}, \citenamefont
  {Hilbig},\ and\ \citenamefont {Schmelcher}}]{mistakidis2019correlated}%
  \BibitemOpen
  \bibfield  {author} {\bibinfo {author} {\bibfnamefont {S.~I.}\ \bibnamefont
  {Mistakidis}}, \bibinfo {author} {\bibfnamefont {L.}~\bibnamefont {Hilbig}},
  \ and\ \bibinfo {author} {\bibfnamefont {P.}~\bibnamefont {Schmelcher}},\
  }\href@noop {} {\bibfield  {journal} {\bibinfo  {journal} {Phys. Rev. A}\
  }\textbf {\bibinfo {volume} {100}},\ \bibinfo {pages} {023620} (\bibinfo
  {year} {2019}{\natexlab{d}})}\BibitemShut {NoStop}%
\bibitem [{\citenamefont {Shchadilova}\ \emph {et~al.}(2016)\citenamefont
  {Shchadilova}, \citenamefont {Schmidt}, \citenamefont {Grusdt},\ and\
  \citenamefont {Demler}}]{shchadilova2016quantum}%
  \BibitemOpen
  \bibfield  {author} {\bibinfo {author} {\bibfnamefont {Y.~E.}\ \bibnamefont
  {Shchadilova}}, \bibinfo {author} {\bibfnamefont {R.}~\bibnamefont
  {Schmidt}}, \bibinfo {author} {\bibfnamefont {F.}~\bibnamefont {Grusdt}}, \
  and\ \bibinfo {author} {\bibfnamefont {E.}~\bibnamefont {Demler}},\
  }\href@noop {} {\bibfield  {journal} {\bibinfo  {journal} {Phys. Rev. Lett.}\
  }\textbf {\bibinfo {volume} {117}},\ \bibinfo {pages} {113002} (\bibinfo
  {year} {2016})}\BibitemShut {NoStop}%
\bibitem [{\citenamefont {Kamar}\ \emph {et~al.}(2019)\citenamefont {Kamar},
  \citenamefont {Kantian},\ and\ \citenamefont
  {Giamarchi}}]{kamar2019dynamics}%
  \BibitemOpen
  \bibfield  {author} {\bibinfo {author} {\bibfnamefont {N.~A.}\ \bibnamefont
  {Kamar}}, \bibinfo {author} {\bibfnamefont {A.}~\bibnamefont {Kantian}}, \
  and\ \bibinfo {author} {\bibfnamefont {T.}~\bibnamefont {Giamarchi}},\
  }\href@noop {} {\bibfield  {journal} {\bibinfo  {journal} {Phys. Rev. A}\
  }\textbf {\bibinfo {volume} {100}},\ \bibinfo {pages} {023614} (\bibinfo
  {year} {2019})}\BibitemShut {NoStop}%
\bibitem [{\citenamefont {Boyanovsky}\ \emph {et~al.}(2019)\citenamefont
  {Boyanovsky}, \citenamefont {Jasnow}, \citenamefont {Wu},\ and\ \citenamefont
  {Coalson}}]{boyanovsky2019dynamics}%
  \BibitemOpen
  \bibfield  {author} {\bibinfo {author} {\bibfnamefont {D.}~\bibnamefont
  {Boyanovsky}}, \bibinfo {author} {\bibfnamefont {D.}~\bibnamefont {Jasnow}},
  \bibinfo {author} {\bibfnamefont {X.-L.}\ \bibnamefont {Wu}}, \ and\ \bibinfo
  {author} {\bibfnamefont {R.~C.}\ \bibnamefont {Coalson}},\ }\href@noop {}
  {\bibfield  {journal} {\bibinfo  {journal} {Phys. Rev. A}\ }\textbf {\bibinfo
  {volume} {100}},\ \bibinfo {pages} {043617} (\bibinfo {year}
  {2019})}\BibitemShut {NoStop}%
\bibitem [{\citenamefont {Li}\ and\ \citenamefont
  {Kuang}(2019)}]{li2019controlling}%
  \BibitemOpen
  \bibfield  {author} {\bibinfo {author} {\bibfnamefont {Z.}~\bibnamefont
  {Li}}\ and\ \bibinfo {author} {\bibfnamefont {L.-M.}\ \bibnamefont {Kuang}},\
  }\href@noop {} {\bibfield  {journal} {\bibinfo  {journal}
  {arXiv:\textbf{1909.03374}}\ } (\bibinfo {year} {2019})}\BibitemShut
  {NoStop}%
\bibitem [{\citenamefont {Pasek}\ and\ \citenamefont
  {Orso}(2019)}]{pasek2019induced}%
  \BibitemOpen
  \bibfield  {author} {\bibinfo {author} {\bibfnamefont {M.}~\bibnamefont
  {Pasek}}\ and\ \bibinfo {author} {\bibfnamefont {G.}~\bibnamefont {Orso}},\
  }\href@noop {} {\bibfield  {journal} {\bibinfo  {journal} {Phys. Rev. B}\
  }\textbf {\bibinfo {volume} {100}},\ \bibinfo {pages} {245419} (\bibinfo
  {year} {2019})}\BibitemShut {NoStop}%
\bibitem [{\citenamefont {Grusdt}\ \emph
  {et~al.}(2017{\natexlab{b}})\citenamefont {Grusdt}, \citenamefont
  {Astrakharchik},\ and\ \citenamefont {Demler}}]{grusdt2017bose}%
  \BibitemOpen
  \bibfield  {author} {\bibinfo {author} {\bibfnamefont {F.}~\bibnamefont
  {Grusdt}}, \bibinfo {author} {\bibfnamefont {G.~E.}\ \bibnamefont
  {Astrakharchik}}, \ and\ \bibinfo {author} {\bibfnamefont {E.}~\bibnamefont
  {Demler}},\ }\href@noop {} {\bibfield  {journal} {\bibinfo  {journal} {New J.
  Phys.}\ }\textbf {\bibinfo {volume} {19}},\ \bibinfo {pages} {103035}
  (\bibinfo {year} {2017}{\natexlab{b}})}\BibitemShut {NoStop}%
\bibitem [{\citenamefont {Knap}\ \emph {et~al.}(2012)\citenamefont {Knap},
  \citenamefont {Shashi}, \citenamefont {Nishida}, \citenamefont {Imambekov},
  \citenamefont {Abanin},\ and\ \citenamefont {Demler}}]{knap2012time}%
  \BibitemOpen
  \bibfield  {author} {\bibinfo {author} {\bibfnamefont {M.}~\bibnamefont
  {Knap}}, \bibinfo {author} {\bibfnamefont {A.}~\bibnamefont {Shashi}},
  \bibinfo {author} {\bibfnamefont {Y.}~\bibnamefont {Nishida}}, \bibinfo
  {author} {\bibfnamefont {A.}~\bibnamefont {Imambekov}}, \bibinfo {author}
  {\bibfnamefont {D.~A.}\ \bibnamefont {Abanin}}, \ and\ \bibinfo {author}
  {\bibfnamefont {E.}~\bibnamefont {Demler}},\ }\href@noop {} {\bibfield
  {journal} {\bibinfo  {journal} {Phys. Rev. X}\ }\textbf {\bibinfo {volume}
  {2}},\ \bibinfo {pages} {041020} (\bibinfo {year} {2012})}\BibitemShut
  {NoStop}%
\bibitem [{\citenamefont {Anderson}(1967)}]{Anderson1967}%
  \BibitemOpen
  \bibfield  {author} {\bibinfo {author} {\bibfnamefont {P.~W.}\ \bibnamefont
  {Anderson}},\ }\href {\doibase 10.1103/PhysRevLett.18.1049} {\bibfield
  {journal} {\bibinfo  {journal} {Phys. Rev. Lett.}\ }\textbf {\bibinfo
  {volume} {18}},\ \bibinfo {pages} {1049} (\bibinfo {year}
  {1967})}\BibitemShut {NoStop}%
\bibitem [{\citenamefont {Mistakidis}\ \emph
  {et~al.}(2019{\natexlab{e}})\citenamefont {Mistakidis}, \citenamefont
  {Grusdt}, \citenamefont {Koutentakis},\ and\ \citenamefont
  {Schmelcher}}]{mistakidis2019dissipative}%
  \BibitemOpen
  \bibfield  {author} {\bibinfo {author} {\bibfnamefont {S.~I.}\ \bibnamefont
  {Mistakidis}}, \bibinfo {author} {\bibfnamefont {F.}~\bibnamefont {Grusdt}},
  \bibinfo {author} {\bibfnamefont {G.~M.}\ \bibnamefont {Koutentakis}}, \ and\
  \bibinfo {author} {\bibfnamefont {P.}~\bibnamefont {Schmelcher}},\
  }\href@noop {} {\bibfield  {journal} {\bibinfo  {journal} {New J. Phys.}\
  }\textbf {\bibinfo {volume} {21}},\ \bibinfo {pages} {103026} (\bibinfo
  {year} {2019}{\natexlab{e}})}\BibitemShut {NoStop}%
\bibitem [{\citenamefont {Lausch}\ \emph {et~al.}(2018)\citenamefont {Lausch},
  \citenamefont {Widera},\ and\ \citenamefont
  {Fleischhauer}}]{lausch2018prethermalization}%
  \BibitemOpen
  \bibfield  {author} {\bibinfo {author} {\bibfnamefont {T.}~\bibnamefont
  {Lausch}}, \bibinfo {author} {\bibfnamefont {A.}~\bibnamefont {Widera}}, \
  and\ \bibinfo {author} {\bibfnamefont {M.}~\bibnamefont {Fleischhauer}},\
  }\href@noop {} {\bibfield  {journal} {\bibinfo  {journal} {Phys. Rev. A}\
  }\textbf {\bibinfo {volume} {97}},\ \bibinfo {pages} {023621} (\bibinfo
  {year} {2018})}\BibitemShut {NoStop}%
\bibitem [{\citenamefont {Burovski}\ \emph {et~al.}(2014)\citenamefont
  {Burovski}, \citenamefont {Cheianov}, \citenamefont {Gamayun},\ and\
  \citenamefont {Lychkovskiy}}]{burovski2014momentum}%
  \BibitemOpen
  \bibfield  {author} {\bibinfo {author} {\bibfnamefont {E.}~\bibnamefont
  {Burovski}}, \bibinfo {author} {\bibfnamefont {V.}~\bibnamefont {Cheianov}},
  \bibinfo {author} {\bibfnamefont {O.}~\bibnamefont {Gamayun}}, \ and\
  \bibinfo {author} {\bibfnamefont {O.}~\bibnamefont {Lychkovskiy}},\
  }\href@noop {} {\bibfield  {journal} {\bibinfo  {journal} {Phys. Rev. A}\
  }\textbf {\bibinfo {volume} {89}},\ \bibinfo {pages} {041601} (\bibinfo
  {year} {2014})}\BibitemShut {NoStop}%
\bibitem [{\citenamefont {Lychkovskiy}\ \emph {et~al.}(2018)\citenamefont
  {Lychkovskiy}, \citenamefont {Gamayun},\ and\ \citenamefont
  {Cheianov}}]{lychkovskiy2018necessary}%
  \BibitemOpen
  \bibfield  {author} {\bibinfo {author} {\bibfnamefont {O.}~\bibnamefont
  {Lychkovskiy}}, \bibinfo {author} {\bibfnamefont {O.}~\bibnamefont
  {Gamayun}}, \ and\ \bibinfo {author} {\bibfnamefont {V.}~\bibnamefont
  {Cheianov}},\ }\href@noop {} {\bibfield  {journal} {\bibinfo  {journal}
  {Phys. Rev. B}\ }\textbf {\bibinfo {volume} {98}},\ \bibinfo {pages} {024307}
  (\bibinfo {year} {2018})}\BibitemShut {NoStop}%
\bibitem [{\citenamefont {Meinert}\ \emph {et~al.}(2017)\citenamefont
  {Meinert}, \citenamefont {Knap}, \citenamefont {Kirilov}, \citenamefont
  {Jag-Lauber}, \citenamefont {Zvonarev}, \citenamefont {Demler},\ and\
  \citenamefont {N{\"a}gerl}}]{meinert2017bloch}%
  \BibitemOpen
  \bibfield  {author} {\bibinfo {author} {\bibfnamefont {F.}~\bibnamefont
  {Meinert}}, \bibinfo {author} {\bibfnamefont {M.}~\bibnamefont {Knap}},
  \bibinfo {author} {\bibfnamefont {E.}~\bibnamefont {Kirilov}}, \bibinfo
  {author} {\bibfnamefont {K.}~\bibnamefont {Jag-Lauber}}, \bibinfo {author}
  {\bibfnamefont {M.~B.}\ \bibnamefont {Zvonarev}}, \bibinfo {author}
  {\bibfnamefont {E.}~\bibnamefont {Demler}}, \ and\ \bibinfo {author}
  {\bibfnamefont {H.-C.}\ \bibnamefont {N{\"a}gerl}},\ }\href@noop {}
  {\bibfield  {journal} {\bibinfo  {journal} {Science}\ }\textbf {\bibinfo
  {volume} {356}},\ \bibinfo {pages} {945} (\bibinfo {year}
  {2017})}\BibitemShut {NoStop}%
\bibitem [{\citenamefont {Knap}\ \emph {et~al.}(2014)\citenamefont {Knap},
  \citenamefont {Mathy}, \citenamefont {Ganahl}, \citenamefont {Zvonarev},\
  and\ \citenamefont {Demler}}]{knap2014quantum}%
  \BibitemOpen
  \bibfield  {author} {\bibinfo {author} {\bibfnamefont {M.}~\bibnamefont
  {Knap}}, \bibinfo {author} {\bibfnamefont {C.~J.}\ \bibnamefont {Mathy}},
  \bibinfo {author} {\bibfnamefont {M.}~\bibnamefont {Ganahl}}, \bibinfo
  {author} {\bibfnamefont {M.~B.}\ \bibnamefont {Zvonarev}}, \ and\ \bibinfo
  {author} {\bibfnamefont {E.}~\bibnamefont {Demler}},\ }\href@noop {}
  {\bibfield  {journal} {\bibinfo  {journal} {Phys. Rev. Lett.}\ }\textbf
  {\bibinfo {volume} {112}},\ \bibinfo {pages} {015302} (\bibinfo {year}
  {2014})}\BibitemShut {NoStop}%
\bibitem [{\citenamefont {Gamayun}\ \emph {et~al.}(2018)\citenamefont
  {Gamayun}, \citenamefont {Lychkovskiy}, \citenamefont {Burovski},
  \citenamefont {Malcomson}, \citenamefont {Cheianov},\ and\ \citenamefont
  {Zvonarev}}]{gamayun2018impact}%
  \BibitemOpen
  \bibfield  {author} {\bibinfo {author} {\bibfnamefont {O.}~\bibnamefont
  {Gamayun}}, \bibinfo {author} {\bibfnamefont {O.}~\bibnamefont
  {Lychkovskiy}}, \bibinfo {author} {\bibfnamefont {E.}~\bibnamefont
  {Burovski}}, \bibinfo {author} {\bibfnamefont {M.}~\bibnamefont {Malcomson}},
  \bibinfo {author} {\bibfnamefont {V.~V.}\ \bibnamefont {Cheianov}}, \ and\
  \bibinfo {author} {\bibfnamefont {M.~B.}\ \bibnamefont {Zvonarev}},\
  }\href@noop {} {\bibfield  {journal} {\bibinfo  {journal} {Phys. Rev. Lett.}\
  }\textbf {\bibinfo {volume} {120}},\ \bibinfo {pages} {220605} (\bibinfo
  {year} {2018})}\BibitemShut {NoStop}%
\bibitem [{\citenamefont {Cai}\ \emph {et~al.}(2010)\citenamefont {Cai},
  \citenamefont {Wang}, \citenamefont {Xie},\ and\ \citenamefont
  {Wang}}]{cai2010interaction}%
  \BibitemOpen
  \bibfield  {author} {\bibinfo {author} {\bibfnamefont {Z.}~\bibnamefont
  {Cai}}, \bibinfo {author} {\bibfnamefont {L.}~\bibnamefont {Wang}}, \bibinfo
  {author} {\bibfnamefont {X.}~\bibnamefont {Xie}}, \ and\ \bibinfo {author}
  {\bibfnamefont {Y.}~\bibnamefont {Wang}},\ }\href@noop {} {\bibfield
  {journal} {\bibinfo  {journal} {Phys. Rev. A}\ }\textbf {\bibinfo {volume}
  {81}},\ \bibinfo {pages} {043602} (\bibinfo {year} {2010})}\BibitemShut
  {NoStop}%
\bibitem [{\citenamefont {Johnson}\ \emph {et~al.}(2011)\citenamefont
  {Johnson}, \citenamefont {Clark}, \citenamefont {Bruderer},\ and\
  \citenamefont {Jaksch}}]{Johnson2011}%
  \BibitemOpen
  \bibfield  {author} {\bibinfo {author} {\bibfnamefont {T.~H.}\ \bibnamefont
  {Johnson}}, \bibinfo {author} {\bibfnamefont {S.~R.}\ \bibnamefont {Clark}},
  \bibinfo {author} {\bibfnamefont {M.}~\bibnamefont {Bruderer}}, \ and\
  \bibinfo {author} {\bibfnamefont {D.}~\bibnamefont {Jaksch}},\ }\href
  {\doibase 10.1103/PhysRevA.84.023617} {\bibfield  {journal} {\bibinfo
  {journal} {Phys. Rev. A}\ }\textbf {\bibinfo {volume} {84}},\ \bibinfo
  {pages} {023617} (\bibinfo {year} {2011})}\BibitemShut {NoStop}%
\bibitem [{\citenamefont {Siegl}\ \emph {et~al.}(2018)\citenamefont {Siegl},
  \citenamefont {Mistakidis},\ and\ \citenamefont {Schmelcher}}]{Siegl2018}%
  \BibitemOpen
  \bibfield  {author} {\bibinfo {author} {\bibfnamefont {P.}~\bibnamefont
  {Siegl}}, \bibinfo {author} {\bibfnamefont {S.~I.}\ \bibnamefont
  {Mistakidis}}, \ and\ \bibinfo {author} {\bibfnamefont {P.}~\bibnamefont
  {Schmelcher}},\ }\href {\doibase 10.1103/PhysRevA.97.053626} {\bibfield
  {journal} {\bibinfo  {journal} {Phys. Rev. A}\ }\textbf {\bibinfo {volume}
  {97}},\ \bibinfo {pages} {053626} (\bibinfo {year} {2018})}\BibitemShut
  {NoStop}%
\bibitem [{\citenamefont {Theel}\ \emph {et~al.}(2020)\citenamefont {Theel},
  \citenamefont {Keiler}, \citenamefont {Mistakidis},\ and\ \citenamefont
  {Schmelcher}}]{theel2020entanglement}%
  \BibitemOpen
  \bibfield  {author} {\bibinfo {author} {\bibfnamefont {F.}~\bibnamefont
  {Theel}}, \bibinfo {author} {\bibfnamefont {K.}~\bibnamefont {Keiler}},
  \bibinfo {author} {\bibfnamefont {S.~I.}\ \bibnamefont {Mistakidis}}, \ and\
  \bibinfo {author} {\bibfnamefont {P.}~\bibnamefont {Schmelcher}},\
  }\href@noop {} {\bibfield  {journal} {\bibinfo  {journal} {New J. Phys.}\
  }\textbf {\bibinfo {volume} {22}},\ \bibinfo {pages} {023027} (\bibinfo
  {year} {2020})}\BibitemShut {NoStop}%
\bibitem [{\citenamefont {Barfknecht}\ \emph {et~al.}(2019)\citenamefont
  {Barfknecht}, \citenamefont {Foerster},\ and\ \citenamefont
  {Zinner}}]{barfknecht2019dynamics}%
  \BibitemOpen
  \bibfield  {author} {\bibinfo {author} {\bibfnamefont {R.~E.}\ \bibnamefont
  {Barfknecht}}, \bibinfo {author} {\bibfnamefont {A.}~\bibnamefont
  {Foerster}}, \ and\ \bibinfo {author} {\bibfnamefont {N.~T.}\ \bibnamefont
  {Zinner}},\ }\href@noop {} {\bibfield  {journal} {\bibinfo  {journal} {Sci.
  Rep.}\ }\textbf {\bibinfo {volume} {9}},\ \bibinfo {pages} {1} (\bibinfo
  {year} {2019})}\BibitemShut {NoStop}%
\bibitem [{\citenamefont {Mistakidis}\ \emph {et~al.}(2018)\citenamefont
  {Mistakidis}, \citenamefont {Katsimiga}, \citenamefont {Kevrekidis},\ and\
  \citenamefont {Schmelcher}}]{Mistakidis2018}%
  \BibitemOpen
  \bibfield  {author} {\bibinfo {author} {\bibfnamefont {S.~I.}\ \bibnamefont
  {Mistakidis}}, \bibinfo {author} {\bibfnamefont {G.~C.}\ \bibnamefont
  {Katsimiga}}, \bibinfo {author} {\bibfnamefont {P.~G.}\ \bibnamefont
  {Kevrekidis}}, \ and\ \bibinfo {author} {\bibfnamefont {P.}~\bibnamefont
  {Schmelcher}},\ }\href {\doibase 10.1088/1367-2630/aabc6a} {\bibfield
  {journal} {\bibinfo  {journal} {New J. Phys.}\ }\textbf {\bibinfo {volume}
  {20}},\ \bibinfo {pages} {043052} (\bibinfo {year} {2018})}\BibitemShut
  {NoStop}%
\bibitem [{\citenamefont {Erdmann}\ \emph {et~al.}(2019)\citenamefont
  {Erdmann}, \citenamefont {Mistakidis},\ and\ \citenamefont
  {Schmelcher}}]{erdmann2019phase}%
  \BibitemOpen
  \bibfield  {author} {\bibinfo {author} {\bibfnamefont {J.}~\bibnamefont
  {Erdmann}}, \bibinfo {author} {\bibfnamefont {S.~I.}\ \bibnamefont
  {Mistakidis}}, \ and\ \bibinfo {author} {\bibfnamefont {P.}~\bibnamefont
  {Schmelcher}},\ }\href@noop {} {\bibfield  {journal} {\bibinfo  {journal}
  {Phys. Rev. A}\ }\textbf {\bibinfo {volume} {99}},\ \bibinfo {pages} {013605}
  (\bibinfo {year} {2019})}\BibitemShut {NoStop}%
\bibitem [{\citenamefont {P{\k{e}}cak}\ \emph {et~al.}(2016)\citenamefont
  {P{\k{e}}cak}, \citenamefont {Gajda},\ and\ \citenamefont
  {Sowi{\'n}ski}}]{pecak2016two}%
  \BibitemOpen
  \bibfield  {author} {\bibinfo {author} {\bibfnamefont {D.}~\bibnamefont
  {P{\k{e}}cak}}, \bibinfo {author} {\bibfnamefont {M.}~\bibnamefont {Gajda}},
  \ and\ \bibinfo {author} {\bibfnamefont {T.}~\bibnamefont {Sowi{\'n}ski}},\
  }\href@noop {} {\bibfield  {journal} {\bibinfo  {journal} {New J. Phys.}\
  }\textbf {\bibinfo {volume} {18}},\ \bibinfo {pages} {013030} (\bibinfo
  {year} {2016})}\BibitemShut {NoStop}%
\bibitem [{\citenamefont {Hamner}\ \emph {et~al.}(2011)\citenamefont {Hamner},
  \citenamefont {Chang}, \citenamefont {Engels},\ and\ \citenamefont
  {Hoefer}}]{hamner2011generation}%
  \BibitemOpen
  \bibfield  {author} {\bibinfo {author} {\bibfnamefont {C.}~\bibnamefont
  {Hamner}}, \bibinfo {author} {\bibfnamefont {J.~J.}\ \bibnamefont {Chang}},
  \bibinfo {author} {\bibfnamefont {P.}~\bibnamefont {Engels}}, \ and\ \bibinfo
  {author} {\bibfnamefont {M.~A.}\ \bibnamefont {Hoefer}},\ }\href@noop {}
  {\bibfield  {journal} {\bibinfo  {journal} {Phys. Rev. Lett.}\ }\textbf
  {\bibinfo {volume} {106}},\ \bibinfo {pages} {065302} (\bibinfo {year}
  {2011})}\BibitemShut {NoStop}%
\bibitem [{\citenamefont {Weller}\ \emph {et~al.}(2008)\citenamefont {Weller},
  \citenamefont {Ronzheimer}, \citenamefont {Gross}, \citenamefont {Esteve},
  \citenamefont {Oberthaler}, \citenamefont {Frantzeskakis}, \citenamefont
  {Theocharis},\ and\ \citenamefont {Kevrekidis}}]{weller2008experimental}%
  \BibitemOpen
  \bibfield  {author} {\bibinfo {author} {\bibfnamefont {A.}~\bibnamefont
  {Weller}}, \bibinfo {author} {\bibfnamefont {J.~P.}\ \bibnamefont
  {Ronzheimer}}, \bibinfo {author} {\bibfnamefont {C.}~\bibnamefont {Gross}},
  \bibinfo {author} {\bibfnamefont {J.}~\bibnamefont {Esteve}}, \bibinfo
  {author} {\bibfnamefont {M.~K.}\ \bibnamefont {Oberthaler}}, \bibinfo
  {author} {\bibfnamefont {D.~J.}\ \bibnamefont {Frantzeskakis}}, \bibinfo
  {author} {\bibfnamefont {G.}~\bibnamefont {Theocharis}}, \ and\ \bibinfo
  {author} {\bibfnamefont {P.~G.}\ \bibnamefont {Kevrekidis}},\ }\href@noop {}
  {\bibfield  {journal} {\bibinfo  {journal} {Phys. Rev. Lett.}\ }\textbf
  {\bibinfo {volume} {101}},\ \bibinfo {pages} {130401} (\bibinfo {year}
  {2008})}\BibitemShut {NoStop}%
\bibitem [{\citenamefont {Kevrekidis}\ \emph {et~al.}(2007)\citenamefont
  {Kevrekidis}, \citenamefont {Frantzeskakis},\ and\ \citenamefont
  {Carretero-Gonz{\'a}lez}}]{kevrekidis2007emergent}%
  \BibitemOpen
  \bibfield  {author} {\bibinfo {author} {\bibfnamefont {P.~G.}\ \bibnamefont
  {Kevrekidis}}, \bibinfo {author} {\bibfnamefont {D.~J.}\ \bibnamefont
  {Frantzeskakis}}, \ and\ \bibinfo {author} {\bibfnamefont {R.}~\bibnamefont
  {Carretero-Gonz{\'a}lez}},\ }\href@noop {} {\emph {\bibinfo {title} {Emergent
  nonlinear phenomena in Bose-Einstein condensates: theory and experiment}}},\
  Vol.~\bibinfo {volume} {45}\ (\bibinfo  {publisher} {Springer Science \&
  Business Media},\ \bibinfo {year} {2007})\BibitemShut {NoStop}%
\bibitem [{\citenamefont {Cao}\ \emph {et~al.}(2017{\natexlab{a}})\citenamefont
  {Cao}, \citenamefont {Bolsinger}, \citenamefont {Mistakidis}, \citenamefont
  {Koutentakis}, \citenamefont {Kr{\"o}nke}, \citenamefont {Schurer},\ and\
  \citenamefont {Schmelcher}}]{cao2017unified}%
  \BibitemOpen
  \bibfield  {author} {\bibinfo {author} {\bibfnamefont {L.}~\bibnamefont
  {Cao}}, \bibinfo {author} {\bibfnamefont {V.}~\bibnamefont {Bolsinger}},
  \bibinfo {author} {\bibfnamefont {S.~I.}\ \bibnamefont {Mistakidis}},
  \bibinfo {author} {\bibfnamefont {G.~M.}\ \bibnamefont {Koutentakis}},
  \bibinfo {author} {\bibfnamefont {S.}~\bibnamefont {Kr{\"o}nke}}, \bibinfo
  {author} {\bibfnamefont {J.}~\bibnamefont {Schurer}}, \ and\ \bibinfo
  {author} {\bibfnamefont {P.}~\bibnamefont {Schmelcher}},\ }\href@noop {}
  {\bibfield  {journal} {\bibinfo  {journal} {J. Chem. Phys.}\ }\textbf
  {\bibinfo {volume} {147}},\ \bibinfo {pages} {044106} (\bibinfo {year}
  {2017}{\natexlab{a}})}\BibitemShut {NoStop}%
\bibitem [{\citenamefont {Cao}\ \emph {et~al.}(2013)\citenamefont {Cao},
  \citenamefont {Kr{\"o}nke}, \citenamefont {Vendrell},\ and\ \citenamefont
  {Schmelcher}}]{cao2013multi}%
  \BibitemOpen
  \bibfield  {author} {\bibinfo {author} {\bibfnamefont {L.}~\bibnamefont
  {Cao}}, \bibinfo {author} {\bibfnamefont {S.}~\bibnamefont {Kr{\"o}nke}},
  \bibinfo {author} {\bibfnamefont {O.}~\bibnamefont {Vendrell}}, \ and\
  \bibinfo {author} {\bibfnamefont {P.}~\bibnamefont {Schmelcher}},\
  }\href@noop {} {\bibfield  {journal} {\bibinfo  {journal} {J. Chem. Phys.}\
  }\textbf {\bibinfo {volume} {139}},\ \bibinfo {pages} {134103} (\bibinfo
  {year} {2013})}\BibitemShut {NoStop}%
\bibitem [{\citenamefont {Kr{\"o}nke}\ \emph {et~al.}(2013)\citenamefont
  {Kr{\"o}nke}, \citenamefont {Cao}, \citenamefont {Vendrell},\ and\
  \citenamefont {Schmelcher}}]{kronke2013non}%
  \BibitemOpen
  \bibfield  {author} {\bibinfo {author} {\bibfnamefont {S.}~\bibnamefont
  {Kr{\"o}nke}}, \bibinfo {author} {\bibfnamefont {L.}~\bibnamefont {Cao}},
  \bibinfo {author} {\bibfnamefont {O.}~\bibnamefont {Vendrell}}, \ and\
  \bibinfo {author} {\bibfnamefont {P.}~\bibnamefont {Schmelcher}},\
  }\href@noop {} {\bibfield  {journal} {\bibinfo  {journal} {New J. Phys.}\
  }\textbf {\bibinfo {volume} {15}},\ \bibinfo {pages} {063018} (\bibinfo
  {year} {2013})}\BibitemShut {NoStop}%
\bibitem [{\citenamefont {Mistakidis}\ \emph
  {et~al.}(2019{\natexlab{f}})\citenamefont {Mistakidis}, \citenamefont
  {Koutentakis}, \citenamefont {Katsimiga}, \citenamefont {Busch},\ and\
  \citenamefont {Schmelcher}}]{mistakidis2019many}%
  \BibitemOpen
  \bibfield  {author} {\bibinfo {author} {\bibfnamefont {S.~I.}\ \bibnamefont
  {Mistakidis}}, \bibinfo {author} {\bibfnamefont {G.~M.}\ \bibnamefont
  {Koutentakis}}, \bibinfo {author} {\bibfnamefont {G.~C.}\ \bibnamefont
  {Katsimiga}}, \bibinfo {author} {\bibfnamefont {T.}~\bibnamefont {Busch}}, \
  and\ \bibinfo {author} {\bibfnamefont {P.}~\bibnamefont {Schmelcher}},\
  }\href@noop {} {\bibfield  {journal} {\bibinfo  {journal}
  {\textbf{arXiv:1911.02011}}\ } (\bibinfo {year}
  {2019}{\natexlab{f}})}\BibitemShut {NoStop}%
\bibitem [{\citenamefont {Mistakidis}\ \emph
  {et~al.}(2019{\natexlab{g}})\citenamefont {Mistakidis}, \citenamefont
  {Volosniev},\ and\ \citenamefont {Schmelcher}}]{mistakidis2019induced}%
  \BibitemOpen
  \bibfield  {author} {\bibinfo {author} {\bibfnamefont {S.~I.}\ \bibnamefont
  {Mistakidis}}, \bibinfo {author} {\bibfnamefont {A.~G.}\ \bibnamefont
  {Volosniev}}, \ and\ \bibinfo {author} {\bibfnamefont {P.}~\bibnamefont
  {Schmelcher}},\ }\href@noop {} {\bibfield  {journal} {\bibinfo  {journal}
  {arXiv:\textbf{1911.05353}}\ } (\bibinfo {year}
  {2019}{\natexlab{g}})}\BibitemShut {NoStop}%
\bibitem [{\citenamefont {Cetina}\ \emph
  {et~al.}(2016{\natexlab{b}})\citenamefont {Cetina}, \citenamefont {Jag},
  \citenamefont {Lous}, \citenamefont {Fritsche}, \citenamefont {Walraven},
  \citenamefont {Grimm}, \citenamefont {Levinsen}, \citenamefont {Parish},
  \citenamefont {Schmidt}, \citenamefont {Knap},\ and\ \citenamefont
  {Demler}}]{cetina2016}%
  \BibitemOpen
  \bibfield  {author} {\bibinfo {author} {\bibfnamefont {M.}~\bibnamefont
  {Cetina}}, \bibinfo {author} {\bibfnamefont {M.}~\bibnamefont {Jag}},
  \bibinfo {author} {\bibfnamefont {R.~S.}\ \bibnamefont {Lous}}, \bibinfo
  {author} {\bibfnamefont {I.}~\bibnamefont {Fritsche}}, \bibinfo {author}
  {\bibfnamefont {J.~T.~M.}\ \bibnamefont {Walraven}}, \bibinfo {author}
  {\bibfnamefont {R.}~\bibnamefont {Grimm}}, \bibinfo {author} {\bibfnamefont
  {J.}~\bibnamefont {Levinsen}}, \bibinfo {author} {\bibfnamefont {M.~M.}\
  \bibnamefont {Parish}}, \bibinfo {author} {\bibfnamefont {R.}~\bibnamefont
  {Schmidt}}, \bibinfo {author} {\bibfnamefont {M.}~\bibnamefont {Knap}}, \
  and\ \bibinfo {author} {\bibfnamefont {E.}~\bibnamefont {Demler}},\ }\href
  {\doibase 10.1126/science.aaf5134} {\bibfield  {journal} {\bibinfo  {journal}
  {Science}\ }\textbf {\bibinfo {volume} {354}},\ \bibinfo {pages} {96}
  (\bibinfo {year} {2016}{\natexlab{b}})}\BibitemShut {NoStop}%
\bibitem [{\citenamefont {Grimm}\ \emph {et~al.}(2000)\citenamefont {Grimm},
  \citenamefont {Weidem{\"u}ller},\ and\ \citenamefont
  {Ovchinnikov}}]{grimm2000}%
  \BibitemOpen
  \bibfield  {author} {\bibinfo {author} {\bibfnamefont {R.}~\bibnamefont
  {Grimm}}, \bibinfo {author} {\bibfnamefont {M.}~\bibnamefont
  {Weidem{\"u}ller}}, \ and\ \bibinfo {author} {\bibfnamefont {Y.~B.}\
  \bibnamefont {Ovchinnikov}},\ }in\ \href {\doibase
  10.1016/S1049-250X(08)60186-X} {\emph {\bibinfo {booktitle} {Advances {{In
  Atomic}}, {{Molecular}}, and {{Optical Physics}}}}},\ Vol.~\bibinfo {volume}
  {42},\ \bibinfo {editor} {edited by\ \bibinfo {editor} {\bibfnamefont
  {B.}~\bibnamefont {Bederson}}\ and\ \bibinfo {editor} {\bibfnamefont
  {H.}~\bibnamefont {Walther}}}\ (\bibinfo  {publisher} {{Academic Press}},\
  \bibinfo {year} {2000})\ pp.\ \bibinfo {pages} {95--170}\BibitemShut
  {NoStop}%
\bibitem [{\citenamefont {Pethick}\ and\ \citenamefont
  {Smith}(2008)}]{pethick2008}%
  \BibitemOpen
  \bibfield  {author} {\bibinfo {author} {\bibfnamefont {C.~J.}\ \bibnamefont
  {Pethick}}\ and\ \bibinfo {author} {\bibfnamefont {H.}~\bibnamefont
  {Smith}},\ }\href@noop {} {\emph {\bibinfo {title} {Bose--Einstein
  condensation in dilute gases}}}\ (\bibinfo  {publisher} {Cambridge university
  press},\ \bibinfo {year} {2008})\BibitemShut {NoStop}%
\bibitem [{\citenamefont {Pitaevskii}\ and\ \citenamefont
  {Stringari}(2003)}]{pitaevskii2003}%
  \BibitemOpen
  \bibfield  {author} {\bibinfo {author} {\bibfnamefont {L.}~\bibnamefont
  {Pitaevskii}}\ and\ \bibinfo {author} {\bibfnamefont {S.}~\bibnamefont
  {Stringari}},\ }\href@noop {} {\emph {\bibinfo {title} {Bose-{{Einstein
  Condensation}}}}}\ (\bibinfo  {publisher} {{Oxford Science Publications}},\
  \bibinfo {address} {{Oxford}},\ \bibinfo {year} {2003})\BibitemShut {NoStop}%
\bibitem [{\citenamefont {Lewenstein}\ \emph {et~al.}(2012)\citenamefont
  {Lewenstein}, \citenamefont {Sanpera},\ and\ \citenamefont
  {Ahufinger}}]{lewenstein2012}%
  \BibitemOpen
  \bibfield  {author} {\bibinfo {author} {\bibfnamefont {M.}~\bibnamefont
  {Lewenstein}}, \bibinfo {author} {\bibfnamefont {A.}~\bibnamefont {Sanpera}},
  \ and\ \bibinfo {author} {\bibfnamefont {V.}~\bibnamefont {Ahufinger}},\
  }\href@noop {} {\emph {\bibinfo {title} {Ultracold Atoms in Optical Lattices:
  Simulating quantum many-body systems}}}\ (\bibinfo  {publisher} {Oxford
  University Press},\ \bibinfo {year} {2012})\BibitemShut {NoStop}%
\bibitem [{\citenamefont {Olshanii}(1998)}]{Olshanii1998}%
  \BibitemOpen
  \bibfield  {author} {\bibinfo {author} {\bibfnamefont {M.}~\bibnamefont
  {Olshanii}},\ }\href {\doibase 10.1103/PhysRevLett.81.938} {\bibfield
  {journal} {\bibinfo  {journal} {Phys. Rev. Lett.}\ }\textbf {\bibinfo
  {volume} {81}},\ \bibinfo {pages} {938} (\bibinfo {year} {1998})}\BibitemShut
  {NoStop}%
\bibitem [{\citenamefont {K{\"o}hler}\ \emph {et~al.}(2006)\citenamefont
  {K{\"o}hler}, \citenamefont {G{\'o}ral},\ and\ \citenamefont
  {Julienne}}]{kohler2006}%
  \BibitemOpen
  \bibfield  {author} {\bibinfo {author} {\bibfnamefont {T.}~\bibnamefont
  {K{\"o}hler}}, \bibinfo {author} {\bibfnamefont {K.}~\bibnamefont
  {G{\'o}ral}}, \ and\ \bibinfo {author} {\bibfnamefont {P.~S.}\ \bibnamefont
  {Julienne}},\ }\href {\doibase 10.1103/RevModPhys.78.1311} {\bibfield
  {journal} {\bibinfo  {journal} {Rev. Mod. Phys.}\ }\textbf {\bibinfo {volume}
  {78}},\ \bibinfo {pages} {1311} (\bibinfo {year} {2006})}\BibitemShut
  {NoStop}%
\bibitem [{\citenamefont {Kim}\ \emph {et~al.}(2006)\citenamefont {Kim},
  \citenamefont {Melezhik},\ and\ \citenamefont {Schmelcher}}]{kim2006}%
  \BibitemOpen
  \bibfield  {author} {\bibinfo {author} {\bibfnamefont {J.~I.}\ \bibnamefont
  {Kim}}, \bibinfo {author} {\bibfnamefont {V.~S.}\ \bibnamefont {Melezhik}}, \
  and\ \bibinfo {author} {\bibfnamefont {P.}~\bibnamefont {Schmelcher}},\
  }\href {\doibase 10.1103/PhysRevLett.97.193203} {\bibfield  {journal}
  {\bibinfo  {journal} {Phys. Rev. Lett.}\ }\textbf {\bibinfo {volume} {97}},\
  \bibinfo {pages} {193203} (\bibinfo {year} {2006})}\BibitemShut {NoStop}%
\bibitem [{\citenamefont {Bersano}\ \emph {et~al.}(2018)\citenamefont
  {Bersano}, \citenamefont {Gokhroo}, \citenamefont {Khamehchi}, \citenamefont
  {D’Ambroise}, \citenamefont {Frantzeskakis}, \citenamefont {Engels},\ and\
  \citenamefont {Kevrekidis}}]{bersano2018three}%
  \BibitemOpen
  \bibfield  {author} {\bibinfo {author} {\bibfnamefont {T.}~\bibnamefont
  {Bersano}}, \bibinfo {author} {\bibfnamefont {V.}~\bibnamefont {Gokhroo}},
  \bibinfo {author} {\bibfnamefont {M.}~\bibnamefont {Khamehchi}}, \bibinfo
  {author} {\bibfnamefont {J.}~\bibnamefont {D’Ambroise}}, \bibinfo {author}
  {\bibfnamefont {D.}~\bibnamefont {Frantzeskakis}}, \bibinfo {author}
  {\bibfnamefont {P.}~\bibnamefont {Engels}}, \ and\ \bibinfo {author}
  {\bibfnamefont {P.}~\bibnamefont {Kevrekidis}},\ }\href@noop {} {\bibfield
  {journal} {\bibinfo  {journal} {Phys. Rev. Lett.}\ }\textbf {\bibinfo
  {volume} {120}},\ \bibinfo {pages} {063202} (\bibinfo {year}
  {2018})}\BibitemShut {NoStop}%
\bibitem [{\citenamefont {Katsimiga}\ \emph {et~al.}(2020)\citenamefont
  {Katsimiga}, \citenamefont {Mistakidis}, \citenamefont {Bersano},
  \citenamefont {Ome}, \citenamefont {Mossman}, \citenamefont {Mukherjee},
  \citenamefont {Schmelcher}, \citenamefont {Engels},\ and\ \citenamefont
  {Kevrekidis}}]{katsimiga2020observation}%
  \BibitemOpen
  \bibfield  {author} {\bibinfo {author} {\bibfnamefont {G.~C.}\ \bibnamefont
  {Katsimiga}}, \bibinfo {author} {\bibfnamefont {S.~I.}\ \bibnamefont
  {Mistakidis}}, \bibinfo {author} {\bibfnamefont {T.~M.}\ \bibnamefont
  {Bersano}}, \bibinfo {author} {\bibfnamefont {M.~K.~H.}\ \bibnamefont {Ome}},
  \bibinfo {author} {\bibfnamefont {S.~M.}\ \bibnamefont {Mossman}}, \bibinfo
  {author} {\bibfnamefont {K.}~\bibnamefont {Mukherjee}}, \bibinfo {author}
  {\bibfnamefont {P.}~\bibnamefont {Schmelcher}}, \bibinfo {author}
  {\bibfnamefont {P.}~\bibnamefont {Engels}}, \ and\ \bibinfo {author}
  {\bibfnamefont {P.~G.}\ \bibnamefont {Kevrekidis}},\ }\href@noop {}
  {\bibfield  {journal} {\bibinfo  {journal} {arXiv preprint arXiv:2003.00259}\
  } (\bibinfo {year} {2020})}\BibitemShut {NoStop}%
\bibitem [{\citenamefont {Tylutki}\ \emph {et~al.}(2017)\citenamefont
  {Tylutki}, \citenamefont {Astrakharchik},\ and\ \citenamefont
  {Recati}}]{tylutki2017}%
  \BibitemOpen
  \bibfield  {author} {\bibinfo {author} {\bibfnamefont {M.}~\bibnamefont
  {Tylutki}}, \bibinfo {author} {\bibfnamefont {G.~E.}\ \bibnamefont
  {Astrakharchik}}, \ and\ \bibinfo {author} {\bibfnamefont {A.}~\bibnamefont
  {Recati}},\ }\href {\doibase 10.1103/PhysRevA.96.063603} {\bibfield
  {journal} {\bibinfo  {journal} {Phys. Rev. A}\ }\textbf {\bibinfo {volume}
  {96}},\ \bibinfo {pages} {063603} (\bibinfo {year} {2017})}\BibitemShut
  {NoStop}%
\bibitem [{\citenamefont {Horodecki}\ \emph {et~al.}(2009)\citenamefont
  {Horodecki}, \citenamefont {Horodecki}, \citenamefont {Horodecki},\ and\
  \citenamefont {Horodecki}}]{horodecki2009}%
  \BibitemOpen
  \bibfield  {author} {\bibinfo {author} {\bibfnamefont {R.}~\bibnamefont
  {Horodecki}}, \bibinfo {author} {\bibfnamefont {P.}~\bibnamefont
  {Horodecki}}, \bibinfo {author} {\bibfnamefont {M.}~\bibnamefont
  {Horodecki}}, \ and\ \bibinfo {author} {\bibfnamefont {K.}~\bibnamefont
  {Horodecki}},\ }\href {\doibase 10.1103/RevModPhys.81.865} {\bibfield
  {journal} {\bibinfo  {journal} {Rev. Mod. Phys.}\ }\textbf {\bibinfo {volume}
  {81}},\ \bibinfo {pages} {865} (\bibinfo {year} {2009})}\BibitemShut
  {NoStop}%
\bibitem [{\citenamefont {Roncaglia}\ \emph {et~al.}(2014)\citenamefont
  {Roncaglia}, \citenamefont {Montorsi},\ and\ \citenamefont
  {Genovese}}]{roncaglia2014}%
  \BibitemOpen
  \bibfield  {author} {\bibinfo {author} {\bibfnamefont {M.}~\bibnamefont
  {Roncaglia}}, \bibinfo {author} {\bibfnamefont {A.}~\bibnamefont {Montorsi}},
  \ and\ \bibinfo {author} {\bibfnamefont {M.}~\bibnamefont {Genovese}},\
  }\href {\doibase 10.1103/PhysRevA.90.062303} {\bibfield  {journal} {\bibinfo
  {journal} {Phys. Rev. A}\ }\textbf {\bibinfo {volume} {90}},\ \bibinfo
  {pages} {062303} (\bibinfo {year} {2014})}\BibitemShut {NoStop}%
\bibitem [{\citenamefont {Giorgini}\ \emph {et~al.}(2008)\citenamefont
  {Giorgini}, \citenamefont {Pitaevskii},\ and\ \citenamefont
  {Stringari}}]{giorgini2008}%
  \BibitemOpen
  \bibfield  {author} {\bibinfo {author} {\bibfnamefont {S.}~\bibnamefont
  {Giorgini}}, \bibinfo {author} {\bibfnamefont {L.~P.}\ \bibnamefont
  {Pitaevskii}}, \ and\ \bibinfo {author} {\bibfnamefont {S.}~\bibnamefont
  {Stringari}},\ }\href {\doibase 10.1103/RevModPhys.80.1215} {\bibfield
  {journal} {\bibinfo  {journal} {Rev. Mod. Phys.}\ }\textbf {\bibinfo {volume}
  {80}},\ \bibinfo {pages} {1215} (\bibinfo {year} {2008})}\BibitemShut
  {NoStop}%
\bibitem [{\citenamefont {K{\"o}hler}\ \emph {et~al.}(2019)\citenamefont
  {K{\"o}hler}, \citenamefont {Keiler}, \citenamefont {Mistakidis},
  \citenamefont {Meyer},\ and\ \citenamefont
  {Schmelcher}}]{kohler2019dynamical}%
  \BibitemOpen
  \bibfield  {author} {\bibinfo {author} {\bibfnamefont {F.}~\bibnamefont
  {K{\"o}hler}}, \bibinfo {author} {\bibfnamefont {K.}~\bibnamefont {Keiler}},
  \bibinfo {author} {\bibfnamefont {S.~I.}\ \bibnamefont {Mistakidis}},
  \bibinfo {author} {\bibfnamefont {H.-D.}\ \bibnamefont {Meyer}}, \ and\
  \bibinfo {author} {\bibfnamefont {P.}~\bibnamefont {Schmelcher}},\
  }\href@noop {} {\bibfield  {journal} {\bibinfo  {journal} {J. Chem. Phys.}\
  }\textbf {\bibinfo {volume} {151}},\ \bibinfo {pages} {054108} (\bibinfo
  {year} {2019})}\BibitemShut {NoStop}%
\bibitem [{\citenamefont {Frenkel}(1932)}]{frenkel1932}%
  \BibitemOpen
  \bibfield  {author} {\bibinfo {author} {\bibfnamefont {J.}~\bibnamefont
  {Frenkel}},\ }\href@noop {} {\emph {\bibinfo {title} {Wave {{Mechanics}}:
  {{Elementary Theory}}}}}\ (\bibinfo  {publisher} {{Clarendon Press Oxford}},\
  \bibinfo {address} {{Oxford}},\ \bibinfo {year} {1932})\BibitemShut {NoStop}%
\bibitem [{\citenamefont {Dirac}(1930)}]{dirac1930}%
  \BibitemOpen
  \bibfield  {author} {\bibinfo {author} {\bibfnamefont {P.~A.}\ \bibnamefont
  {Dirac}},\ }\href@noop {} {\bibfield  {journal} {\bibinfo  {journal} {Proc.
  Camb. Phil. Soc.}\ }\textbf {\bibinfo {volume} {26}} (\bibinfo {year}
  {1930})}\BibitemShut {NoStop}%
\bibitem [{\citenamefont {Koutentakis}\ \emph {et~al.}(2019)\citenamefont
  {Koutentakis}, \citenamefont {Mistakidis},\ and\ \citenamefont
  {Schmelcher}}]{Koutentakis2019}%
  \BibitemOpen
  \bibfield  {author} {\bibinfo {author} {\bibfnamefont {G.~M.}\ \bibnamefont
  {Koutentakis}}, \bibinfo {author} {\bibfnamefont {S.~I.}\ \bibnamefont
  {Mistakidis}}, \ and\ \bibinfo {author} {\bibfnamefont {P.}~\bibnamefont
  {Schmelcher}},\ }\href {\doibase 10.1088/1367-2630/ab14ba} {\bibfield
  {journal} {\bibinfo  {journal} {New J. Phys.}\ }\textbf {\bibinfo {volume}
  {21}},\ \bibinfo {pages} {053005} (\bibinfo {year} {2019})}\BibitemShut
  {NoStop}%
\bibitem [{\citenamefont {Sakmann}\ \emph {et~al.}(2008)\citenamefont
  {Sakmann}, \citenamefont {Streltsov}, \citenamefont {Alon},\ and\
  \citenamefont {Cederbaum}}]{sakmann2008}%
  \BibitemOpen
  \bibfield  {author} {\bibinfo {author} {\bibfnamefont {K.}~\bibnamefont
  {Sakmann}}, \bibinfo {author} {\bibfnamefont {A.~I.}\ \bibnamefont
  {Streltsov}}, \bibinfo {author} {\bibfnamefont {O.~E.}\ \bibnamefont {Alon}},
  \ and\ \bibinfo {author} {\bibfnamefont {L.~S.}\ \bibnamefont {Cederbaum}},\
  }\href {\doibase 10.1103/PhysRevA.78.023615} {\bibfield  {journal} {\bibinfo
  {journal} {Phys. Rev. A}\ }\textbf {\bibinfo {volume} {78}},\ \bibinfo
  {pages} {023615} (\bibinfo {year} {2008})}\BibitemShut {NoStop}%
\bibitem [{\citenamefont {Naraschewski}\ and\ \citenamefont
  {Glauber}(1999)}]{naraschewski1999}%
  \BibitemOpen
  \bibfield  {author} {\bibinfo {author} {\bibfnamefont {M.}~\bibnamefont
  {Naraschewski}}\ and\ \bibinfo {author} {\bibfnamefont {R.~J.}\ \bibnamefont
  {Glauber}},\ }\href {\doibase 10.1103/PhysRevA.59.4595} {\bibfield  {journal}
  {\bibinfo  {journal} {Phys. Rev. A}\ }\textbf {\bibinfo {volume} {59}},\
  \bibinfo {pages} {4595} (\bibinfo {year} {1999})}\BibitemShut {NoStop}%
\bibitem [{\citenamefont {Tavares}\ \emph {et~al.}(2017)\citenamefont
  {Tavares}, \citenamefont {Fritsch}, \citenamefont {Telles}, \citenamefont
  {Hussein}, \citenamefont {Impens}, \citenamefont {Kaiser},\ and\
  \citenamefont {Bagnato}}]{tavares2017}%
  \BibitemOpen
  \bibfield  {author} {\bibinfo {author} {\bibfnamefont {P.~E.}\ \bibnamefont
  {Tavares}}, \bibinfo {author} {\bibfnamefont {A.~R.}\ \bibnamefont
  {Fritsch}}, \bibinfo {author} {\bibfnamefont {G.~D.}\ \bibnamefont {Telles}},
  \bibinfo {author} {\bibfnamefont {M.~S.}\ \bibnamefont {Hussein}}, \bibinfo
  {author} {\bibfnamefont {F.}~\bibnamefont {Impens}}, \bibinfo {author}
  {\bibfnamefont {R.}~\bibnamefont {Kaiser}}, \ and\ \bibinfo {author}
  {\bibfnamefont {V.~S.}\ \bibnamefont {Bagnato}},\ }\href@noop {} {\bibfield
  {journal} {\bibinfo  {journal} {Proc. Nat. Acad. Sc.}\ }\textbf {\bibinfo
  {volume} {114}},\ \bibinfo {pages} {12691} (\bibinfo {year}
  {2017})}\BibitemShut {NoStop}%
\bibitem [{\citenamefont {Nguyen}\ \emph {et~al.}(2019)\citenamefont {Nguyen},
  \citenamefont {Tsatsos}, \citenamefont {Luo}, \citenamefont {Lode},
  \citenamefont {Telles}, \citenamefont {Bagnato},\ and\ \citenamefont
  {Hulet}}]{nguyen2019}%
  \BibitemOpen
  \bibfield  {author} {\bibinfo {author} {\bibfnamefont {J.~H.~V.}\
  \bibnamefont {Nguyen}}, \bibinfo {author} {\bibfnamefont {M.~C.}\
  \bibnamefont {Tsatsos}}, \bibinfo {author} {\bibfnamefont {D.}~\bibnamefont
  {Luo}}, \bibinfo {author} {\bibfnamefont {A.~U.~J.}\ \bibnamefont {Lode}},
  \bibinfo {author} {\bibfnamefont {G.~D.}\ \bibnamefont {Telles}}, \bibinfo
  {author} {\bibfnamefont {V.~S.}\ \bibnamefont {Bagnato}}, \ and\ \bibinfo
  {author} {\bibfnamefont {R.~G.}\ \bibnamefont {Hulet}},\ }\href {\doibase
  10.1103/PhysRevX.9.011052} {\bibfield  {journal} {\bibinfo  {journal} {Phys.
  Rev. X}\ }\textbf {\bibinfo {volume} {9}},\ \bibinfo {pages} {011052}
  (\bibinfo {year} {2019})}\BibitemShut {NoStop}%
\bibitem [{\citenamefont {Fasshauer}\ and\ \citenamefont
  {Lode}(2016)}]{fasshauer2016}%
  \BibitemOpen
  \bibfield  {author} {\bibinfo {author} {\bibfnamefont {E.}~\bibnamefont
  {Fasshauer}}\ and\ \bibinfo {author} {\bibfnamefont {A.~U.~J.}\ \bibnamefont
  {Lode}},\ }\href {\doibase 10.1103/PhysRevA.93.033635} {\bibfield  {journal}
  {\bibinfo  {journal} {Phys. Rev. A}\ }\textbf {\bibinfo {volume} {93}},\
  \bibinfo {pages} {033635} (\bibinfo {year} {2016})}\BibitemShut {NoStop}%
\bibitem [{\citenamefont {Yu}\ \emph {et~al.}(2009)\citenamefont {Yu},
  \citenamefont {Bruun},\ and\ \citenamefont {Baym}}]{yu2009}%
  \BibitemOpen
  \bibfield  {author} {\bibinfo {author} {\bibfnamefont {Z.}~\bibnamefont
  {Yu}}, \bibinfo {author} {\bibfnamefont {G.~M.}\ \bibnamefont {Bruun}}, \
  and\ \bibinfo {author} {\bibfnamefont {G.}~\bibnamefont {Baym}},\ }\href
  {\doibase 10.1103/PhysRevA.80.023615} {\bibfield  {journal} {\bibinfo
  {journal} {Phys. Rev. A}\ }\textbf {\bibinfo {volume} {80}},\ \bibinfo
  {pages} {023615} (\bibinfo {year} {2009})}\BibitemShut {NoStop}%
\bibitem [{\citenamefont {Catani}\ \emph
  {et~al.}(2009{\natexlab{b}})\citenamefont {Catani}, \citenamefont
  {Barontini}, \citenamefont {Lamporesi}, \citenamefont {Rabatti},
  \citenamefont {Thalhammer}, \citenamefont {Minardi}, \citenamefont
  {Stringari},\ and\ \citenamefont {Inguscio}}]{catani2009}%
  \BibitemOpen
  \bibfield  {author} {\bibinfo {author} {\bibfnamefont {J.}~\bibnamefont
  {Catani}}, \bibinfo {author} {\bibfnamefont {G.}~\bibnamefont {Barontini}},
  \bibinfo {author} {\bibfnamefont {G.}~\bibnamefont {Lamporesi}}, \bibinfo
  {author} {\bibfnamefont {F.}~\bibnamefont {Rabatti}}, \bibinfo {author}
  {\bibfnamefont {G.}~\bibnamefont {Thalhammer}}, \bibinfo {author}
  {\bibfnamefont {F.}~\bibnamefont {Minardi}}, \bibinfo {author} {\bibfnamefont
  {S.}~\bibnamefont {Stringari}}, \ and\ \bibinfo {author} {\bibfnamefont
  {M.}~\bibnamefont {Inguscio}},\ }\href {\doibase
  10.1103/PhysRevLett.103.140401} {\bibfield  {journal} {\bibinfo  {journal}
  {Phys. Rev. Lett.}\ }\textbf {\bibinfo {volume} {103}},\ \bibinfo {pages}
  {140401} (\bibinfo {year} {2009}{\natexlab{b}})}\BibitemShut {NoStop}%
\bibitem [{\citenamefont {Erdmann}\ \emph {et~al.}(2018)\citenamefont
  {Erdmann}, \citenamefont {Mistakidis},\ and\ \citenamefont
  {Schmelcher}}]{erdmann2018}%
  \BibitemOpen
  \bibfield  {author} {\bibinfo {author} {\bibfnamefont {J.}~\bibnamefont
  {Erdmann}}, \bibinfo {author} {\bibfnamefont {S.~I.}\ \bibnamefont
  {Mistakidis}}, \ and\ \bibinfo {author} {\bibfnamefont {P.}~\bibnamefont
  {Schmelcher}},\ }\href {\doibase 10.1103/PhysRevA.98.053614} {\bibfield
  {journal} {\bibinfo  {journal} {Phys. Rev. A}\ }\textbf {\bibinfo {volume}
  {98}},\ \bibinfo {pages} {053614} (\bibinfo {year} {2018})}\BibitemShut
  {NoStop}%
\bibitem [{\citenamefont {Abraham}\ and\ \citenamefont
  {Bonitz}(2014)}]{abraham2014}%
  \BibitemOpen
  \bibfield  {author} {\bibinfo {author} {\bibfnamefont {J.~W.}\ \bibnamefont
  {Abraham}}\ and\ \bibinfo {author} {\bibfnamefont {M.}~\bibnamefont
  {Bonitz}},\ }\href {\doibase 10.1002/ctpp.201300066} {\bibfield  {journal}
  {\bibinfo  {journal} {Contr. Pl. Phys.}\ }\textbf {\bibinfo {volume} {54}},\
  \bibinfo {pages} {27} (\bibinfo {year} {2014})}\BibitemShut {NoStop}%
\bibitem [{\citenamefont {Koutentakis}\ \emph {et~al.}(2017)\citenamefont
  {Koutentakis}, \citenamefont {Mistakidis},\ and\ \citenamefont
  {Schmelcher}}]{koutentakis2017}%
  \BibitemOpen
  \bibfield  {author} {\bibinfo {author} {\bibfnamefont {G.~M.}\ \bibnamefont
  {Koutentakis}}, \bibinfo {author} {\bibfnamefont {S.~I.}\ \bibnamefont
  {Mistakidis}}, \ and\ \bibinfo {author} {\bibfnamefont {P.}~\bibnamefont
  {Schmelcher}},\ }\href {\doibase 10.1103/PhysRevA.95.013617} {\bibfield
  {journal} {\bibinfo  {journal} {Phys. Rev. A}\ }\textbf {\bibinfo {volume}
  {95}},\ \bibinfo {pages} {013617} (\bibinfo {year} {2017})}\BibitemShut
  {NoStop}%
\bibitem [{\citenamefont {Ronzheimer}\ \emph {et~al.}(2013)\citenamefont
  {Ronzheimer}, \citenamefont {Schreiber}, \citenamefont {Braun}, \citenamefont
  {Hodgman}, \citenamefont {Langer}, \citenamefont {McCulloch}, \citenamefont
  {{Heidrich-Meisner}}, \citenamefont {Bloch},\ and\ \citenamefont
  {Schneider}}]{ronzheimer2013}%
  \BibitemOpen
  \bibfield  {author} {\bibinfo {author} {\bibfnamefont {J.~P.}\ \bibnamefont
  {Ronzheimer}}, \bibinfo {author} {\bibfnamefont {M.}~\bibnamefont
  {Schreiber}}, \bibinfo {author} {\bibfnamefont {S.}~\bibnamefont {Braun}},
  \bibinfo {author} {\bibfnamefont {S.~S.}\ \bibnamefont {Hodgman}}, \bibinfo
  {author} {\bibfnamefont {S.}~\bibnamefont {Langer}}, \bibinfo {author}
  {\bibfnamefont {I.~P.}\ \bibnamefont {McCulloch}}, \bibinfo {author}
  {\bibfnamefont {F.}~\bibnamefont {{Heidrich-Meisner}}}, \bibinfo {author}
  {\bibfnamefont {I.}~\bibnamefont {Bloch}}, \ and\ \bibinfo {author}
  {\bibfnamefont {U.}~\bibnamefont {Schneider}},\ }\href {\doibase
  10.1103/PhysRevLett.110.205301} {\bibfield  {journal} {\bibinfo  {journal}
  {Phys. Rev. Lett.}\ }\textbf {\bibinfo {volume} {110}},\ \bibinfo {pages}
  {205301} (\bibinfo {year} {2013})}\BibitemShut {NoStop}%
\bibitem [{\citenamefont {Abraham}\ \emph
  {et~al.}(2012{\natexlab{a}})\citenamefont {Abraham}, \citenamefont {Balzer},
  \citenamefont {Hochstuhl},\ and\ \citenamefont {Bonitz}}]{abraham2012}%
  \BibitemOpen
  \bibfield  {author} {\bibinfo {author} {\bibfnamefont {J.~W.}\ \bibnamefont
  {Abraham}}, \bibinfo {author} {\bibfnamefont {K.}~\bibnamefont {Balzer}},
  \bibinfo {author} {\bibfnamefont {D.}~\bibnamefont {Hochstuhl}}, \ and\
  \bibinfo {author} {\bibfnamefont {M.}~\bibnamefont {Bonitz}},\ }\href
  {\doibase 10.1103/PhysRevB.86.125112} {\bibfield  {journal} {\bibinfo
  {journal} {Phys. Rev. B}\ }\textbf {\bibinfo {volume} {86}},\ \bibinfo
  {pages} {125112} (\bibinfo {year} {2012}{\natexlab{a}})}\BibitemShut
  {NoStop}%
\bibitem [{\citenamefont {Bauch}\ \emph {et~al.}(2010)\citenamefont {Bauch},
  \citenamefont {Hochstuhl}, \citenamefont {Balzer},\ and\ \citenamefont
  {Bonitz}}]{bauch2010quantum}%
  \BibitemOpen
  \bibfield  {author} {\bibinfo {author} {\bibfnamefont {S.}~\bibnamefont
  {Bauch}}, \bibinfo {author} {\bibfnamefont {D.}~\bibnamefont {Hochstuhl}},
  \bibinfo {author} {\bibfnamefont {K.}~\bibnamefont {Balzer}}, \ and\ \bibinfo
  {author} {\bibfnamefont {M.}~\bibnamefont {Bonitz}},\ }in\ \href@noop {}
  {\emph {\bibinfo {booktitle} {J. Phys.: Conf. Series}}},\ Vol.\ \bibinfo
  {volume} {220}\ (\bibinfo {organization} {IOP Publishing},\ \bibinfo {year}
  {2010})\ p.\ \bibinfo {pages} {012013}\BibitemShut {NoStop}%
\bibitem [{\citenamefont {Bauch}\ \emph {et~al.}(2009)\citenamefont {Bauch},
  \citenamefont {Balzer}, \citenamefont {Henning},\ and\ \citenamefont
  {Bonitz}}]{bauch2009quantum}%
  \BibitemOpen
  \bibfield  {author} {\bibinfo {author} {\bibfnamefont {S.}~\bibnamefont
  {Bauch}}, \bibinfo {author} {\bibfnamefont {K.}~\bibnamefont {Balzer}},
  \bibinfo {author} {\bibfnamefont {C.}~\bibnamefont {Henning}}, \ and\
  \bibinfo {author} {\bibfnamefont {M.}~\bibnamefont {Bonitz}},\ }\href@noop {}
  {\bibfield  {journal} {\bibinfo  {journal} {Phys. Rev. B}\ }\textbf {\bibinfo
  {volume} {80}},\ \bibinfo {pages} {054515} (\bibinfo {year}
  {2009})}\BibitemShut {NoStop}%
\bibitem [{\citenamefont {Abraham}\ \emph
  {et~al.}(2012{\natexlab{b}})\citenamefont {Abraham}, \citenamefont {Balzer},
  \citenamefont {Hochstuhl},\ and\ \citenamefont
  {Bonitz}}]{abraham2012quantum}%
  \BibitemOpen
  \bibfield  {author} {\bibinfo {author} {\bibfnamefont {J.~W.}\ \bibnamefont
  {Abraham}}, \bibinfo {author} {\bibfnamefont {K.}~\bibnamefont {Balzer}},
  \bibinfo {author} {\bibfnamefont {D.}~\bibnamefont {Hochstuhl}}, \ and\
  \bibinfo {author} {\bibfnamefont {M.}~\bibnamefont {Bonitz}},\ }\href@noop {}
  {\bibfield  {journal} {\bibinfo  {journal} {Phys. Rev. B}\ }\textbf {\bibinfo
  {volume} {86}},\ \bibinfo {pages} {125112} (\bibinfo {year}
  {2012}{\natexlab{b}})}\BibitemShut {NoStop}%
\bibitem [{\citenamefont {Hara}\ \emph {et~al.}(2011)\citenamefont {Hara},
  \citenamefont {Takasu}, \citenamefont {Yamaoka}, \citenamefont {Doyle},\ and\
  \citenamefont {Takahashi}}]{hara2011quantum}%
  \BibitemOpen
  \bibfield  {author} {\bibinfo {author} {\bibfnamefont {H.}~\bibnamefont
  {Hara}}, \bibinfo {author} {\bibfnamefont {Y.}~\bibnamefont {Takasu}},
  \bibinfo {author} {\bibfnamefont {Y.}~\bibnamefont {Yamaoka}}, \bibinfo
  {author} {\bibfnamefont {J.~M.}\ \bibnamefont {Doyle}}, \ and\ \bibinfo
  {author} {\bibfnamefont {Y.}~\bibnamefont {Takahashi}},\ }\href@noop {}
  {\bibfield  {journal} {\bibinfo  {journal} {Phys. Rev. Lett.}\ }\textbf
  {\bibinfo {volume} {106}},\ \bibinfo {pages} {205304} (\bibinfo {year}
  {2011})}\BibitemShut {NoStop}%
\bibitem [{\citenamefont {Khramov}\ \emph {et~al.}(2014)\citenamefont
  {Khramov}, \citenamefont {Hansen}, \citenamefont {Dowd}, \citenamefont {Roy},
  \citenamefont {Makrides}, \citenamefont {Petrov}, \citenamefont
  {Kotochigova},\ and\ \citenamefont {Gupta}}]{khramov2014ultracold}%
  \BibitemOpen
  \bibfield  {author} {\bibinfo {author} {\bibfnamefont {A.}~\bibnamefont
  {Khramov}}, \bibinfo {author} {\bibfnamefont {A.}~\bibnamefont {Hansen}},
  \bibinfo {author} {\bibfnamefont {W.}~\bibnamefont {Dowd}}, \bibinfo {author}
  {\bibfnamefont {R.~J.}\ \bibnamefont {Roy}}, \bibinfo {author} {\bibfnamefont
  {C.}~\bibnamefont {Makrides}}, \bibinfo {author} {\bibfnamefont
  {A.}~\bibnamefont {Petrov}}, \bibinfo {author} {\bibfnamefont
  {S.}~\bibnamefont {Kotochigova}}, \ and\ \bibinfo {author} {\bibfnamefont
  {S.}~\bibnamefont {Gupta}},\ }\href@noop {} {\bibfield  {journal} {\bibinfo
  {journal} {Phys. Rev. Lett.}\ }\textbf {\bibinfo {volume} {112}},\ \bibinfo
  {pages} {033201} (\bibinfo {year} {2014})}\BibitemShut {NoStop}%
\bibitem [{\citenamefont {W{\l}odzy{\'n}ski}\ \emph {et~al.}(2020)\citenamefont
  {W{\l}odzy{\'n}ski}, \citenamefont {P{\k{e}}cak},\ and\ \citenamefont
  {Sowi{\'n}ski}}]{wlodzynski2020geometry}%
  \BibitemOpen
  \bibfield  {author} {\bibinfo {author} {\bibfnamefont {D.}~\bibnamefont
  {W{\l}odzy{\'n}ski}}, \bibinfo {author} {\bibfnamefont {D.}~\bibnamefont
  {P{\k{e}}cak}}, \ and\ \bibinfo {author} {\bibfnamefont {T.}~\bibnamefont
  {Sowi{\'n}ski}},\ }\href@noop {} {\bibfield  {journal} {\bibinfo  {journal}
  {Phys. Rev. A}\ }\textbf {\bibinfo {volume} {101}},\ \bibinfo {pages}
  {023604} (\bibinfo {year} {2020})}\BibitemShut {NoStop}%
\bibitem [{\citenamefont {P{\k{e}}cak}\ and\ \citenamefont
  {Sowi{\'n}ski}(2019)}]{pkecak2019intercomponent}%
  \BibitemOpen
  \bibfield  {author} {\bibinfo {author} {\bibfnamefont {D.}~\bibnamefont
  {P{\k{e}}cak}}\ and\ \bibinfo {author} {\bibfnamefont {T.}~\bibnamefont
  {Sowi{\'n}ski}},\ }\href@noop {} {\bibfield  {journal} {\bibinfo  {journal}
  {Phys. Rev. A}\ }\textbf {\bibinfo {volume} {99}},\ \bibinfo {pages} {043612}
  (\bibinfo {year} {2019})}\BibitemShut {NoStop}%
\bibitem [{\citenamefont {Tajima}\ and\ \citenamefont
  {Uchino}(2018)}]{tajima2018many}%
  \BibitemOpen
  \bibfield  {author} {\bibinfo {author} {\bibfnamefont {H.}~\bibnamefont
  {Tajima}}\ and\ \bibinfo {author} {\bibfnamefont {S.}~\bibnamefont
  {Uchino}},\ }\href@noop {} {\bibfield  {journal} {\bibinfo  {journal} {New J.
  Phys.}\ }\textbf {\bibinfo {volume} {20}},\ \bibinfo {pages} {073048}
  (\bibinfo {year} {2018})}\BibitemShut {NoStop}%
\bibitem [{\citenamefont {Field}\ \emph {et~al.}(2020)\citenamefont {Field},
  \citenamefont {Levinsen},\ and\ \citenamefont {Parish}}]{field2020fate}%
  \BibitemOpen
  \bibfield  {author} {\bibinfo {author} {\bibfnamefont {B.}~\bibnamefont
  {Field}}, \bibinfo {author} {\bibfnamefont {J.}~\bibnamefont {Levinsen}}, \
  and\ \bibinfo {author} {\bibfnamefont {M.~M.}\ \bibnamefont {Parish}},\
  }\href@noop {} {\bibfield  {journal} {\bibinfo  {journal} {Phys. Rev. A}\
  }\textbf {\bibinfo {volume} {101}},\ \bibinfo {pages} {013623} (\bibinfo
  {year} {2020})}\BibitemShut {NoStop}%
\bibitem [{\citenamefont {Cao}\ \emph {et~al.}(2017{\natexlab{b}})\citenamefont
  {Cao}, \citenamefont {Mistakidis}, \citenamefont {Deng},\ and\ \citenamefont
  {Schmelcher}}]{cao2017a}%
  \BibitemOpen
  \bibfield  {author} {\bibinfo {author} {\bibfnamefont {L.}~\bibnamefont
  {Cao}}, \bibinfo {author} {\bibfnamefont {S.~I.}\ \bibnamefont {Mistakidis}},
  \bibinfo {author} {\bibfnamefont {X.}~\bibnamefont {Deng}}, \ and\ \bibinfo
  {author} {\bibfnamefont {P.}~\bibnamefont {Schmelcher}},\ }\href {\doibase
  10.1016/j.chemphys.2016.08.026} {\bibfield  {journal} {\bibinfo  {journal}
  {Chem. Phys.}\ }\textbf {\bibinfo {volume} {482}},\ \bibinfo {pages} {303}
  (\bibinfo {year} {2017}{\natexlab{b}})}\BibitemShut {NoStop}%
\bibitem [{\citenamefont {Kiehn}\ \emph {et~al.}(2019)\citenamefont {Kiehn},
  \citenamefont {Mistakidis}, \citenamefont {Katsimiga},\ and\ \citenamefont
  {Schmelcher}}]{kiehn2019spontaneous}%
  \BibitemOpen
  \bibfield  {author} {\bibinfo {author} {\bibfnamefont {H.}~\bibnamefont
  {Kiehn}}, \bibinfo {author} {\bibfnamefont {S.~I.}\ \bibnamefont
  {Mistakidis}}, \bibinfo {author} {\bibfnamefont {G.~C.}\ \bibnamefont
  {Katsimiga}}, \ and\ \bibinfo {author} {\bibfnamefont {P.}~\bibnamefont
  {Schmelcher}},\ }\href@noop {} {\bibfield  {journal} {\bibinfo  {journal}
  {Phys. Rev. A}\ }\textbf {\bibinfo {volume} {100}},\ \bibinfo {pages}
  {023613} (\bibinfo {year} {2019})}\BibitemShut {NoStop}%
\bibitem [{\citenamefont {Scott}\ \emph {et~al.}(1998)\citenamefont {Scott},
  \citenamefont {Ballagh},\ and\ \citenamefont {Burnett}}]{scott1998formation}%
  \BibitemOpen
  \bibfield  {author} {\bibinfo {author} {\bibfnamefont {T.~F.}\ \bibnamefont
  {Scott}}, \bibinfo {author} {\bibfnamefont {R.~J.}\ \bibnamefont {Ballagh}},
  \ and\ \bibinfo {author} {\bibfnamefont {K.}~\bibnamefont {Burnett}},\
  }\href@noop {} {\bibfield  {journal} {\bibinfo  {journal} {J. Phys. B: At.
  Mol. and Opt. Phys.}\ }\textbf {\bibinfo {volume} {31}},\ \bibinfo {pages}
  {L329} (\bibinfo {year} {1998})}\BibitemShut {NoStop}%
\end{thebibliography}%

\end{document}